\documentclass[twocolumn,trackchanges]{aastex701}
\usepackage{xcolor}
\usepackage[utf8]{inputenc}
\DeclareUnicodeCharacter{2212}{\ensuremath{-}}
\usepackage{multirow}

\begin{document}

\title{Radar observations of Europa in 2011--2024: New insights into radar scattering properties}

\correspondingauthor{Jean-Luc Margot}
\email{jlm@epss.ucla.edu}

\author[orcid=0009-0008-8444-6988,gname='Tunhui', sname='Xie']{Tunhui Xie}
\affiliation{University of California, Los Angeles, Department of Physics \& Astronomy, Los Angeles, CA 90095}
\email[show]{tunhuixie@g.ucla.edu}  

\author[orcid=0000-0001-9798-1797,gname='Jean-Luc', sname='Margot']{Jean-Luc Margot} 
\affiliation{University of California, Los Angeles, Department of Earth, Planetary, and Space Sciences, Los Angeles, CA 90095}
\affiliation{University of California, Los Angeles, Department of Physics \& Astronomy, Los Angeles, CA 90095}
\email{jlm@epss.ucla.edu}

\author[orcid=0000-0002-8652-3704,gname='Sebastiano', sname='Padovan']{Sebastiano Padovan}
\affiliation{Radio Occultation Group, EUMETSAT, 64295 Darmstadt, Germany}
\email{sebastiano.padovan@external.eumetsat.int}

\author[gname='Frank', sname='Ghigo']{Frank Ghigo}
\affiliation{Green Bank Observatory, National Radio Astronomy Observatory, Green Bank, WV 24944}
\email{fghigo@nrao.edu}

\author[orcid=0000-0002-7045-9277,gname='Will', sname='Armentrout']{Will Armentrout}
\affiliation{Green Bank Observatory, National Radio Astronomy Observatory, Green Bank, WV 24944}
\email{warmentr@nrao.edu}

\author[gname='Joseph', sname='Jao']{Joseph S. Jao}
\affiliation{Jet Propulsion Laboratory, California Institute of Technology, Pasadena, CA, USA}
\email{joseph.s.jao@jpl.nasa.gov}

\author[gname='Jon', sname='Giorgini']{Jon D. Giorgini}
\affiliation{Jet Propulsion Laboratory, California Institute of Technology, Pasadena, CA, USA}
\email{Jon.D.Giorgini@jpl.nasa.gov}

\author[orcid=0009-0003-8984-388X,gname=Joseph,sname=Lazio]{T.~Joseph~W.~Lazio}
\affiliation{Dept.~Climate and Space Sciences and Engineering, University of Michigan, Ann Arbor, MI 48109}
\email{jlazio@umich.edu}

\begin{abstract}
Three of Jupiter’s Galilean satellites – Europa, Ganymede, and Callisto – are of particular scientific interest due to their icy shells and suspected subsurface oceans. However, the radar properties of the icy satellites have not been measured since observations in 1987–1991. Because radio waves can penetrate pure ice to considerable depths, radar observations provide a powerful means of characterizing the subsurface properties of the icy shells of these satellites, offering key insights into planetary evolution. We have observed Europa using the Goldstone 3.5-cm Solar System Radar and the Green Bank Telescope (GBT) in 2011–2024 in order to address a longstanding gap in the radar studies of these moons. In this paper, we present the most longitudinally comprehensive set of radar measurements of Europa to date and describe its disk-integrated radar properties. On the basis of monostatic Goldstone data, we find radar albedo values in two circular polarizations of $\hat\sigma_{\rm OC}$ = 0.92 $\pm$ 0.11 and $\hat\sigma_{\rm SC}$ = 1.35 $\pm$ 0.13 (unweighted mean and root-mean-square dispersion), with a circular polarization ratio of $\mu_c$ = 1.44 $\pm$ 0.12 (weighted mean and root-mean-square dispersion). Values obtained bistatically at the GBT are similar. The $\mu_c$ values suggest a leading-vs-trailing side dichotomy in radar scattering properties on Europa.  Our results support the existence of the coherent backscatter opposition effect (CBOE), currently the most widely accepted physical mechanism that explains the unusual radar scattering properties of the icy Galilean satellites. Because we observed Europa with a bistatic configuration, we can place a lower bound on the width of Europa’s CBOE peak equal to 36 arcsec, which provides an upper bound of 32~m ($\sim$1000 wavelengths) on the penetrating depth of X-band radar waves at Europa.
\end{abstract}

\section{Introduction}
\label{sec:introduction}

\begin{figure*}[ht!]
\plotone{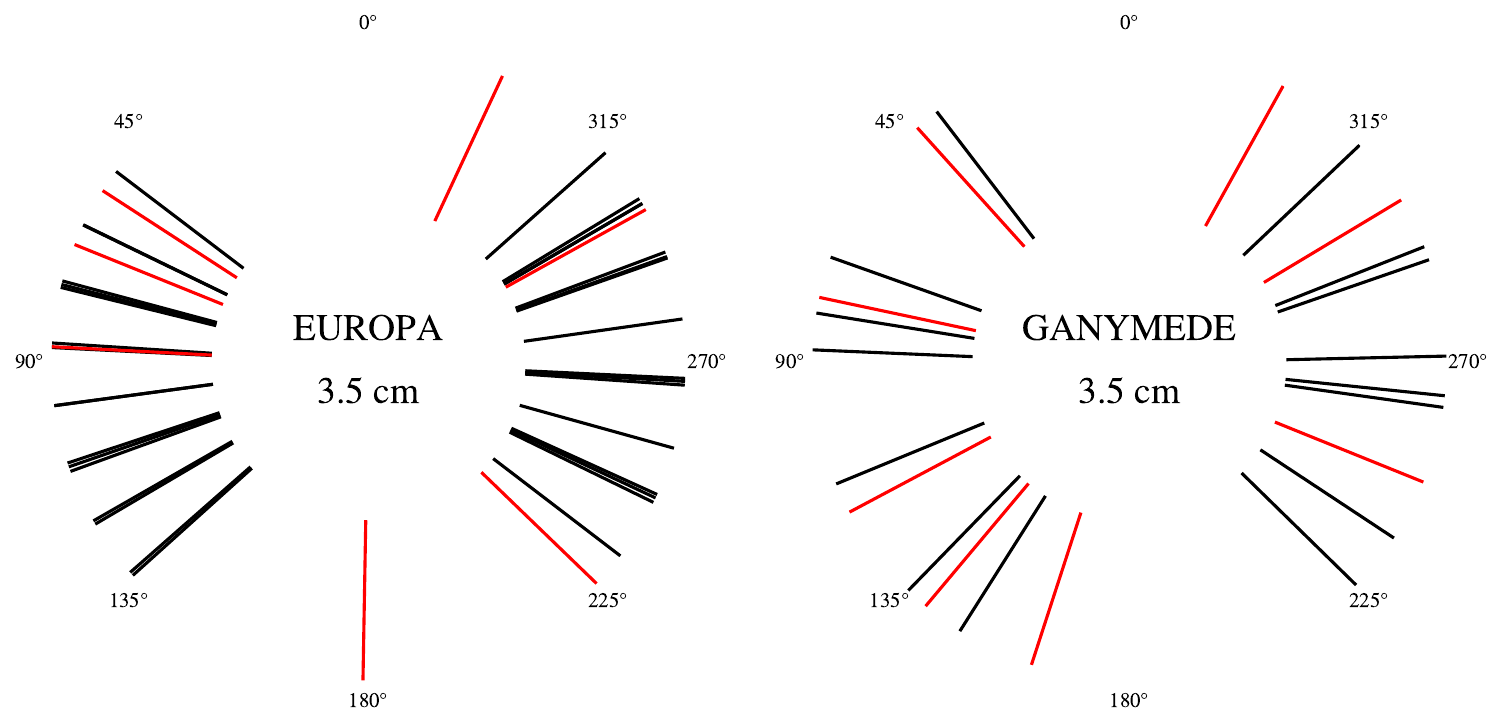}
\caption{Subradar point (west) longitudes of Europa and Ganymede at the epochs of 3.5 cm radar observations. The black lines represent our 2011--2024 observations, whereas the red lines represent the observations obtained in 1987--1991 -- the most recent and extensive set of published 3.5 cm radar results for the Galilean satellites \citep{Ostro1992} prior to this work.
\label{fig:subradarlongitudes}}
\end{figure*}

The icy Galilean satellites -- Europa, Ganymede, and Callisto -- are of particular scientific interest due to the strong likelihood of subsurface oceans beneath their crusts \citep{kive04,nimm16}. Therefore, they became prime targets in the search for extraterrestrial life, as water is a key component of life as we know it. Their surfaces also reveal fascinating geological activity such as the dark, cratered regions on Ganymede and the chaos terrain on Europa that is possibly due to water within the icy crust  \citep{voyager1_results, Schmidt2011-qr}. These geological features offer crucial insights into planetary evolution, shedding light on the history of the Solar System and the dynamic interactions between surface ice and interior oceans, which are key factors in assessing the moons' habitability. Recognizing the significance of the moons, NASA launched the Europa Clipper
mission \citep{papp24} in October 2024 to explore Europa and investigate its habitability.
Likewise, ESA launched the Jupiter Icy Moons Explorer (JUICE)
mission \citep{wita22} in April 2023 to study the icy satellites in detail.

Ground-based radar observations of the icy Galilean satellites have been conducted since 1974 \citep[e.g.,][]{camp77,camp78,ostr80,Ostro1992}.
By transmitting radio waves and measuring their backscattered echoes, radar astronomers are able to study planetary surface properties. Since radio waves can penetrate pure ice to considerable
depth \citep{Thompson1990,Warren1984}, the radar echoes of the satellites also provide insights into the subsurface structure of their icy crusts. The first few radar observations of the satellites in the 1970s revealed strikingly unusual radar signatures. Observations across multiple radio wavelengths (3.5 cm, 13 cm, and 70 cm) demonstrated that these satellites have radar albedos $\hat\sigma$ -- the radar cross section $\sigma$ normalized by projected target area -- much greater than the typical value of
$\sim$0.1 observed for terrestrial planets and asteroids. Additionally, when radar waves are circularly polarized, the circular polarization ratio $\mu_c$ -- the ratio of echo power in the same sense (SC) of circular polarization as transmitted to that in the opposite (OC) sense -- are found to exceed unity, significantly surpassing the $\sim$0.1 values recorded for terrestrial planets \citep{camp78,OSTRO1990335, Ostro1992, BLACK_70}. These extraordinary results prompted further efforts to thoroughly document and interpret the radar properties of the icy Galilean satellites. \citet{Moore2009} reviewed these efforts.

  Ground-based observations of Saturn’s icy satellites, such as Enceladus, Tethys, and Rhea, also reveal
  similarly anomalous signatures. Like the icy Galilean moons, these targets exhibit radar albedos higher than those of the terrestrial planets and circular polarization ratios exceeding unity \citep{BLACK_Saturn}.

Currently, the best physical explanation for the unusual radar signatures of the icy satellites is the coherent backscatter opposition effect (CBOE), which is the constructive interference of two waves experiencing forward scattering events and traveling through the icy medium along the same path but in opposite
directions \citep{hapk90,hapke_nature,Hofgartner2023}. The constructive interference happens when the two waves emerge from the medium traveling back towards the receiver at a transmitter-target-receiver angle ($\beta$)
near zero \citep{hapk90}. CBOE predicts that when $\beta$ equals zero, the intensity of the received light in the same circular polarization as transmitted is up to twice of what is measured at large $\beta$ angles. The CBOE enhancement in the opposite sense of circular polarization, however, is believed to be much smaller \citep{Mishchenko:92}. Therefore, the circular polarization ratio $\mu_c$ is expected to decrease with increasing $\beta$ angles and this trend has been observed in the laboratory \citep{hapke_nature}. The angular width of the CBOE peak is $\sim \lambda/2\pi L$, where $\lambda$ is the wavelength and $L$ is the
mean depth into the medium
where a photon diffuses before being absorbed.
The distance $L$ can also be approximated as $(L_SL_A/3)^{1/2}$, where $L_S$ is the elastic mean free path, or the mean distance a photon travels between scatterings, and $L_A$ is the inelastic mean free path, or the mean distance a photon travels before being absorbed \citep{mackintosh1988}. Therefore, if we know the angular width and $\lambda$, we can constrain the $L$ parameter.

The last major investigation of the radar properties of the icy Galilean satellites was a 2001 reanalysis \citep{BLACK2001167} of the 1987--1991 observations of \citet{Ostro1992}.
Between 2011 and 2024, we observed Europa and Ganymede at 3.5 cm wavelength with the Goldstone Solar System Radar, which is part of NASA's Deep Space Network (DSN) in Goldstone, California, and the Green Bank Telescope (GBT). Our observations are more numerous and cover a broader range of longitudes than previously published results for both Europa and Ganymede (Figure \ref{fig:subradarlongitudes}), allowing our study to fill a significant gap in the radar studies of the icy Galilean satellites. In this paper, we first determine Europa's radar properties, such as disk-integrated radar albedos, circular polarization ratios, and scattering law exponents.  Because of our broad rotation phase coverage, we then
examine the relationship between Europa's radar properties and geological units.
Moreover, because we observed bistatically with the GBT, we set constraints on the angular width of Europa's CBOE peak, which puts an upper bound on the mean depth photons can diffuse inside the icy shell before absorption.

Section \ref{sec:observations} describes how we made the observations. Section \ref{sec:data_reduction} describes how we reduced the data to single-date, dual-polarization power spectra. Section \ref{sec:results} presents the disk-integrated radar properties calculated from the spectra. Section \ref{sec:discussion} analyzes the results.
Section \ref{sec:conclusion} summarizes our conclusions.

\section{Observations} \label{sec:observations}

In this work, we analyzed a subset of radar observations of Europa obtained with the 70 m antenna (DSS-14) from the Deep Space Network (DSN) at Goldstone (35.24° N, $-$116.89° E) and the 100 m Green Bank Telescope in West Virginia (38.24° N, $-$79.84° E).
The subset includes all epochs for which the raw voltage data from both telescopes are available.
Transmissions were monochromatic at 8560 MHz (3.5 cm wavelength) to enable measurements of the spin states of the Galilean satellites with radar speckle tracking \citep{marg12jgr,marg21}.  We analyzed 33 epochs, ranging from 2011 to 2024. For each observation, we used DSS-14 to transmit a right circularly polarized continuous wave signal with a power of $\sim$450~kW. A time-variable Doppler correction was applied to the transmitted signal so that echoes from the sub-radar point were received at a frequency of 8560 MHz ($\lambda$ = 3.5 cm) at the GBT. A time-variable differential Doppler correction was applied to the received signal at Goldstone to also achieve reception at 8560 MHz. At both telescopes, a frequency offset of $\sim$10 kHz was introduced in the frequency downconversion chain to prevent the Doppler-broadened echo from Europa to overlap with 0 frequency (DC).  Anti-aliasing (lowpass) filters were used prior to sampling the complex voltages with 4-bit or 8-bit quantization.  The echoes were recorded with the portable fast sampling (PFS) data-taking system \citep{radar_data_taking_system} or its RADARDAS successor after 2020.   For data collected before 2020, the sampling rate was 5 MHz, and for data collected after 2020, the sampling rate was 6.25 MHz.  For both the Goldstone and GBT receivers, channels 1 and 2 were set to receive the OC and SC signals, respectively.

The geometrical circumstances at the observation epochs are listed in Table \ref{tab:observation_geometry}, including subradar point latitudes and west longitudes, round-trip light-times (RTT), limb-to-limb bandwidths $B_{\rm ll}$ (Section \ref{sec:data_reduction}), and transmitter-target-receiver $\beta$ angles for Goldstone-to-Goldstone observations and Goldstone-to-GBT observations.
      Additional observational circumstances for Goldstone-to-Goldstone and Goldstone-to-GBT observations are listed in Tables \ref{tab:Goldstone_observation} and \ref{tab:gbt_observation}, respectively, including data-taking durations, average transmitter powers, system temperatures ($\rm T_{\rm sys}$)
      and elevation angles of the transmitter and receiver. 

We approximated on-source system temperatures at the GBT by averaging the system temperatures of the radio noise baseline detected when scanning across a nearby radio source to verify the accuracy of telescope pointing with Astrid's ``peak'' procedure (\url{https://gbtdocs.readthedocs.io/en/latest/references/astrid.html}). These scans were conducted immediately prior to data collection and targeted radio source calibrators that are within five degrees of Europa. Because the observations were scheduled specifically when Europa was at large angular distances ($>$ 2--3 arcmin) from Jupiter, the radio noise contribution due to Jupiter is negligible ($<<$1 K) and the pointing scan values provide excellent estimates. However, for the 2011 NOV 07 and 2023 OCT 30 epochs, system temperature information is unavailable. These epochs, therefore, are excluded from our analysis.

On-source system temperatures at Goldstone were measured by monitoring a power meter and scaling the reading to the system temperatures determined at zenith prior to the track. The zenith system temperatures were determined by comparing the power levels obtained when the low-noise amplifier was connected to (1) the antenna pointing at zenith and (2) a dummy load of known temperature.
For the two dates (2011 NOV 16 and 23) marked with an asterisk in Table \ref{tab:Goldstone_observation}, the Goldstone on-source system temperature was not measured, and only the zenith y-factor or the zenith system temperature is available. 
For those two epochs, we estimated the on-source system temperatures using the method described in Appendix \ref{app:sysT_estimation}.

Because of its active surface, the gain of the GBT antenna is fairly constant as a function of elevation. With a 71\% aperture efficiency and a telescope gain of 2.0~K/Jy \citep[][Table 3]{gbt_proposers_guide}, we find an antenna gain of 77.57 dBi. 
At Goldstone, the antenna gain (for both transmit and receive) is elevation dependent. To account for these variations, we adopted the gain equation from the Deep Space Network Telecommunications Link Design Handbook \citep{jpl_dsn_handbook}: 
\begin{equation}
  G(\theta)=G_0-G_1(\theta-\gamma)^2-\frac{A_{\rm ZEN}}{\sin(\theta)},
 \label{eq:gain_function}
\end{equation}
where $G$ is the gain in dBi, $\theta$ is the elevation angle in degrees, $G_0$, $G_1$, and $\gamma$ are constants taken from Table A-1 of the handbook ($G_0=73.17$ dBi for transmission at 7145 MHz, $G_0=74.55$ dBi for reception at 8420 MHz, $G_1=0.000285$ dBi, and $\gamma=38.35$ degrees), and $A_{\rm ZEN}$ is the zenith atmospheric attenuation taken as 0.04, assuming a weather cumulative distribution (CD) of 0.5. A CD value of 0.0 corresponds to the lowest-loss condition of the atmosphere and a CD value of 0.90 corresponds to very cloudy weather.

We adjusted the $G_0$ value to account for the fact that the values listed in the DSN handbook apply to the telemetry receiver, which uses a different
Cassegrain feedcone than the cone used for X-band radar observations.  Illumination of the subreflector
with different horns at different locations may result in antenna gain differences.
Inquiries made to a dozen DSN antenna experts suggest that the $G_0$ value applicable to X-band radar observations is not currently known.  However, the antenna gain
used 
at the time of \citet{Ostro1992}'s measurements was based on
an ``extensive series of drift scans on point radio sources for which reliable flux densities are available''.
These authors wrote that a ``typical'' antenna gain for DSS-14 at 8560 MHz is 74.0 dBi for zenith angles less than 70 degrees.  We found that a $G_0$ value of 74.23 dBi produces an average gain of 74.0 dBi over the antenna elevation angles spanned during \citet{Ostro1992}'s observations of Europa, and we used this value to calculate the elevation-dependent antenna gain.  
The final estimated gain values for Goldstone are listed in Table \ref{tab:Goldstone_observation}.

\begin{deluxetable*}{ccrcccc}
\tabletypesize{\footnotesize}
\tablewidth{\textwidth}
\tablecaption{Geometry of the Observations \label{tab:observation_geometry}}
\tablehead{
  \colhead{Date} & \colhead{SRP} & \colhead{SRP} & \colhead{RTT} & \colhead{$B_{\rm ll}$} & \colhead{$\beta$} &  \colhead{$\beta$} \\ [-7pt]
  \colhead{} & \colhead{Lat} & \colhead{W Lon} & \colhead{} & \colhead{} & \colhead{DSN-to-DSN} & \colhead{DSN-to-GBT} \\ [-7pt]
  \colhead{(UTC)} & \colhead{(deg)} & \colhead{(deg)} & \colhead{(s)} & \colhead{(Hz)} & \colhead{(arcsec)} & \colhead{(arcsec)}
}
\startdata
2011 SEP 24 & 3.4 & 301.3 & 4126.3 & 3649.5 & 5.17  &  5.97 \\
2011 SEP 27 & 3.4 & 244.9 & 4095.4 & 3653.6 & 24.6  &  25.4 \\
2011 SEP 29 & 3.4 & 87.2 & 4081.1 & 3652.3 & 14.9  &  15.7 \\
2011 OCT 01 & 3.4 & 289.7 & 4066.5 & 3650.8 & 11.6  &  12.5 \\
2011 OCT 03 & 3.4 & 132.0 & 4047.6 & 3655.4 & 30.1  &  30.9 \\
2011 OCT 06 & 3.4 & 75.6 & 4031.7 & 3652.0 & 13.4  &  14.3 \\
2011 OCT 10 & 3.4 & 120.5 & 4005.5 & 3655.3 & 29.1  &  29.9 \\
2011 OCT 13 & 3.3 & 64.1 & 3995.4 & 3651.7 & 12.1  &  13.0 \\
2011 OCT 15 & 3.3 & 266.7 & 3984.9 & 3653.3 & 23.5  &  24.3 \\
2011 OCT 17 & 3.3 & 109.1 & 3977.1 & 3654.9 & 27.1  &  27.9 \\
2011 OCT 19 & 3.3 & 311.6 & 3975.6 & 3650.6 & 10.3  &  11.1 \\
2011 OCT 20 & 3.3 & 52.8 & 3973.1 & 3651.2 & 10.4  &  11.2 \\
2011 OCT 24 & 3.3 & 97.7 & 3963.2 & 3654.4 & 24.3  &  25.1 \\
2011 OCT 26 & 3.3 & 300.3 & 3964.3 & 3651.4 & 14.1  &  15.0 \\
2011 OCT 29 & 3.3 & 244.0 & 3960.2 & 3655.3 & 32.0  &  32.8 \\
2011 OCT 31 & 3.3 & 86.3 & 3964.3 & 3653.6 & 20.7  &  21.5 \\
2011 NOV 04 & 3.3 & 131.3 & 3967.9 & 3656.5 & 34.2  &  35.0 \\
2011 NOV 05 & 3.2 & 232.6 & 3970.4 & 3656.0 & 34.9  &  35.7 \\
2011 NOV 07 & 3.2 & 74.9 & 3980.4 & 3652.7 & 16.7  &  17.5 \\
2011 NOV 11 & 3.2 & 119.8 & 3992.4 & 3655.6 & 30.3  &  31.0 \\
2011 NOV 16 & 3.2 & 266.0 & 4020.0 & 3653.2 & 22.6  &  23.3 \\
2011 NOV 18 & 3.2 & 108.3 & 4031.5 & 3654.5 & 24.6  &  25.4 \\
2011 NOV 23 & 3.1 & 254.5 & 4067.2 & 3653.6 & 23.9  &  24.6 \\
2023 OCT 18 & 3.7 & 289.4 & 4007.8 & 3650.2 & 15.5  &  16.4 \\
2023 OCT 25 & 3.7 & 278.0 & 3983.6 & 3651.3 & 20.4  &  21.3 \\
2023 OCT 30 & 3.7 & 64.1 & 3977.5 & 3650.6 & 14.1  &  15.0 \\
2023 NOV 10 & 3.6 & 97.7 & 3985.1 & 3652.8 & 24.3  &  25.2 \\
2023 NOV 12 & 3.6 & 300.3 & 3993.5 & 3649.8 & 13.5  &  14.4 \\
2024 DEC 03 & 2.8 & 245.6 & 4080.9 & 3656.9 & 31.9  &  33.0 \\
2024 DEC 07 & 2.8 & 290.5 & 4082.9 & 3653.7 & 17.4  &  18.5 \\
2024 DEC 12 & 2.8 & 76.1 & 4087.6 & 3654.5 & 18.5  &  19.6 \\
2024 DEC 21 & 2.8 & 267.3 & 4115.3 & 3654.8 & 23.6  &  24.7 \\
2024 DEC 23 & 2.8 & 109.9 & 4123.8 & 3656.3 & 27.6  &  28.7 \\
\enddata
\tablecomments{Observation epochs in UTC, subradar point (SRP) latitudes and west longitudes in degrees, round-trip light-times (RTT)
  in seconds, limb-to-limb bandwidths ($B_{\rm ll}$) in Hertz, the transmitter-target-receiver $\beta$ angles for Goldstone-to-Goldstone observations and the $\beta$ angles for Goldstone-to-GBT observations in arcsec.}
\end{deluxetable*}

\begin{deluxetable*}{llccccccc}
  \tabletypesize{\footnotesize}
  \tablewidth{\textwidth}
\tablecaption{Goldstone-to-Goldstone Observations of Europa \label{tab:Goldstone_observation}}
\tablehead{
  \colhead{Date} & \colhead{Start - Stop Time} & \colhead{Avg Tx Power} & \colhead{$\rm T_{\rm sys}$ OC} & \colhead{$\rm T_{\rm sys}$ SC} & \colhead{${\rm El}_{\rm tx}$} & \colhead{${\rm Gain}_{\rm tx}$} & \colhead{${\rm El}_{\rm rcv}$} & \colhead{${\rm Gain}_{\rm rcv}$}\\ [-7pt]
\colhead{(UTC)} & \colhead{(UTC)} & \colhead{(kW)} & \colhead{(K)} & \colhead{(K)} & \colhead{(deg)} & \colhead{(dBi)} & \colhead{(deg)} & \colhead{(dBi)}
}
\startdata
2011 SEP 24  &  111555  -  120206 & 454.3 & 15.3  &  15.2 & 66.6 & 73.96 & 58.2 & 74.07 \\
2011 SEP 27  &  111608  -  115905 & 458.4 & 15.7  &  15.7 & 66.4 & 73.96 & 57.9 & 74.07 \\
2011 SEP 29  &  103710  -  114505 & 441.8 & 15.0  &  15.1 & 66.4 & 73.96 & 57.9 & 74.07 \\
2011 OCT 01  &  104000  -  113000 & 456.7 & 15.2  &  15.2 & 66.1 & 73.97 & 57.6 & 74.08 \\
2011 OCT 03  &  104040  -  112540 & 454.3 & 15.2  &  15.4 & 66.1 & 73.97 & 57.5 & 74.08 \\
2011 OCT 06  &  103610  -  111001 & 448.6 & 15.8  &  15.9 & 65.8 & 73.97 & 57.2 & 74.08 \\
2011 OCT 10  &  103400  -  110148 & 446.4 & 14.6  &  14.5 & 65.5 & 73.98 & 56.8 & 74.09 \\
2011 OCT 13  &  101700  -  104700 & 435.7 & 15.4  &  15.4 & 65.2 & 73.98 & 56.4 & 74.09 \\
2011 OCT 15  &  100900  -  103900 & 457.8 & 14.7  &  14.6 & 65.0 & 73.98 & 56.1 & 74.09 \\
2011 OCT 17  &  100100  -  103100 & 457.0 & 14.6  &  14.4 & 64.9 & 73.98 & 56.0 & 74.09 \\
2011 OCT 19  &  095300  -  102020 & 458.0 & 14.9  &  14.9 & 64.6 & 73.99 & 55.7 & 74.10 \\
2011 OCT 20  &  094900  -  101900 & 454.0 & 14.1  &  13.9 & 64.6 & 73.99 & 55.6 & 74.10 \\
2011 OCT 24  &  093400  -  100400 & 460.8 & 14.4  &  14.2 & 64.2 & 74.00 & 55.1 & 74.10 \\
2011 OCT 26  &  092400  -  095400 & 461.0 & 15.3  &  15.0 & 64.0 & 74.00 & 54.8 & 74.10 \\
2011 OCT 29  &  092100  -  093420 & 461.6 & 14.9  &  14.9 & 63.7 & 74.00 & 54.4 & 74.11 \\
2011 OCT 31  &  090600  -  093600 & 460.6 & 14.8  &  14.5 & 63.6 & 74.00 & 54.3 & 74.11 \\
2011 NOV 04  &  085100  -  092100 & 462.2 & 15.6  &  14.9 & 63.2 & 74.01 & 53.8 & 74.11 \\
2011 NOV 05  &  084345  -  091345 & 460.5 & 15.3  &  15.4 & 63.0 & 74.01 & 53.6 & 74.11 \\
2011 NOV 07  &  085030  -  090308 & 462.7 & 14.4  &  14.7 & 63.0 & 74.01 & 53.5 & 74.11 \\
2011 NOV 11  &  082830  -  084830 & 463.3 & 15.5  &  15.5 & 62.6 & 74.02 & 53.0 & 74.12 \\
2011 NOV 16*  &  080000  -  083000 & 462.6 & 14.9  &  14.8 & 62.2 & 74.02 & 52.3 & 74.12 \\
2011 NOV 18  &  075500  -  082459 & 462.8 & 14.3  &  14.2 & 62.1 & 74.02 & 52.3 & 74.12 \\
2011 NOV 23*  &  074100  -  075600 & 461.7 & 15.1  &  15.0 & 61.7 & 74.03 & 51.7 & 74.13 \\
2023 OCT 18  &  095700  -  101200 & 382.5 & 14.3  &  14.7 & 68.6 & 73.93 & 62.2 & 74.02 \\
2023 OCT 25  &  093000  -  094500 & 372.4 & 15.1  &  15.1 & 68.2 & 73.93 & 61.4 & 74.03 \\
2023 OCT 30  &  091000  -  092500 & 355.9 & 15.1  &  15.1 & 67.9 & 73.94 & 60.9 & 74.04 \\
2023 NOV 10  &  082700  -  084200 & 400.0 & 14.8  &  14.9 & 67.1 & 73.95 & 59.6 & 74.05 \\
2023 NOV 12  &  081500  -  083000 & 400.3 & 14.6  &  14.8 & 66.9 & 73.95 & 59.2 & 74.06 \\
2024 DEC 03  &  053705  -  064500 & 354.9 & 16.4  &  16.3 & 54.7 & 74.10 & 67.6 & 73.94 \\
2024 DEC 07  &  052200  -  063000 & 387.1 & 16.1  &  15.9 & 55.2 & 74.10 & 68.1 & 73.93 \\
2024 DEC 12  &  050700  -  061500 & 394.4 & 16.3  &  16.1 & 55.7 & 74.10 & 68.5 & 73.93 \\
2024 DEC 21  &  042130  -  053000 & 390.4 & 16.5  &  15.8 & 56.7 & 74.09 & 69.5 & 73.91 \\
2024 DEC 23  &  042120  -  053000 & 398.5 & 18.1  &  17.9 & 56.8 & 74.09 & 69.6 & 73.91 \\
\enddata
\tablecomments{Goldstone observation epochs in UTC, start and stop times (hhmmss) of data collection in UTC, average transmitted powers in kilowatts, OC and SC system temperatures in kelvins, transmitter elevations in degrees, transmitter antenna gains in dBi, receiver elevations in degrees, and receiver gains in dBi. For the two epochs with an asterisk, system temperatures were estimated using the method described in Appendix \ref{app:sysT_estimation}. 
See text for calculation of gain values.}
\end{deluxetable*}

\begin{deluxetable*}{llccc}[!t]
  \tabletypesize{\footnotesize}
\tablewidth{0pt}
\tablecaption{Goldstone-to-GBT Observations of Europa \label{tab:gbt_observation}}
\tablehead{
  \colhead{Date} & \colhead{Start - Stop Time} & \colhead{$\rm T_{\rm sys}$ OC} & \colhead{$\rm T_{\rm sys}$ SC} & \colhead{${\rm El}_{\rm rcv}$} \\ [-7pt]
\colhead{(UTC)} & \colhead{(UTC)} & \colhead{(K)} & \colhead{(K)} & \colhead{(deg)}
}
\startdata
2011 SEP 24  &  111555  -  111915  &  27.6  &  27.2  &  29.7 \\
2011 SEP 27  &  105225  -  120035  &  27.8  &  27.3  &  29.5 \\
2011 SEP 29  &  103710  -  114505  &  27.9  &  27.4  &  29.4 \\
2011 OCT 01  &  104000  -  113000  &  27.2  &  26.6  &  29.1 \\
2011 OCT 03  &  104040  -  112540  &  28.7  &  28.2  &  29.0 \\
2011 OCT 06  &  103610  -  111024  &  27.0  &  26.5  &  28.7 \\
2011 OCT 10  &  103400  -  110207  &  27.8  &  27.3  &  28.2 \\
2011 OCT 13  &  101700  -  104700  &  38.4  &  38.9  &  27.8 \\
2011 OCT 15  &  100900  -  103900  &  26.8  &  26.2  &  27.5 \\
2011 OCT 17  &  100100  -  103100  &  27.7  &  27.4  &  27.3 \\
2011 OCT 19  &  095300  -  102300  &  28.2  &  27.8  &  27.0 \\
2011 OCT 20  &  094900  -  101900  &  28.4  &  27.9  &  27.0 \\
2011 OCT 24  &  093400  -  100400  &  29.5  &  29.2  &  26.4 \\
2011 OCT 26  &  092400  -  095400  &  29.0  &  28.6  &  26.1 \\
2011 OCT 29  &  092100  -  093200  &  33.4  &  33.1  &  25.7 \\
2011 OCT 31  &  090600  -  093600  &  27.2  &  26.7  &  25.5 \\
2011 NOV 04  &  085100  -  092100  &  28.1  &  27.8  &  25.0 \\
2011 NOV 05  &  084345  -  091345  &  28.4  &  28.0  &  24.8 \\
2011 NOV 07  &  085030  -  090312  &  N/A  &  N/A  &  24.6 \\
2011 NOV 11  &  082830  -  084830  &  29.0  &  29.3  &  24.1 \\
2011 NOV 16  &  080000  -  083000  &  41.9  &  42.6  &  23.4 \\
2011 NOV 18  &  075500  -  082500  &  27.9  &  27.4  &  23.4 \\
2011 NOV 23  &  074100  -  075600  &  32.1  &  31.8  &  22.7 \\
2023 OCT 18  &  100229  -  101729  &  27.4  &  31.5  &  34.4 \\
2023 OCT 25  &  093000  -  094500  &  25.4  &  29.4  &  33.4 \\
2023 OCT 30  &  091000  -  092500  &  N/A  &  N/A  &  32.9 \\
2023 NOV 10  &  082700  -  084200  &  41.1  &  45.2  &  31.4 \\ 
2024 DEC 21  &  050200  -  051541  &  37.1  &  33.1  &  66.6 \\
2024 DEC 23  &  045900  -  053000  &  36.4  &  32.4  &  66.5 \\
\enddata
\tablecomments{GBT observation epochs in UTC, start and stop times (hhmmss) of data collection in UTC, OC and SC system temperatures in kelvins, and elevations of the receiver in degrees. Transmitter quantities are listed in Table \ref{tab:Goldstone_observation}. See text for system temperature estimates. For the 2011 NOV 07 and 2023 OCT 30 epochs, system temperatures are unavailable and listed as ``N/A". Although receiver elevations are listed here for completeness, we used a constant antenna gain value for the GBT, as explained in Section \ref{sec:observations}.}
\end{deluxetable*}

\section{Data Reduction} \label{sec:data_reduction}

For each epoch, we first downsampled the complex voltages to a sampling frequency of 62.5 kHz using the \textit{resample\_poly} function from SciPy~\citep{scipy}. The DC offsets were then computed and subtracted from the signals, and the echoes were frequency-shifted by $\sim$10 kHz in order to center the radar echoes on 0 Hz. We subsequently applied Fast Fourier Transforms (FFT) to the downsampled data and formed power spectra with a frequency resolution of 10 Hz.  Consecutive spectra obtained over the duration of the receive windows (Tables~\ref{tab:Goldstone_observation} and \ref{tab:gbt_observation}) were summed to produce the final power spectra.

We smoothed the spectra using a Savitzky–Golay filter, which fits low-degree polynomials to moving windows of data points. Specifically, we used the \textit{savgol\_filter} function from Scipy \citep{scipy} with a kernel size of $\sim$360 Hz, which corresponds to 10\% of Europa's limb-to-limb bandwidth $B_{\rm ll}$, consistent with the approach used by \citet{Ostro1992}.
The limb-to-limb bandwidth is $8\pi r \sin(\delta)/\lambda P$, where $r$ is the radius of Europa, $P$ is the rotation period, and $\delta$ is the angle between the apparent spin vector and the line of sight.
The Savitzky-Golay filter was chosen for its ability to preserve the overall shape and fine features of the spectra.

\begin{figure}[ht!]
\plotone{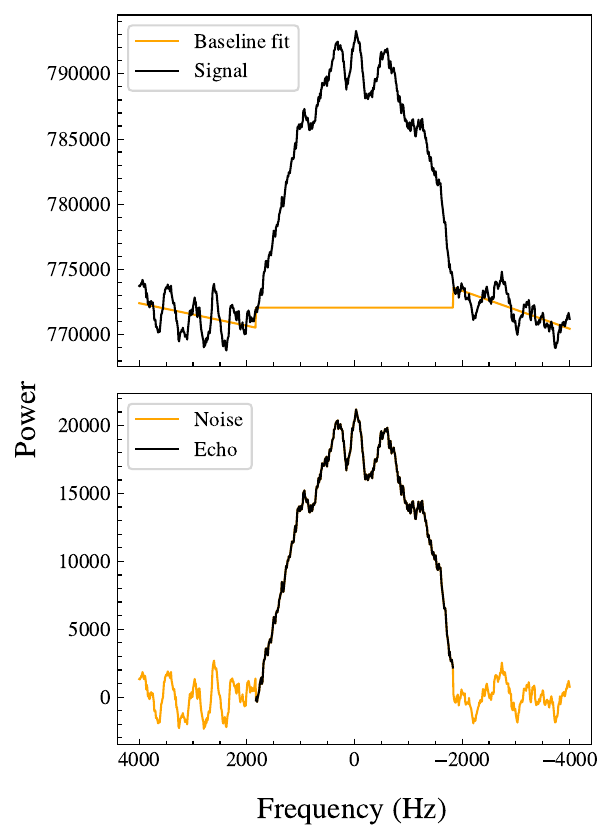}
\caption{(Top) Power spectrum of the 2011 OCT 03 Goldstone SC echo after smoothing by the Savitzky-Golay filter.  Also shown is the fitted noise baseline exhibiting discontinuities at the two frequencies corresponding to the approaching and receding limbs (see text).  (Bottom) Power spectrum of the smoothed echo after baseline subtraction.  The y-axis represents echo power in arbitrary units.
\label{fig:broken_linear_fit}}
\end{figure}

Removal of the noise baseline was complicated by the fact that the baseline was noisy and exhibited different slopes at frequencies above and below the target's frequency.  We experimented with different fitting methods on the receiver's noise baseline and found out that a broken linear fit provided the most reliable approach (details can be found in Appendix \ref{app:baseline_fitting}). In the broken linear fit, straight lines were fit separately to the baselines on the left and right sides of the echo, extending to $\pm$4000~Hz, 
and we extended the baseline beneath the echo with a constant noise power equal to the average of the end point of the left fit and the start point of the right fit.  The extent of the noise baseline provided a pure noise bandwidth comparable to the bandwidth of the echo ($\sim$3600 Hz).  However, for a few epochs, unexpected radio frequency interference appeared in the baseline near the echo; in these cases, we used a narrower extent of $\pm$3000~Hz to avoid biasing the fit. The fitted baselines were then subtracted from the smoothed spectra. An example of the fitted noise baseline and the final spectrum is shown in Figure \ref{fig:broken_linear_fit}.

\section{Results} \label{sec:results}
\begin{figure}
\plotone{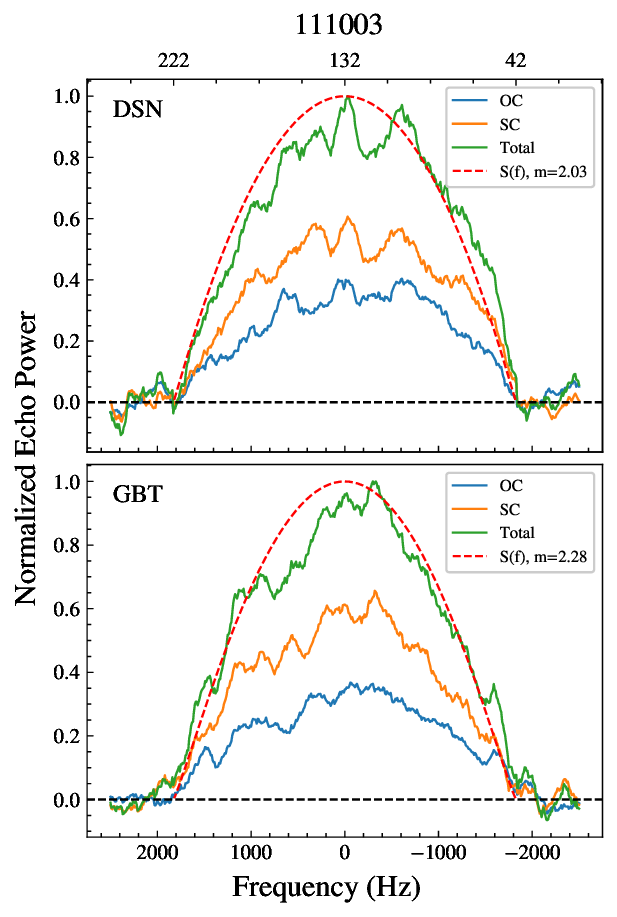}
\caption{Echo power spectra for Goldstone (aka DSN, top) and GBT (bottom) at epoch 2011 OCT 03. 
The blue, orange, and green lines represent the OC, SC, and total (OC+SC) echo power, respectively. 
The power is normalized to the total echo power’s maximum. The red dashed line represents the scattering law S(f) with the fitted exponent m. 
The number at the center just above the top panel is the subradar west longitude in degrees, whereas the numbers on the left and right are the west longitudes of the target’s approaching (left) and receding (right) limbs, respectively. 
Note that frequency on the x-axis increases from right to left.
The complete figure set (33 images) is available in the online journal.
}
\label{fig:example_result}
\end{figure}

Figure \ref{fig:example_result} shows the single-date, dual-polarization echo power spectra for a representative epoch (2011 OCT 3). Power spectra for all epochs can be found in Appendices \ref{app:dsn_spectra} and \ref{app:gbt_spectra}
as well as in a figure set (33 images) in the online version of this article.

\figsetstart
\figsetnum{3}
\figsettitle{Goldstone and GBT Spectra}

\figsetgrpstart
\figsetgrpnum{3.1}
\figsetgrptitle{DSN and GBT Spectra at Epoch 111003}
\figsetplot{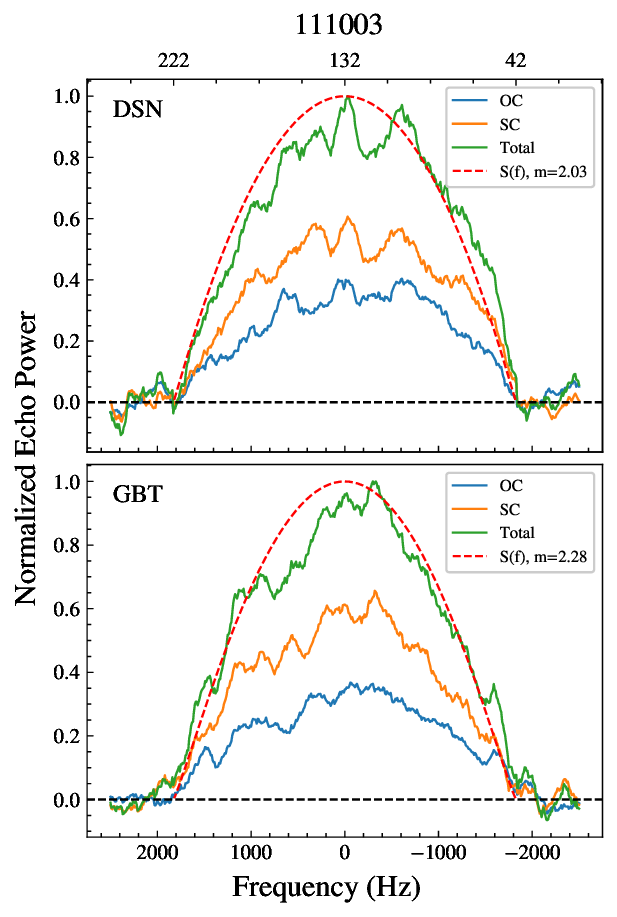}
\figsetgrpnote{Echo power spectra for Goldstone (aka DSN, top) and GBT (bottom) at epoch 2011 OCT 03.
The blue, orange, and green lines represent the OC, SC, and total (OC+SC) echo power, respectively. 
The power is normalized to the total echo power’s maximum. The red dashed line represents the scattering law S(f) with the fitted exponent m. 
The number at the center just above the top panel is the subradar west longitude in degrees, whereas the numbers on the left and right are the west longitudes of the target’s approaching (left) and receding (right) limbs, respectively. 
Note that frequency on the x-axis increases from right to left.}
\figsetgrpend

\figsetgrpstart
\figsetgrpnum{3.2}
\figsetgrptitle{DSN and GBT Spectra at Epoch 110927}
\figsetplot{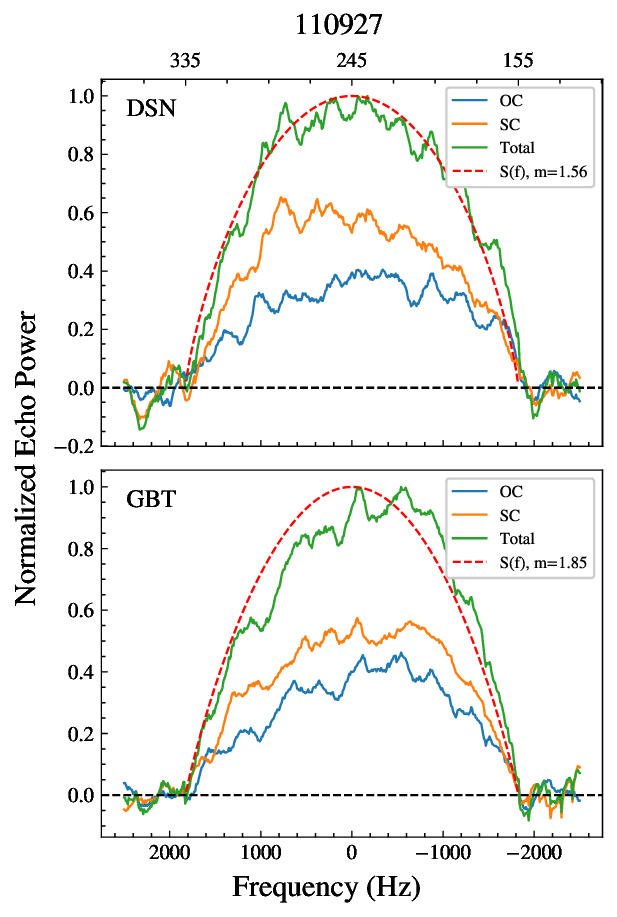}
\figsetgrpnote{Echo power spectra for Goldstone (aka DSN, top) and GBT (bottom) at the epoch indicated at the top of the plot (YYMMDD). 
The blue, orange, and green lines represent the OC, SC, and total (OC+SC) echo power, respectively. 
The power is normalized to the total echo power’s maximum. The red dashed line represents the scattering law S(f) with the fitted exponent m. 
The number at the center just above the top panel is the subradar west longitude in degrees, whereas the numbers on the left and right are the west longitudes of the target’s approaching (left) and receding (right) limbs, respectively. 
Note that frequency on the x-axis increases from right to left.}
\figsetgrpend

\figsetgrpstart
\figsetgrpnum{3.3}
\figsetgrptitle{DSN and GBT Spectra at Epoch 110929}
\figsetplot{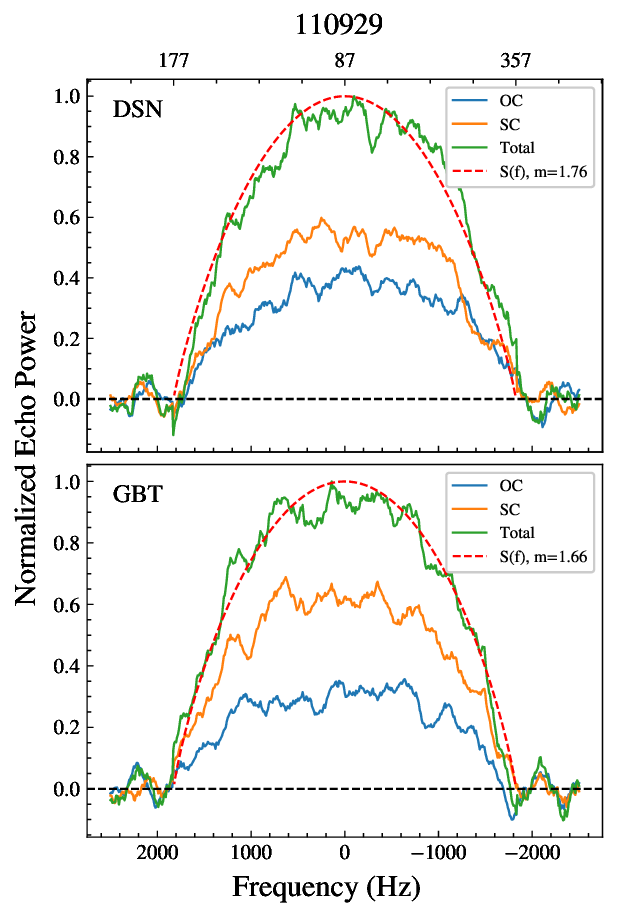}
\figsetgrpnote{Echo power spectra for Goldstone (aka DSN, top) and GBT (bottom) at the epoch indicated at the top of the plot (YYMMDD). 
The blue, orange, and green lines represent the OC, SC, and total (OC+SC) echo power, respectively. 
The power is normalized to the total echo power’s maximum. The red dashed line represents the scattering law S(f) with the fitted exponent m. 
The number at the center just above the top panel is the subradar west longitude in degrees, whereas the numbers on the left and right are the west longitudes of the target’s approaching (left) and receding (right) limbs, respectively. 
Note that frequency on the x-axis increases from right to left.}
\figsetgrpend

\figsetgrpstart
\figsetgrpnum{3.4}
\figsetgrptitle{DSN and GBT Spectra at Epoch 111001}
\figsetplot{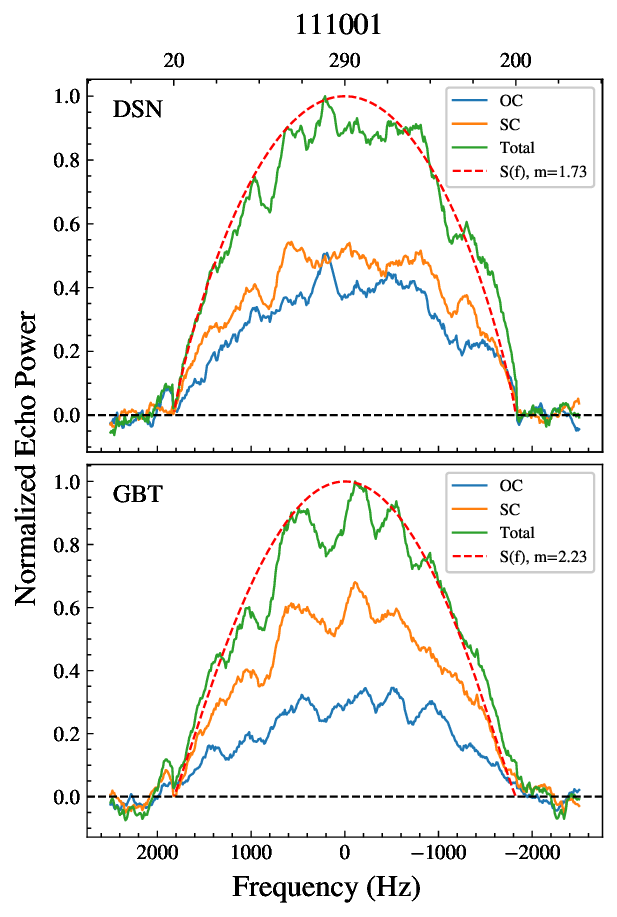}
\figsetgrpnote{Echo power spectra for Goldstone (aka DSN, top) and GBT (bottom) at the epoch indicated at the top of the plot (YYMMDD). 
The blue, orange, and green lines represent the OC, SC, and total (OC+SC) echo power, respectively. 
The power is normalized to the total echo power’s maximum. The red dashed line represents the scattering law S(f) with the fitted exponent m. 
The number at the center just above the top panel is the subradar west longitude in degrees, whereas the numbers on the left and right are the west longitudes of the target’s approaching (left) and receding (right) limbs, respectively. 
Note that frequency on the x-axis increases from right to left.}
\figsetgrpend

\figsetgrpstart
\figsetgrpnum{3.5}
\figsetgrptitle{DSN and GBT Spectra at Epoch 110924}
\figsetplot{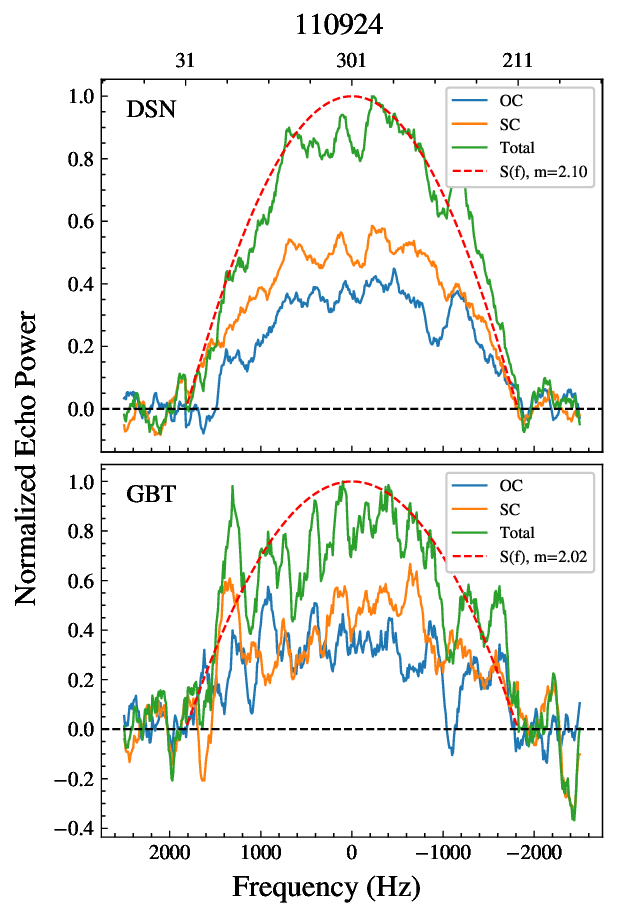}
\figsetgrpnote{Echo power spectra for Goldstone (aka DSN, top) and GBT (bottom) at the epoch indicated at the top of the plot (YYMMDD). 
The blue, orange, and green lines represent the OC, SC, and total (OC+SC) echo power, respectively. 
The power is normalized to the total echo power’s maximum. The red dashed line represents the scattering law S(f) with the fitted exponent m. 
The number at the center just above the top panel is the subradar west longitude in degrees, whereas the numbers on the left and right are the west longitudes of the target’s approaching (left) and receding (right) limbs, respectively. 
Note that frequency on the x-axis increases from right to left.}
\figsetgrpend

\figsetgrpstart
\figsetgrpnum{3.6}
\figsetgrptitle{DSN and GBT Spectra at Epoch 111006}
\figsetplot{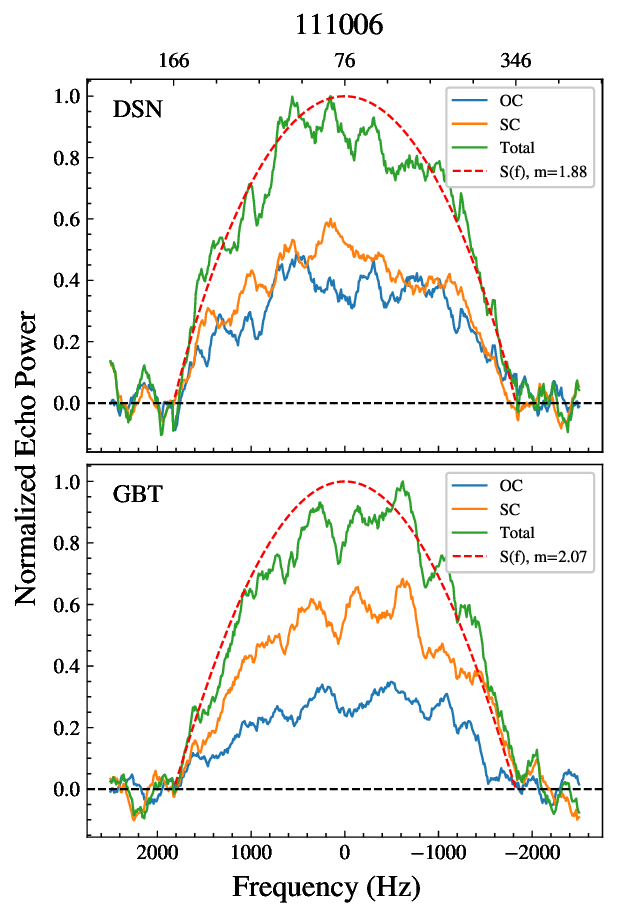}
\figsetgrpnote{Echo power spectra for Goldstone (aka DSN, top) and GBT (bottom) at the epoch indicated at the top of the plot (YYMMDD). 
The blue, orange, and green lines represent the OC, SC, and total (OC+SC) echo power, respectively. 
The power is normalized to the total echo power’s maximum. The red dashed line represents the scattering law S(f) with the fitted exponent m. 
The number at the center just above the top panel is the subradar west longitude in degrees, whereas the numbers on the left and right are the west longitudes of the target’s approaching (left) and receding (right) limbs, respectively. 
Note that frequency on the x-axis increases from right to left.}
\figsetgrpend

\figsetgrpstart
\figsetgrpnum{3.7}
\figsetgrptitle{DSN and GBT Spectra at Epoch 111010}
\figsetplot{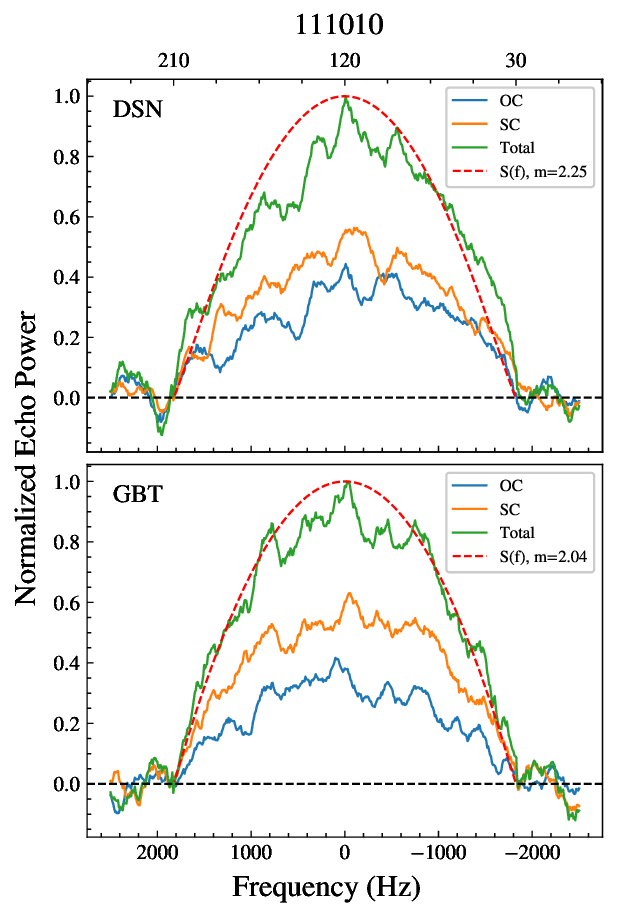}
\figsetgrpnote{Echo power spectra for Goldstone (aka DSN, top) and GBT (bottom) at the epoch indicated at the top of the plot (YYMMDD). 
The blue, orange, and green lines represent the OC, SC, and total (OC+SC) echo power, respectively. 
The power is normalized to the total echo power’s maximum. The red dashed line represents the scattering law S(f) with the fitted exponent m. 
The number at the center just above the top panel is the subradar west longitude in degrees, whereas the numbers on the left and right are the west longitudes of the target’s approaching (left) and receding (right) limbs, respectively. 
Note that frequency on the x-axis increases from right to left.}
\figsetgrpend

\figsetgrpstart
\figsetgrpnum{3.8}
\figsetgrptitle{DSN and GBT Spectra at Epoch 111013}
\figsetplot{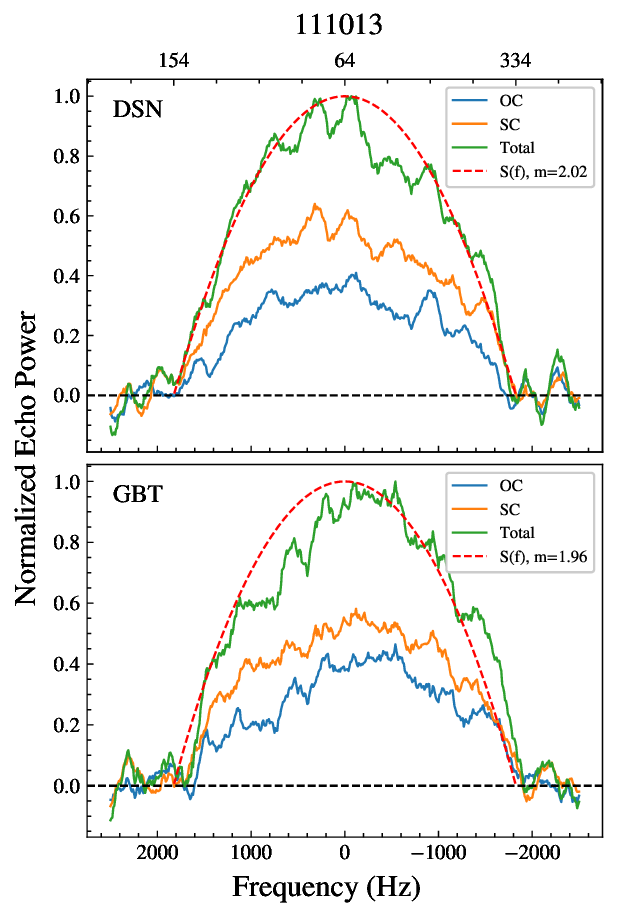}
\figsetgrpnote{Echo power spectra for Goldstone (aka DSN, top) and GBT (bottom) at the epoch indicated at the top of the plot (YYMMDD). 
The blue, orange, and green lines represent the OC, SC, and total (OC+SC) echo power, respectively. 
The power is normalized to the total echo power’s maximum. The red dashed line represents the scattering law S(f) with the fitted exponent m. 
The number at the center just above the top panel is the subradar west longitude in degrees, whereas the numbers on the left and right are the west longitudes of the target’s approaching (left) and receding (right) limbs, respectively. 
Note that frequency on the x-axis increases from right to left.}
\figsetgrpend

\figsetgrpstart
\figsetgrpnum{3.9}
\figsetgrptitle{DSN and GBT Spectra at Epoch 111015}
\figsetplot{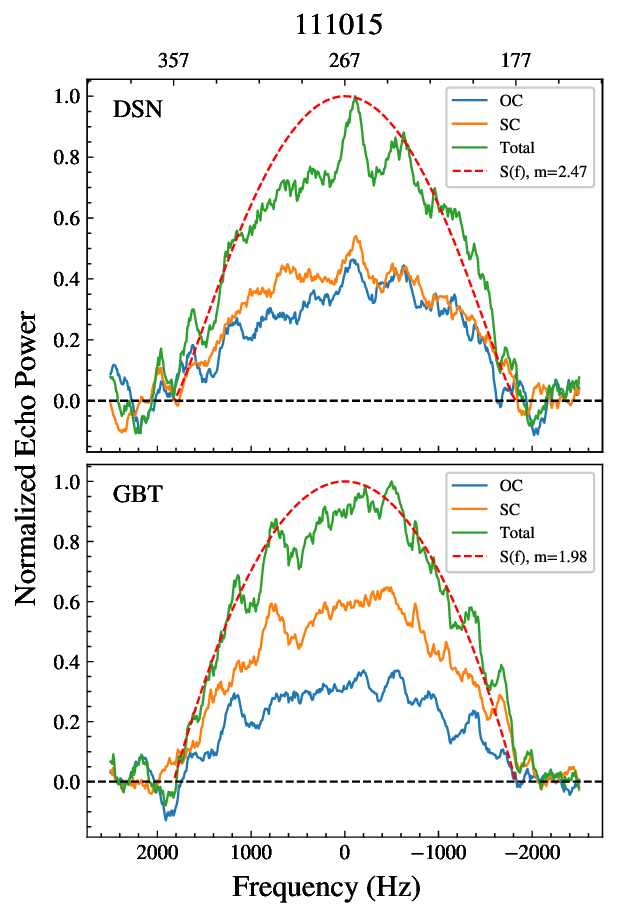}
\figsetgrpnote{Echo power spectra for Goldstone (aka DSN, top) and GBT (bottom) at the epoch indicated at the top of the plot (YYMMDD). 
The blue, orange, and green lines represent the OC, SC, and total (OC+SC) echo power, respectively. 
The power is normalized to the total echo power’s maximum. The red dashed line represents the scattering law S(f) with the fitted exponent m. 
The number at the center just above the top panel is the subradar west longitude in degrees, whereas the numbers on the left and right are the west longitudes of the target’s approaching (left) and receding (right) limbs, respectively. 
Note that frequency on the x-axis increases from right to left.}
\figsetgrpend

\figsetgrpstart
\figsetgrpnum{3.10}
\figsetgrptitle{DSN and GBT Spectra at Epoch 111017}
\figsetplot{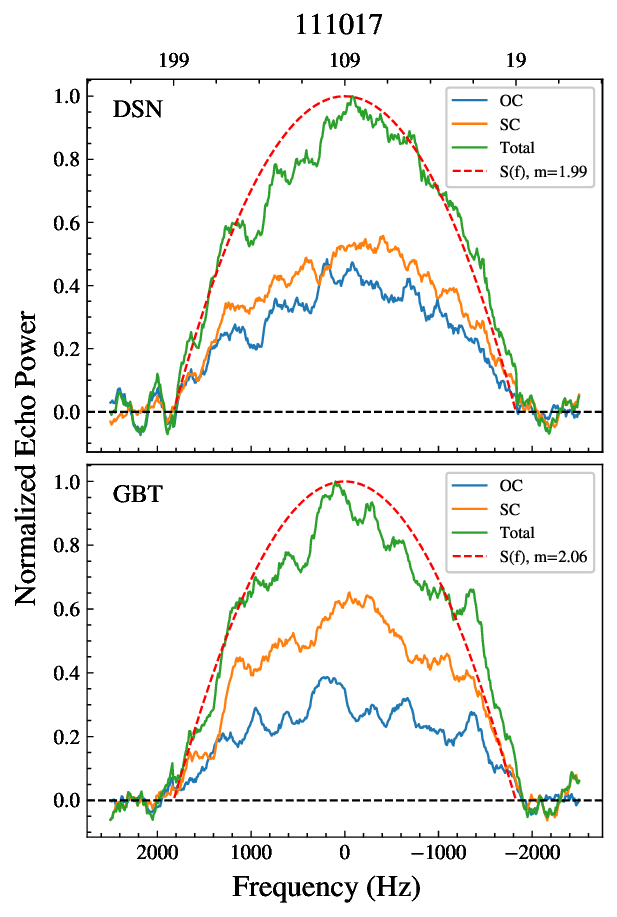}
\figsetgrpnote{Echo power spectra for Goldstone (aka DSN, top) and GBT (bottom) at the epoch indicated at the top of the plot (YYMMDD). 
The blue, orange, and green lines represent the OC, SC, and total (OC+SC) echo power, respectively. 
The power is normalized to the total echo power’s maximum. The red dashed line represents the scattering law S(f) with the fitted exponent m. 
The number at the center just above the top panel is the subradar west longitude in degrees, whereas the numbers on the left and right are the west longitudes of the target’s approaching (left) and receding (right) limbs, respectively. 
Note that frequency on the x-axis increases from right to left.}
\figsetgrpend

\figsetgrpstart
\figsetgrpnum{3.11}
\figsetgrptitle{DSN and GBT Spectra at Epoch 111019}
\figsetplot{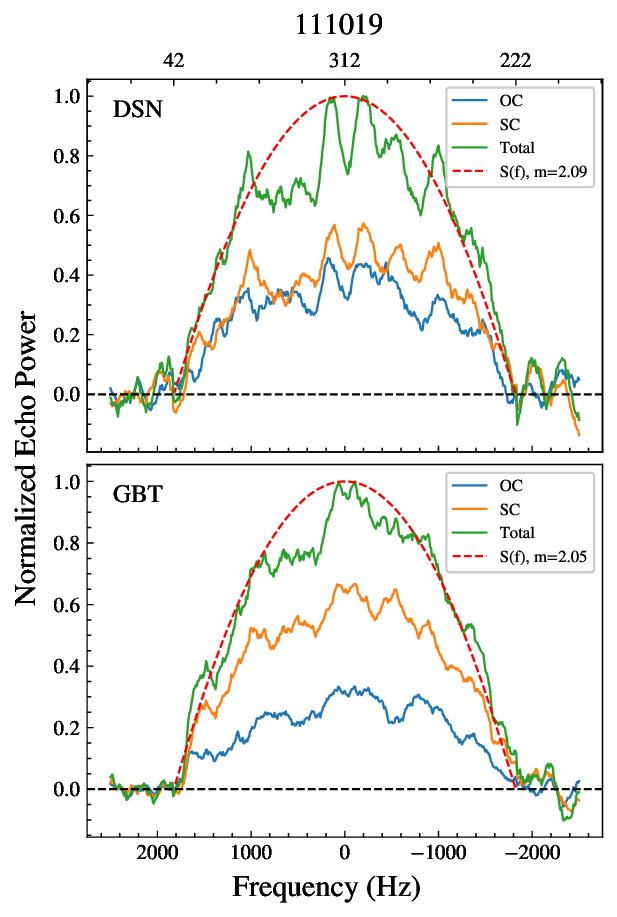}
\figsetgrpnote{Echo power spectra for Goldstone (aka DSN, top) and GBT (bottom) at the epoch indicated at the top of the plot (YYMMDD). 
The blue, orange, and green lines represent the OC, SC, and total (OC+SC) echo power, respectively. 
The power is normalized to the total echo power’s maximum. The red dashed line represents the scattering law S(f) with the fitted exponent m. 
The number at the center just above the top panel is the subradar west longitude in degrees, whereas the numbers on the left and right are the west longitudes of the target’s approaching (left) and receding (right) limbs, respectively. 
Note that frequency on the x-axis increases from right to left.}
\figsetgrpend

\figsetgrpstart
\figsetgrpnum{3.12}
\figsetgrptitle{DSN and GBT Spectra at Epoch 111020}
\figsetplot{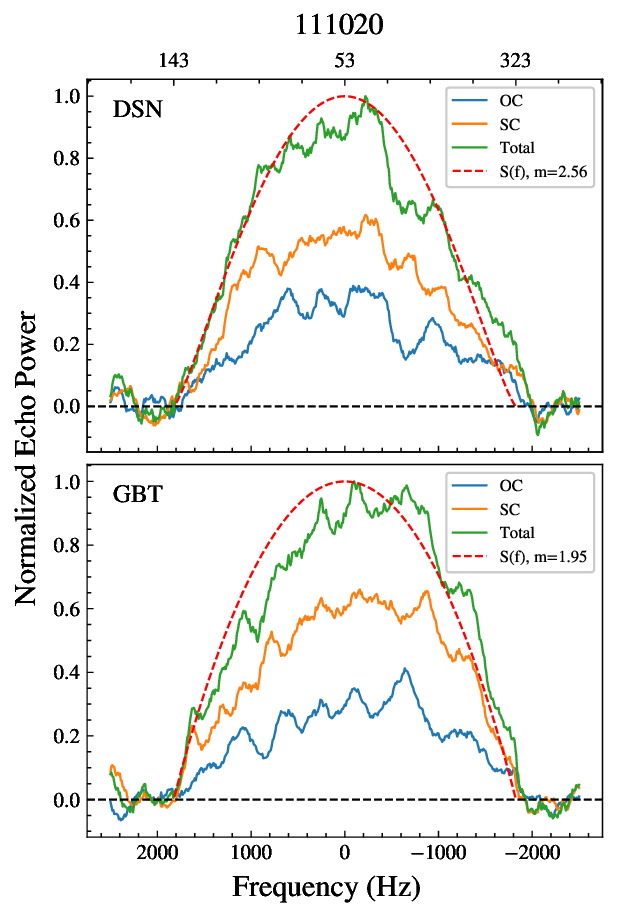}
\figsetgrpnote{Echo power spectra for Goldstone (aka DSN, top) and GBT (bottom) at the epoch indicated at the top of the plot (YYMMDD). 
The blue, orange, and green lines represent the OC, SC, and total (OC+SC) echo power, respectively. 
The power is normalized to the total echo power’s maximum. The red dashed line represents the scattering law S(f) with the fitted exponent m. 
The number at the center just above the top panel is the subradar west longitude in degrees, whereas the numbers on the left and right are the west longitudes of the target’s approaching (left) and receding (right) limbs, respectively. 
Note that frequency on the x-axis increases from right to left.}
\figsetgrpend

\figsetgrpstart
\figsetgrpnum{3.13}
\figsetgrptitle{DSN and GBT Spectra at Epoch 111024}
\figsetplot{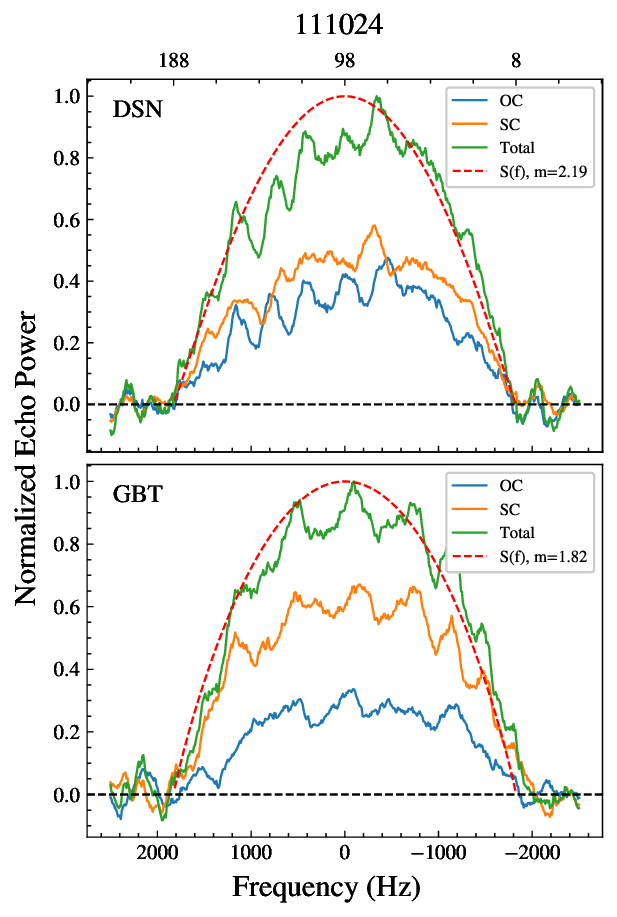}
\figsetgrpnote{Echo power spectra for Goldstone (aka DSN, top) and GBT (bottom) at the epoch indicated at the top of the plot (YYMMDD). 
The blue, orange, and green lines represent the OC, SC, and total (OC+SC) echo power, respectively. 
The power is normalized to the total echo power’s maximum. The red dashed line represents the scattering law S(f) with the fitted exponent m. 
The number at the center just above the top panel is the subradar west longitude in degrees, whereas the numbers on the left and right are the west longitudes of the target’s approaching (left) and receding (right) limbs, respectively. 
Note that frequency on the x-axis increases from right to left.}
\figsetgrpend

\figsetgrpstart
\figsetgrpnum{3.14}
\figsetgrptitle{DSN and GBT Spectra at Epoch 111026}
\figsetplot{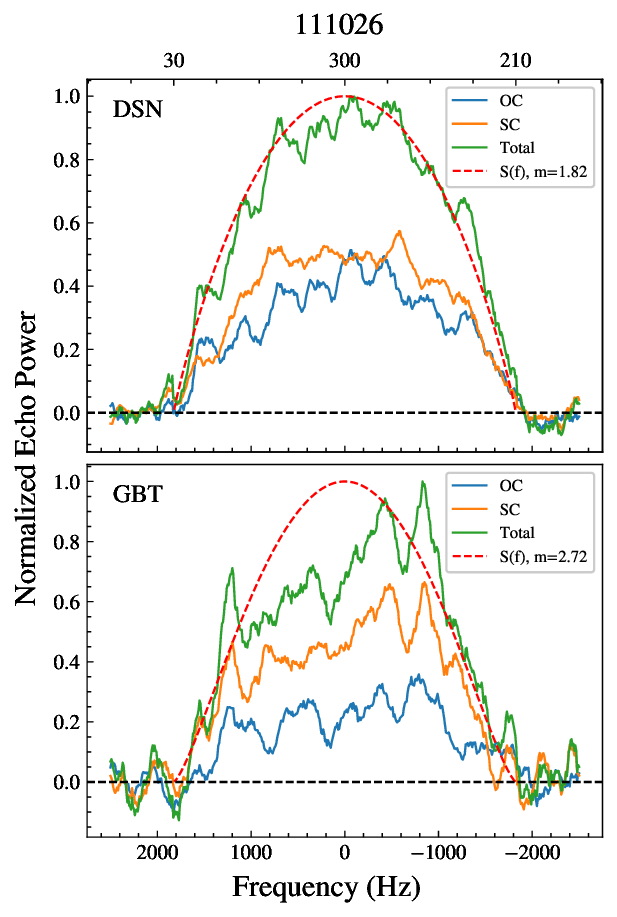}
\figsetgrpnote{Echo power spectra for Goldstone (aka DSN, top) and GBT (bottom) at the epoch indicated at the top of the plot (YYMMDD). 
The blue, orange, and green lines represent the OC, SC, and total (OC+SC) echo power, respectively. 
The power is normalized to the total echo power’s maximum. The red dashed line represents the scattering law S(f) with the fitted exponent m. 
The number at the center just above the top panel is the subradar west longitude in degrees, whereas the numbers on the left and right are the west longitudes of the target’s approaching (left) and receding (right) limbs, respectively. 
Note that frequency on the x-axis increases from right to left.}
\figsetgrpend

\figsetgrpstart
\figsetgrpnum{3.15}
\figsetgrptitle{DSN and GBT Spectra at Epoch 111029}
\figsetplot{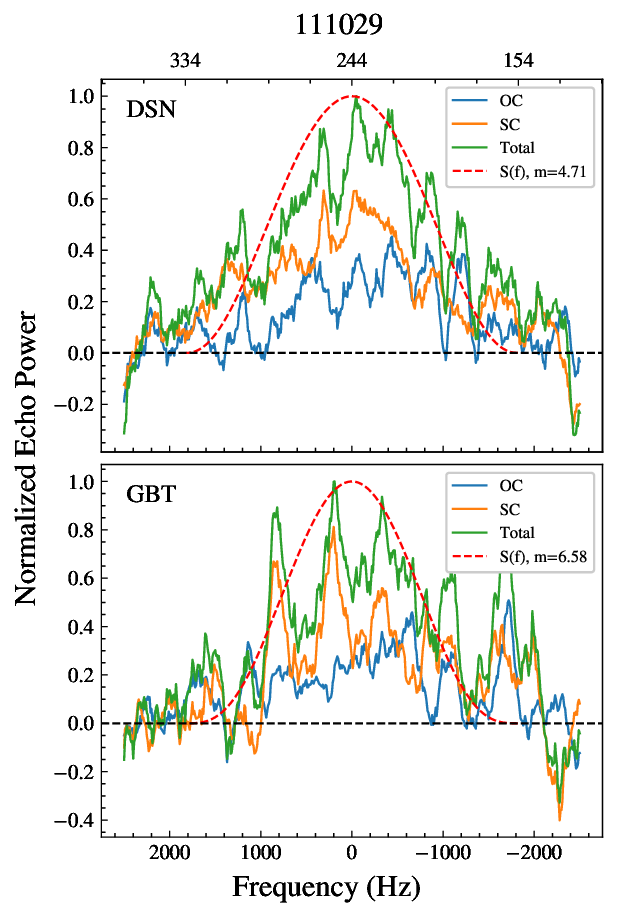}
\figsetgrpnote{Echo power spectra for Goldstone (aka DSN, top) and GBT (bottom) at the epoch indicated at the top of the plot (YYMMDD). 
The blue, orange, and green lines represent the OC, SC, and total (OC+SC) echo power, respectively. 
The power is normalized to the total echo power’s maximum. The red dashed line represents the scattering law S(f) with the fitted exponent m. 
The number at the center just above the top panel is the subradar west longitude in degrees, whereas the numbers on the left and right are the west longitudes of the target’s approaching (left) and receding (right) limbs, respectively. 
Note that frequency on the x-axis increases from right to left.}
\figsetgrpend

\figsetgrpstart
\figsetgrpnum{3.16}
\figsetgrptitle{DSN and GBT Spectra at Epoch 111031}
\figsetplot{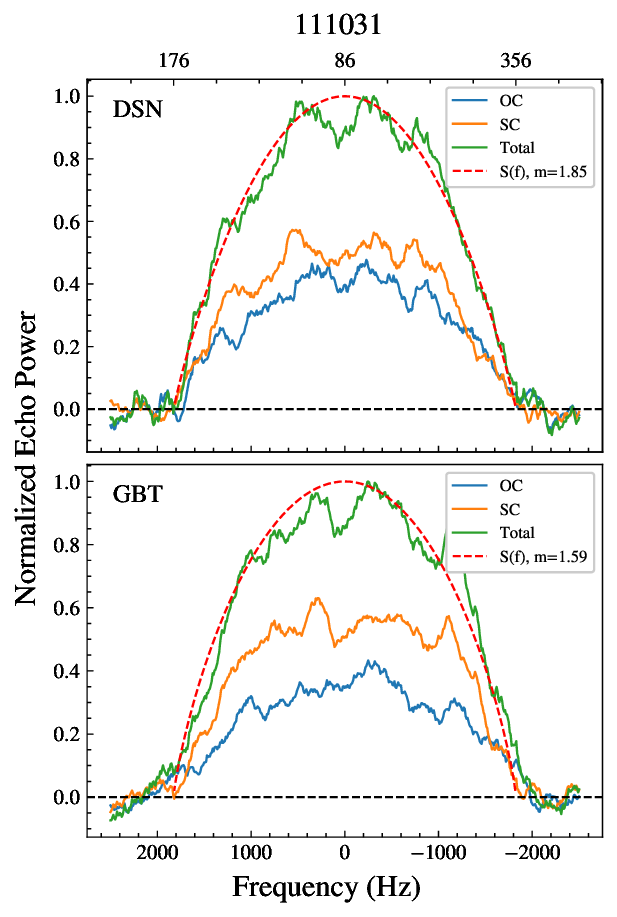}
\figsetgrpnote{Echo power spectra for Goldstone (aka DSN, top) and GBT (bottom) at the epoch indicated at the top of the plot (YYMMDD). 
The blue, orange, and green lines represent the OC, SC, and total (OC+SC) echo power, respectively. 
The power is normalized to the total echo power’s maximum. The red dashed line represents the scattering law S(f) with the fitted exponent m. 
The number at the center just above the top panel is the subradar west longitude in degrees, whereas the numbers on the left and right are the west longitudes of the target’s approaching (left) and receding (right) limbs, respectively. 
Note that frequency on the x-axis increases from right to left.}
\figsetgrpend

\figsetgrpstart
\figsetgrpnum{3.17}
\figsetgrptitle{DSN and GBT Spectra at Epoch 111104}
\figsetplot{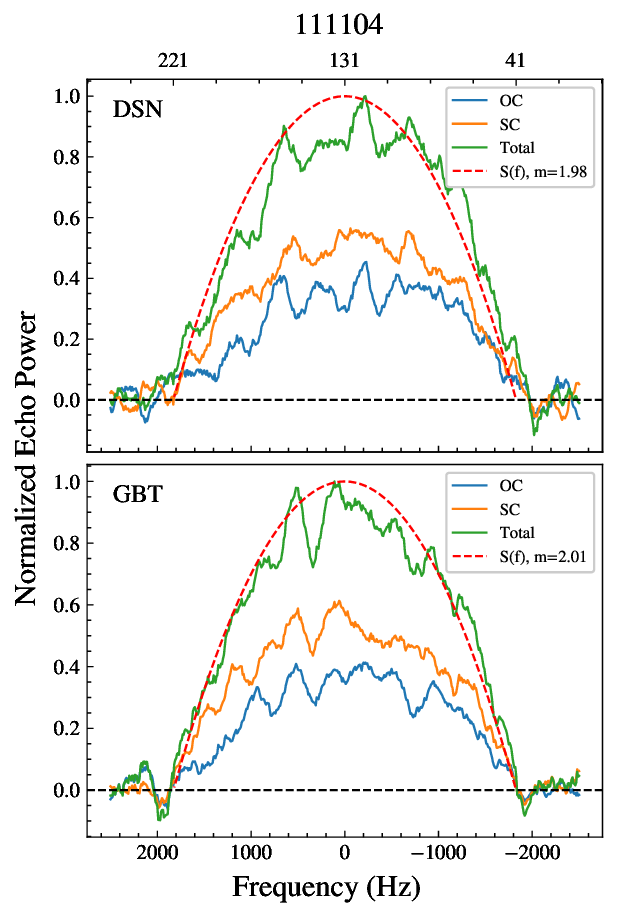}
\figsetgrpnote{Echo power spectra for Goldstone (aka DSN, top) and GBT (bottom) at the epoch indicated at the top of the plot (YYMMDD). 
The blue, orange, and green lines represent the OC, SC, and total (OC+SC) echo power, respectively. 
The power is normalized to the total echo power’s maximum. The red dashed line represents the scattering law S(f) with the fitted exponent m. 
The number at the center just above the top panel is the subradar west longitude in degrees, whereas the numbers on the left and right are the west longitudes of the target’s approaching (left) and receding (right) limbs, respectively. 
Note that frequency on the x-axis increases from right to left.}
\figsetgrpend

\figsetgrpstart
\figsetgrpnum{3.18}
\figsetgrptitle{DSN and GBT Spectra at Epoch 111105}
\figsetplot{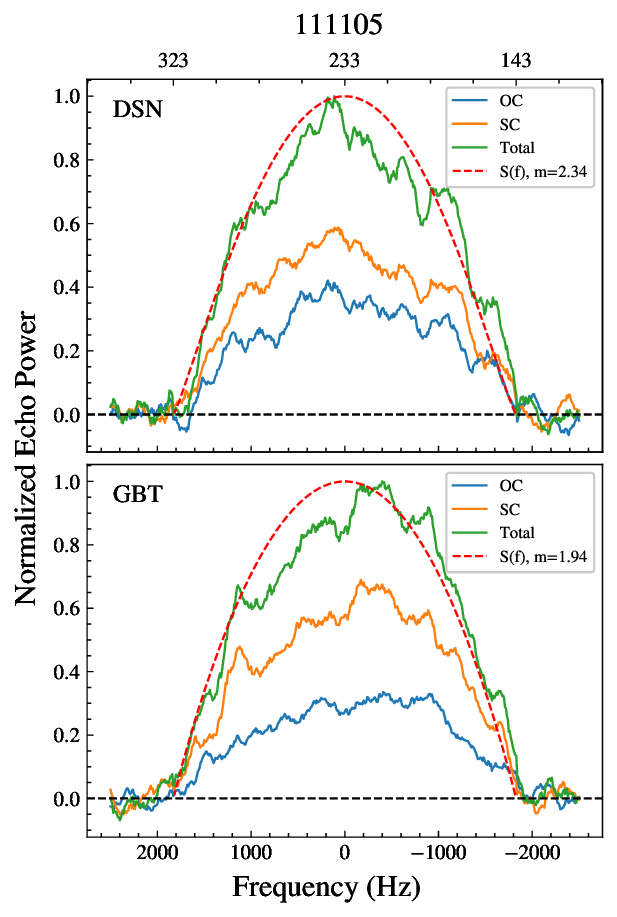}
\figsetgrpnote{Echo power spectra for Goldstone (aka DSN, top) and GBT (bottom) at the epoch indicated at the top of the plot (YYMMDD). 
The blue, orange, and green lines represent the OC, SC, and total (OC+SC) echo power, respectively. 
The power is normalized to the total echo power’s maximum. The red dashed line represents the scattering law S(f) with the fitted exponent m. 
The number at the center just above the top panel is the subradar west longitude in degrees, whereas the numbers on the left and right are the west longitudes of the target’s approaching (left) and receding (right) limbs, respectively. 
Note that frequency on the x-axis increases from right to left.}
\figsetgrpend

\figsetgrpstart
\figsetgrpnum{3.19}
\figsetgrptitle{DSN and GBT Spectra at Epoch 111107}
\figsetplot{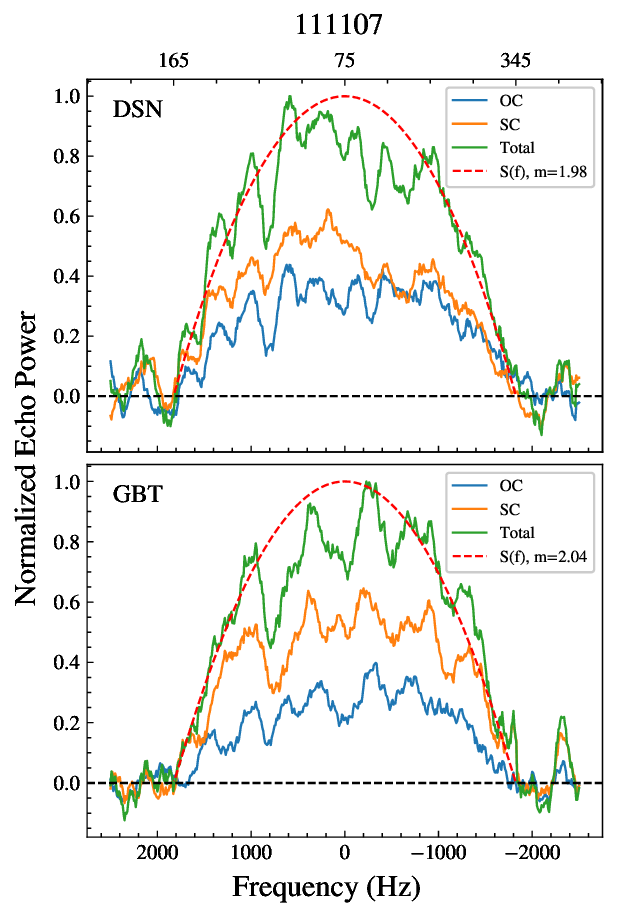}
\figsetgrpnote{Echo power spectra for Goldstone (aka DSN, top) and GBT (bottom) at the epoch indicated at the top of the plot (YYMMDD). 
The blue, orange, and green lines represent the OC, SC, and total (OC+SC) echo power, respectively. 
The power is normalized to the total echo power’s maximum. The red dashed line represents the scattering law S(f) with the fitted exponent m. 
The number at the center just above the top panel is the subradar west longitude in degrees, whereas the numbers on the left and right are the west longitudes of the target’s approaching (left) and receding (right) limbs, respectively. 
Note that frequency on the x-axis increases from right to left.}
\figsetgrpend

\figsetgrpstart
\figsetgrpnum{3.20}
\figsetgrptitle{DSN and GBT Spectra at Epoch 111111}
\figsetplot{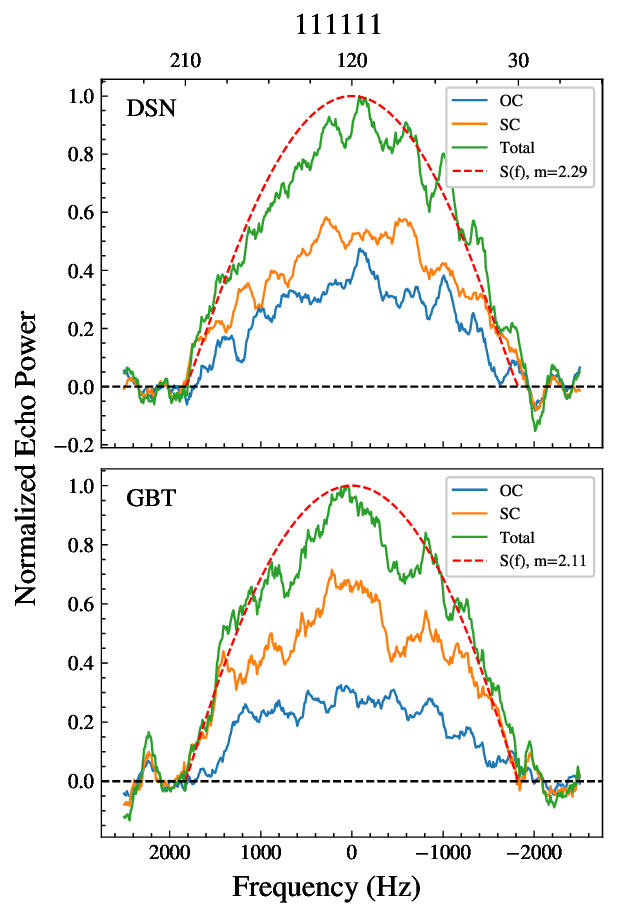}
\figsetgrpnote{Echo power spectra for Goldstone (aka DSN, top) and GBT (bottom) at the epoch indicated at the top of the plot (YYMMDD). 
The blue, orange, and green lines represent the OC, SC, and total (OC+SC) echo power, respectively. 
The power is normalized to the total echo power’s maximum. The red dashed line represents the scattering law S(f) with the fitted exponent m. 
The number at the center just above the top panel is the subradar west longitude in degrees, whereas the numbers on the left and right are the west longitudes of the target’s approaching (left) and receding (right) limbs, respectively. 
Note that frequency on the x-axis increases from right to left.}
\figsetgrpend

\figsetgrpstart
\figsetgrpnum{3.21}
\figsetgrptitle{DSN and GBT Spectra at Epoch 111116}
\figsetplot{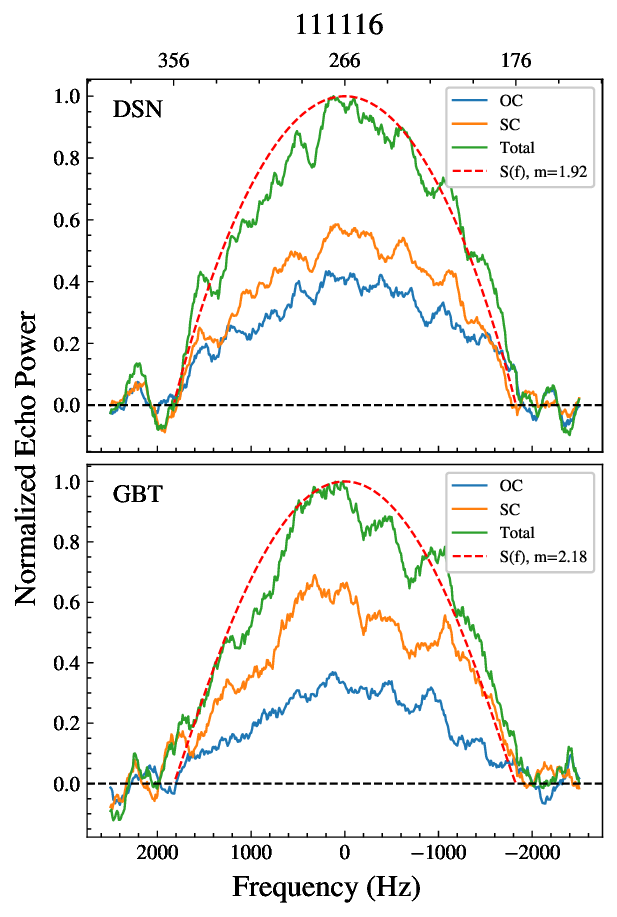}
\figsetgrpnote{Echo power spectra for Goldstone (aka DSN, top) and GBT (bottom) at the epoch indicated at the top of the plot (YYMMDD). 
The blue, orange, and green lines represent the OC, SC, and total (OC+SC) echo power, respectively. 
The power is normalized to the total echo power’s maximum. The red dashed line represents the scattering law S(f) with the fitted exponent m. 
The number at the center just above the top panel is the subradar west longitude in degrees, whereas the numbers on the left and right are the west longitudes of the target’s approaching (left) and receding (right) limbs, respectively. 
Note that frequency on the x-axis increases from right to left.}
\figsetgrpend

\figsetgrpstart
\figsetgrpnum{3.22}
\figsetgrptitle{DSN and GBT Spectra at Epoch 111118}
\figsetplot{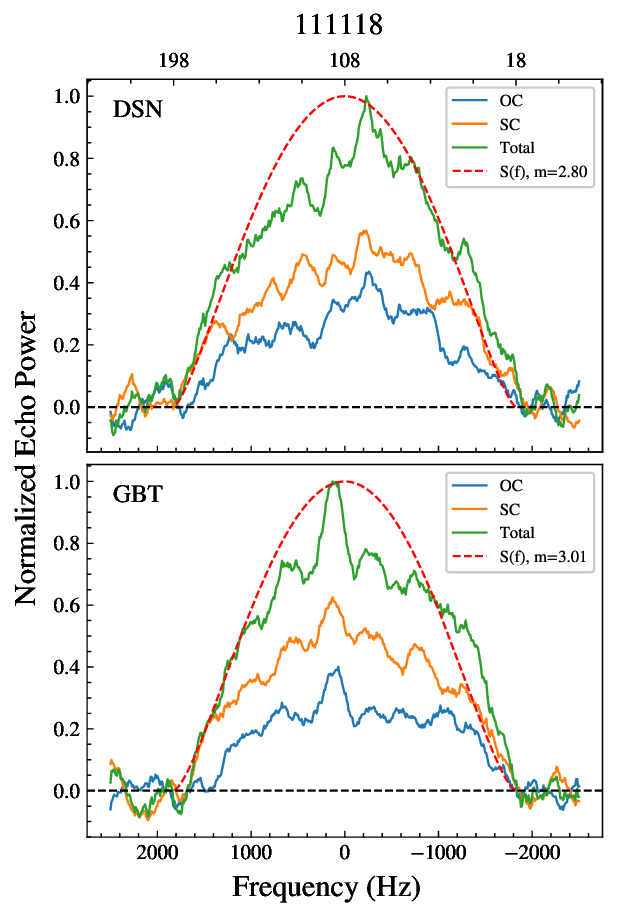}
\figsetgrpnote{Echo power spectra for Goldstone (aka DSN, top) and GBT (bottom) at the epoch indicated at the top of the plot (YYMMDD). 
The blue, orange, and green lines represent the OC, SC, and total (OC+SC) echo power, respectively. 
The power is normalized to the total echo power’s maximum. The red dashed line represents the scattering law S(f) with the fitted exponent m. 
The number at the center just above the top panel is the subradar west longitude in degrees, whereas the numbers on the left and right are the west longitudes of the target’s approaching (left) and receding (right) limbs, respectively. 
Note that frequency on the x-axis increases from right to left.}
\figsetgrpend

\figsetgrpstart
\figsetgrpnum{3.23}
\figsetgrptitle{DSN and GBT Spectra at Epoch 111123}
\figsetplot{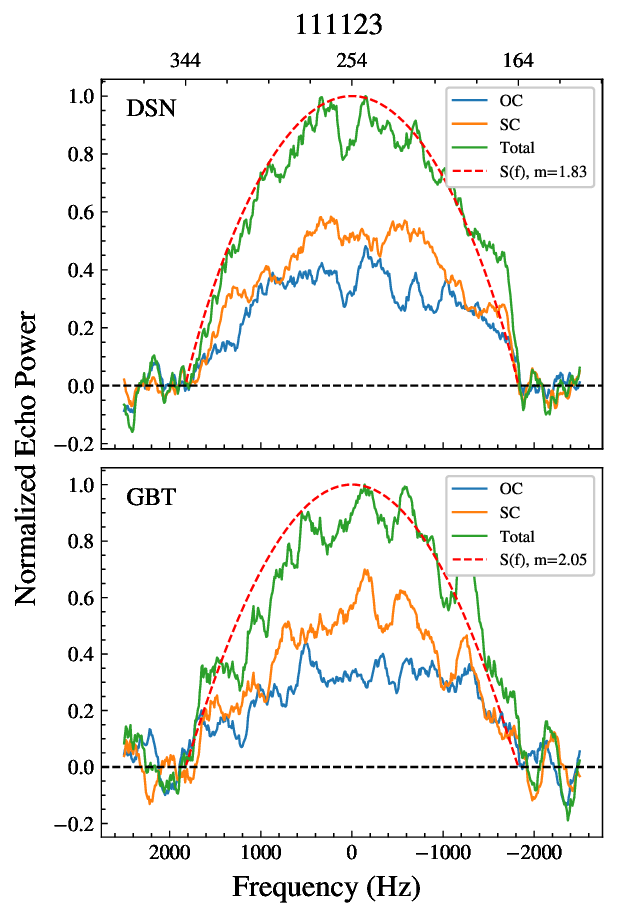}
\figsetgrpnote{Echo power spectra for Goldstone (aka DSN, top) and GBT (bottom) at the epoch indicated at the top of the plot (YYMMDD). 
The blue, orange, and green lines represent the OC, SC, and total (OC+SC) echo power, respectively. 
The power is normalized to the total echo power’s maximum. The red dashed line represents the scattering law S(f) with the fitted exponent m. 
The number at the center just above the top panel is the subradar west longitude in degrees, whereas the numbers on the left and right are the west longitudes of the target’s approaching (left) and receding (right) limbs, respectively. 
Note that frequency on the x-axis increases from right to left.}
\figsetgrpend

\figsetgrpstart
\figsetgrpnum{3.24}
\figsetgrptitle{DSN and GBT Spectra at Epoch 231018}
\figsetplot{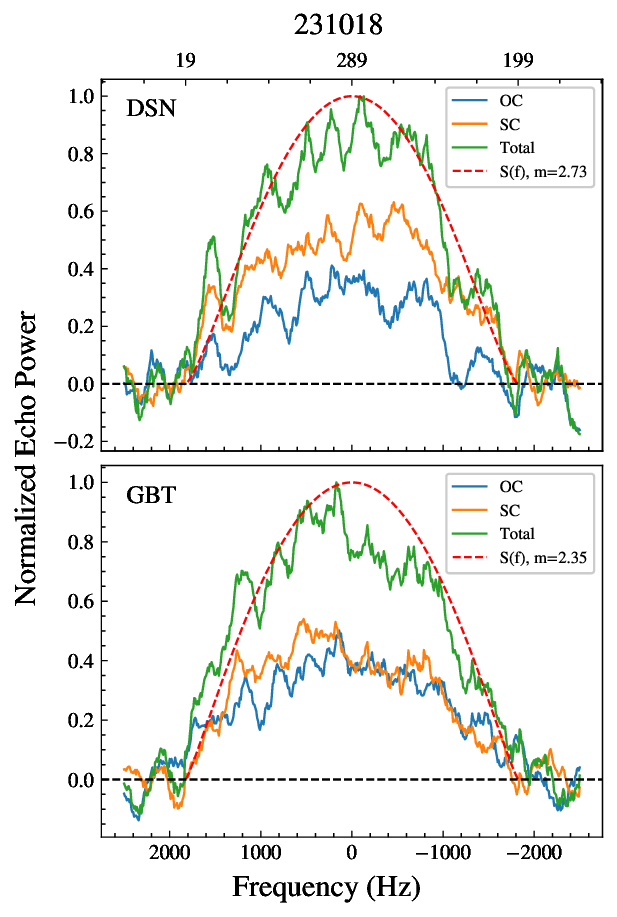}
\figsetgrpnote{Echo power spectra for Goldstone (aka DSN, top) and GBT (bottom) at the epoch indicated at the top of the plot (YYMMDD). 
The blue, orange, and green lines represent the OC, SC, and total (OC+SC) echo power, respectively. 
The power is normalized to the total echo power’s maximum. The red dashed line represents the scattering law S(f) with the fitted exponent m. 
The number at the center just above the top panel is the subradar west longitude in degrees, whereas the numbers on the left and right are the west longitudes of the target’s approaching (left) and receding (right) limbs, respectively. 
Note that frequency on the x-axis increases from right to left.}
\figsetgrpend

\figsetgrpstart
\figsetgrpnum{3.25}
\figsetgrptitle{DSN and GBT Spectra at Epoch 231025}
\figsetplot{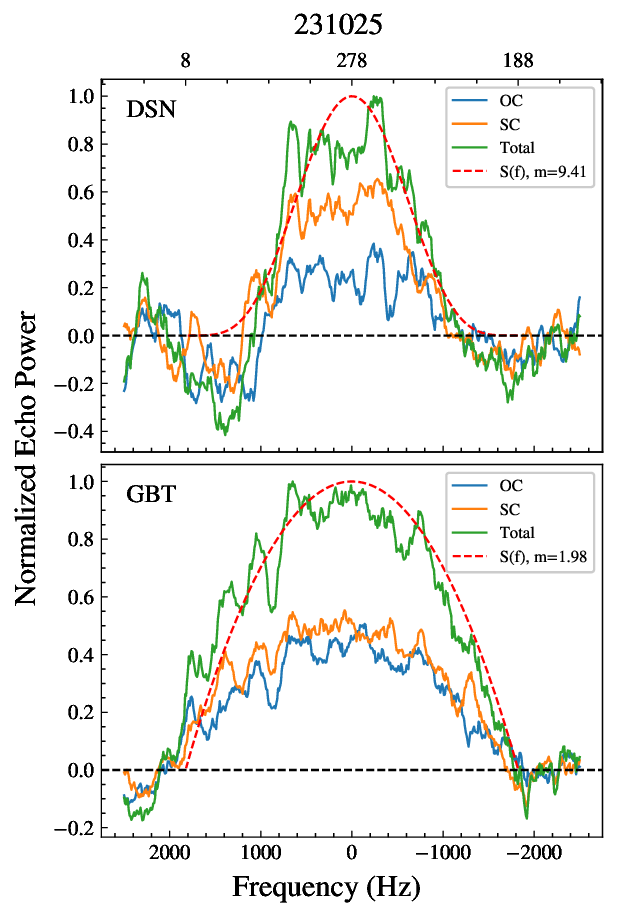}
\figsetgrpnote{Echo power spectra for Goldstone (aka DSN, top) and GBT (bottom) at the epoch indicated at the top of the plot (YYMMDD). 
The blue, orange, and green lines represent the OC, SC, and total (OC+SC) echo power, respectively. 
The power is normalized to the total echo power’s maximum. The red dashed line represents the scattering law S(f) with the fitted exponent m. 
The number at the center just above the top panel is the subradar west longitude in degrees, whereas the numbers on the left and right are the west longitudes of the target’s approaching (left) and receding (right) limbs, respectively. 
Note that frequency on the x-axis increases from right to left.}
\figsetgrpend

\figsetgrpstart
\figsetgrpnum{3.26}
\figsetgrptitle{DSN and GBT Spectra at Epoch 231030}
\figsetplot{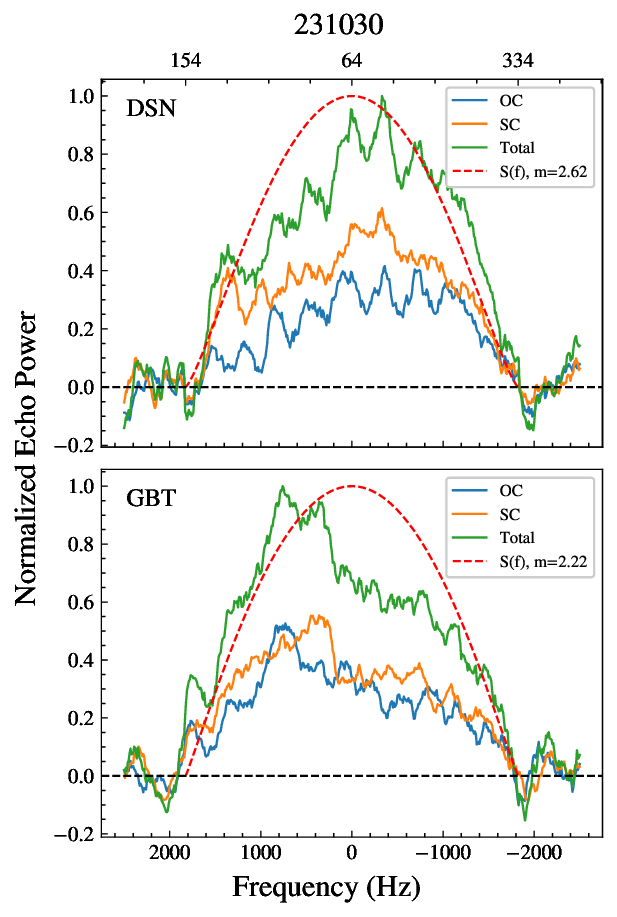}
\figsetgrpnote{Echo power spectra for Goldstone (aka DSN, top) and GBT (bottom) at the epoch indicated at the top of the plot (YYMMDD). 
The blue, orange, and green lines represent the OC, SC, and total (OC+SC) echo power, respectively. 
The power is normalized to the total echo power’s maximum. The red dashed line represents the scattering law S(f) with the fitted exponent m. 
The number at the center just above the top panel is the subradar west longitude in degrees, whereas the numbers on the left and right are the west longitudes of the target’s approaching (left) and receding (right) limbs, respectively. 
Note that frequency on the x-axis increases from right to left.}
\figsetgrpend

\figsetgrpstart
\figsetgrpnum{3.27}
\figsetgrptitle{DSN and GBT Spectra at Epoch 231110}
\figsetplot{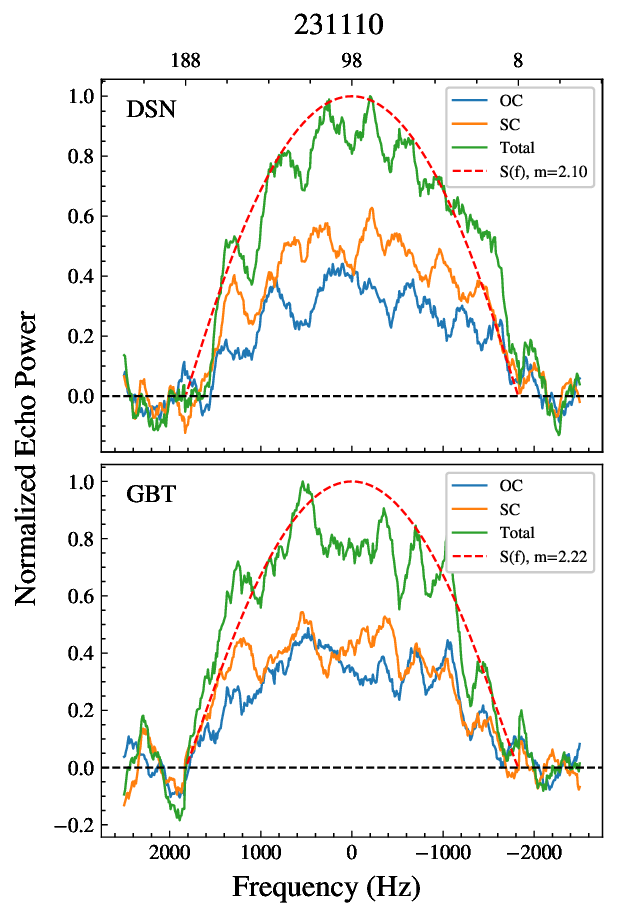}
\figsetgrpnote{Echo power spectra for Goldstone (aka DSN, top) and GBT (bottom) at the epoch indicated at the top of the plot (YYMMDD). 
The blue, orange, and green lines represent the OC, SC, and total (OC+SC) echo power, respectively. 
The power is normalized to the total echo power’s maximum. The red dashed line represents the scattering law S(f) with the fitted exponent m. 
The number at the center just above the top panel is the subradar west longitude in degrees, whereas the numbers on the left and right are the west longitudes of the target’s approaching (left) and receding (right) limbs, respectively. 
Note that frequency on the x-axis increases from right to left.}
\figsetgrpend

\figsetgrpstart
\figsetgrpnum{3.28}
\figsetgrptitle{DSN and GBT Spectra at Epoch 231112}
\figsetplot{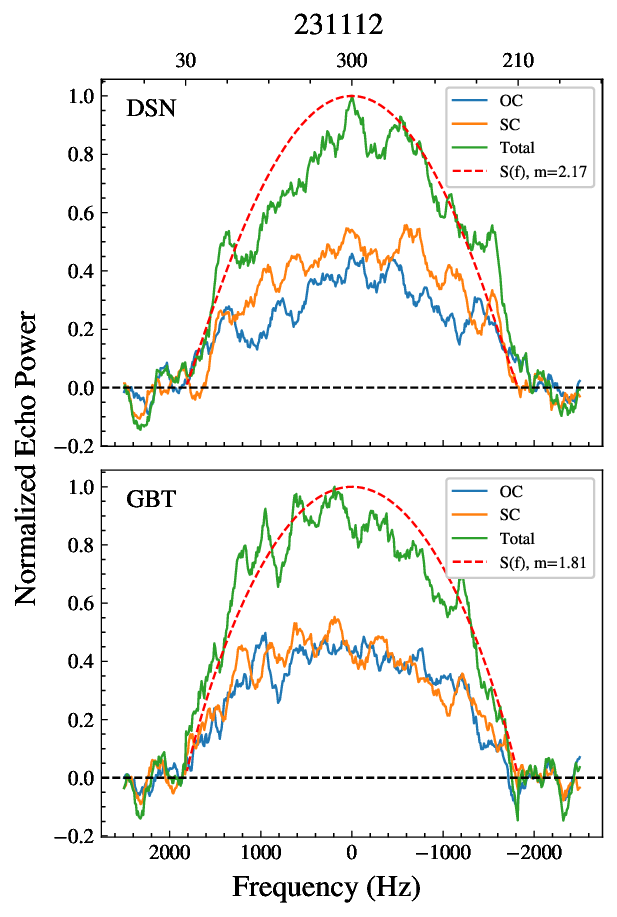}
\figsetgrpnote{Echo power spectra for Goldstone (aka DSN, top) and GBT (bottom) at the epoch indicated at the top of the plot (YYMMDD). 
The blue, orange, and green lines represent the OC, SC, and total (OC+SC) echo power, respectively. 
The power is normalized to the total echo power’s maximum. The red dashed line represents the scattering law S(f) with the fitted exponent m. 
The number at the center just above the top panel is the subradar west longitude in degrees, whereas the numbers on the left and right are the west longitudes of the target’s approaching (left) and receding (right) limbs, respectively. 
Note that frequency on the x-axis increases from right to left.}
\figsetgrpend

\figsetgrpstart
\figsetgrpnum{3.29}
\figsetgrptitle{DSN and GBT Spectra at Epoch 241203}
\figsetplot{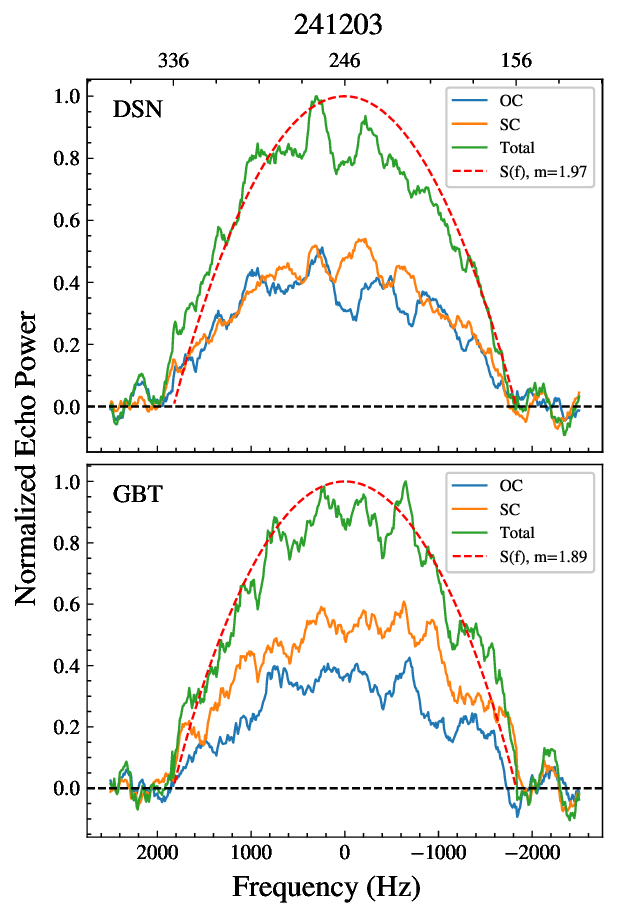}
\figsetgrpnote{Echo power spectra for Goldstone (aka DSN, top) and GBT (bottom) at the epoch indicated at the top of the plot (YYMMDD). 
The blue, orange, and green lines represent the OC, SC, and total (OC+SC) echo power, respectively. 
The power is normalized to the total echo power’s maximum. The red dashed line represents the scattering law S(f) with the fitted exponent m. 
The number at the center just above the top panel is the subradar west longitude in degrees, whereas the numbers on the left and right are the west longitudes of the target’s approaching (left) and receding (right) limbs, respectively. 
Note that frequency on the x-axis increases from right to left.}
\figsetgrpend

\figsetgrpstart
\figsetgrpnum{3.30}
\figsetgrptitle{DSN and GBT Spectra at Epoch 241207}
\figsetplot{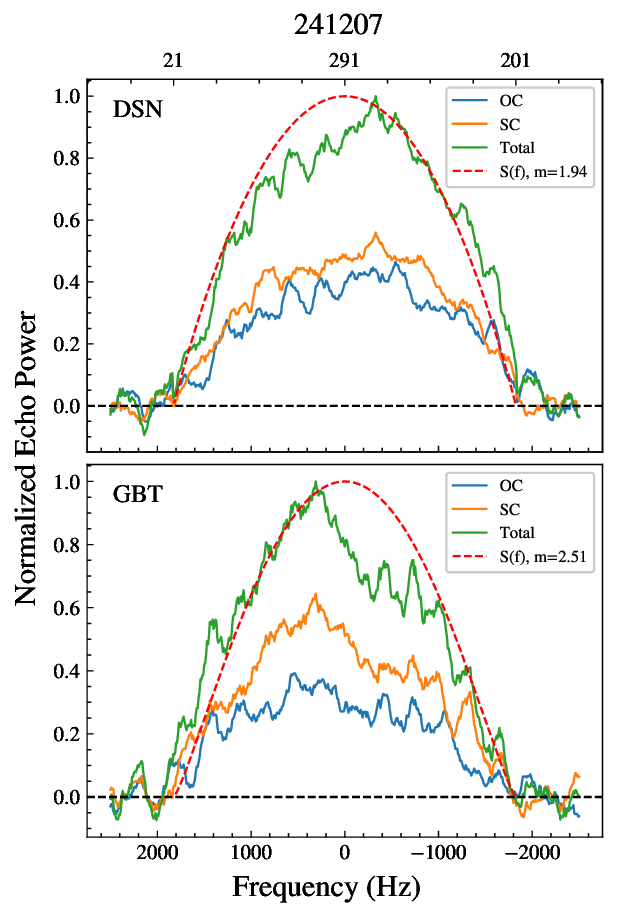}
\figsetgrpnote{Echo power spectra for Goldstone (aka DSN, top) and GBT (bottom) at the epoch indicated at the top of the plot (YYMMDD). 
The blue, orange, and green lines represent the OC, SC, and total (OC+SC) echo power, respectively. 
The power is normalized to the total echo power’s maximum. The red dashed line represents the scattering law S(f) with the fitted exponent m. 
The number at the center just above the top panel is the subradar west longitude in degrees, whereas the numbers on the left and right are the west longitudes of the target’s approaching (left) and receding (right) limbs, respectively. 
Note that frequency on the x-axis increases from right to left.}
\figsetgrpend

\figsetgrpstart
\figsetgrpnum{3.31}
\figsetgrptitle{DSN and GBT Spectra at Epoch 241212}
\figsetplot{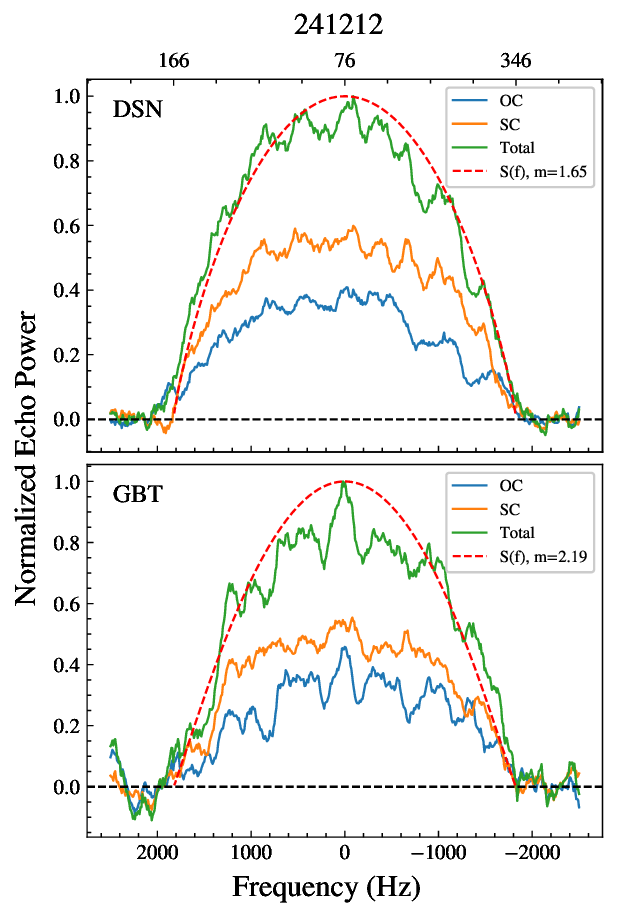}
\figsetgrpnote{Echo power spectra for Goldstone (aka DSN, top) and GBT (bottom) at the epoch indicated at the top of the plot (YYMMDD). 
The blue, orange, and green lines represent the OC, SC, and total (OC+SC) echo power, respectively. 
The power is normalized to the total echo power’s maximum. The red dashed line represents the scattering law S(f) with the fitted exponent m. 
The number at the center just above the top panel is the subradar west longitude in degrees, whereas the numbers on the left and right are the west longitudes of the target’s approaching (left) and receding (right) limbs, respectively. 
Note that frequency on the x-axis increases from right to left.}
\figsetgrpend

\figsetgrpstart
\figsetgrpnum{3.32}
\figsetgrptitle{DSN and GBT Spectra at Epoch 241221}
\figsetplot{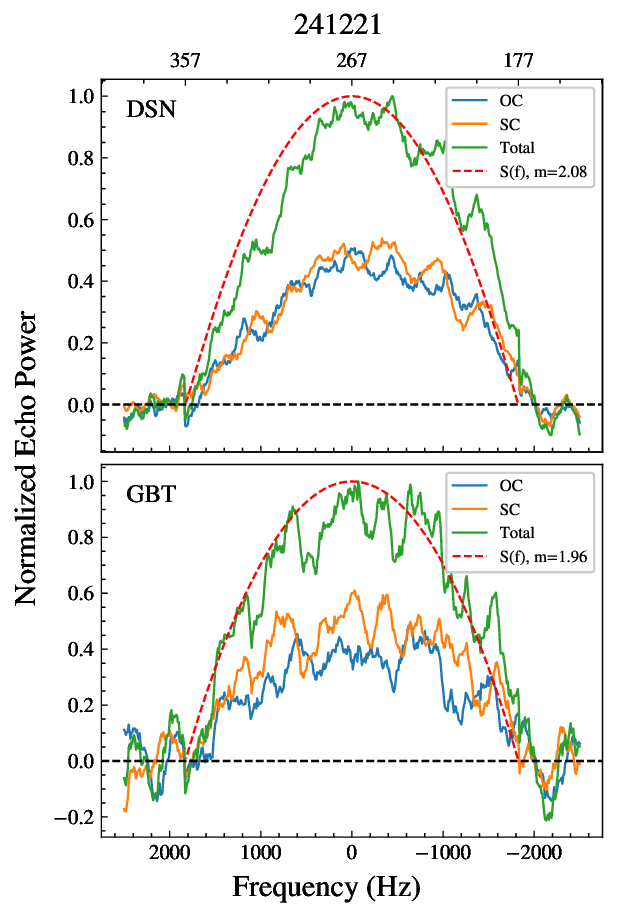}
\figsetgrpnote{Echo power spectra for Goldstone (aka DSN, top) and GBT (bottom) at the epoch indicated at the top of the plot (YYMMDD). 
The blue, orange, and green lines represent the OC, SC, and total (OC+SC) echo power, respectively. 
The power is normalized to the total echo power’s maximum. The red dashed line represents the scattering law S(f) with the fitted exponent m. 
The number at the center just above the top panel is the subradar west longitude in degrees, whereas the numbers on the left and right are the west longitudes of the target’s approaching (left) and receding (right) limbs, respectively. 
Note that frequency on the x-axis increases from right to left.}
\figsetgrpend

\figsetgrpstart
\figsetgrpnum{3.33}
\figsetgrptitle{DSN and GBT Spectra at Epoch 241223}
\figsetplot{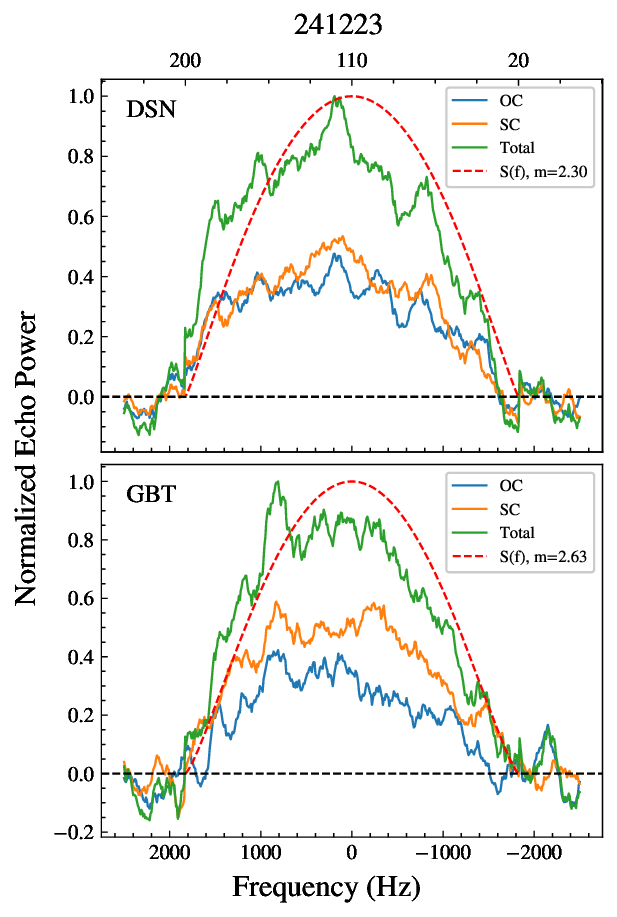}
\figsetgrpnote{Echo power spectra for Goldstone (aka DSN, top) and GBT (bottom) at the epoch indicated at the top of the plot (YYMMDD). 
The blue, orange, and green lines represent the OC, SC, and total (OC+SC) echo power, respectively. 
The power is normalized to the total echo power’s maximum. The red dashed line represents the scattering law S(f) with the fitted exponent m. 
The number at the center just above the top panel is the subradar west longitude in degrees, whereas the numbers on the left and right are the west longitudes of the target’s approaching (left) and receding (right) limbs, respectively. 
Note that frequency on the x-axis increases from right to left.}
\figsetgrpend

\figsetend

The area under the echo and above the noise baseline in the power spectrum is a direct measurement of the {\em radar echo power}, or power received due to backscattering from the target, $P_{\rm received}$.

We divided this power by fluctuations in the radio noise power within the target's limb-to-limb bandwidth, $k T_{\rm sys} B_{\rm ll} / \sqrt{\Delta f \Delta t}$, to obtain the signal-to-noise ratios (SNR) of the echoes at each epoch (Tables \ref{tab:Goldstone_results} and \ref{tab:gbt_results}).
Here $\Delta f$ and $\Delta t$ are the (smoothed) frequency resolution of the spectra and total integration time at each epoch, respectively. 
${\rm SNR}_{\rm OC}$ and ${\rm SNR}_{\rm SC}$ are the SNR for the OC and SC signals, respectively, and ${\rm SNR}_{\rm tot}={\rm SNR}_{\rm OC}+{\rm SNR}_{\rm SC}$.
The quantity ${\rm SNR}_{\rm tot}$ enables identification of the epochs with the highest overall SNR. Although one could define a total SNR by summing the  SNRs in each polarization in quadrature, here we choose instead the simple sum of the SNRs in each polarization, mirroring the conventional calculation for the total radar albedo ($\hat\sigma_{\rm tot}$), which is described later in this section.

The radar echo power can be calculated with the standard radar equation \citep{RevModPhys.65.1235}, which involves the {\em radar cross section} $\sigma$, i.e., the projected area of a perfectly reflective metal sphere that would give the same echo power as the target. The standard radar equation can be written as 
\begin{equation}
\sigma=\frac{P_{\rm received}(4\pi)^3R^4}{P_{\rm tx}G_{\rm tx}G_{\rm rcv}\lambda^2},
\end{equation}
where $R$ is the distance to the target, $P_{\rm tx}$ is the transmitted power, $G_{\rm tx}$ is the antenna gain of the transmitter, $G_{\rm rcv}$ is the antenna gain of the receiver, and $\lambda$ is the wavelength. We calculated $R$  from the ephemeris prediction of the round trip time at each epoch, and $P_{\rm tx}$ is the average transmitted power as listed in Table \ref{tab:Goldstone_observation}. 

After determining the SC and OC cross sections, we also calculated the corresponding {\em radar albedos} $\hat\sigma$, i.e., the radar cross sections divided by the projected area of the target, $\hat\sigma = \sigma/\pi r^2$, where r is the radius of the satellite. The total radar albedo $\hat\sigma_{\rm tot}$ is equal to the sum of $\hat\sigma_{\rm OC}$ and $\hat\sigma_{\rm SC}$.

Important information about scattering processes can be gained by calculating the {\em circular polarization ratio}, $\mu_c = \hat\sigma_{\rm SC}/\hat\sigma_{\rm OC}$.
A smooth surface that produces predominantly specular, single-bounce reflections returns most of the echo power in the opposite-sense circular polarization, resulting in a low circular polarization ratio, as is commonly observed for terrestrial planets \citep{RevModPhys.65.1235}. In contrast, multiple scattering events, including double bounce reflections and scattering by wavelength-scale structures, yield more power in the same-sense circular polarization, and therefore a higher circular polarization ratio.  The series of polarization-preserving, forward-scattering events hypothesized in the CBOE are consistent with a high circular polarization ratio.

Our measurements of the radar albedos and circular polarization ratios are listed in Tables \ref{tab:Goldstone_results} and \ref{tab:gbt_results}. 
The uncertainties in the radar albedo estimates are primarily driven by systematic errors associated with antenna pointing calibration, gain calibration, system temperature calibration, and transmitter power calibration. In 1980s and 1990s radar astronomy practice, investigators often assigned systematic uncertainties to radar albedo estimates of approximately 20\%.
Systematic uncertainties associated with data presented in this work may be slightly better 
due to improvements in pointing performance and calibration techniques, but we will use 20\% for now as a conservative estimate.
In addition to systematic uncertainties, our measurements are also affected by statistical uncertainties arising from
receiver thermal noise, atmospheric fluctuations, small variations in antenna pointing and transmitter performance, etc. To estimate the statistical uncertainty for each epoch, we divided the full power spectrum -- obtained by summing over the consecutive spectra collected during the data receive window (Tables \ref{tab:Goldstone_observation} and \ref{tab:gbt_observation}) -- into 10 independent spectra, thereby generating 10 independent measurements for each epoch. We then calculated the radar albedo from each of the 10 spectra. The mean of these 10 albedo measurements was found to agree very closely with the albedo derived from the full power spectrum, differing by no more than
$\sim$0.01. Therefore, this mean is an excellent estimate of the albedo value for each epoch. The statistical uncertainty of the albedo was then taken to be the standard error of the mean, defined as $\sigma_{\rm stat}=\sigma_{\rm sample}/\sqrt{N}$, where $\sigma_{\rm sample}$ is the sample standard deviation of the 10 albedo measurements and $N=10$. 
The total uncertainty for each albedo estimate was then calculated by adding the systematic and statistical uncertainties in quadrature: $\sigma_{\rm tot}=\sqrt{\sigma_{\rm sys}^2+\sigma_{\rm stat}^2}$.

Uncertainties in the circular polarization ratio were estimated using the standard error propagation formula:
\begin{equation} \label{eq:error_propagation}
\Delta{\mu_c}=\mu_c\sqrt{\left(\frac{\Delta\hat\sigma_{\rm sc}}{\hat\sigma_{\rm sc}}\right)^2+\left(\frac{\Delta\hat\sigma_{\rm oc}}{\hat\sigma_{\rm oc}}\right)^2-2\frac{{\rm cov}(\hat\sigma_{\rm sc},\hat\sigma_{\rm oc})}{\hat\sigma_{\rm sc}\hat\sigma_{\rm oc}}},
\end{equation}
where $\Delta\hat\sigma_{\rm sc}$ and $\Delta\hat\sigma_{\rm oc}$ are the statistical uncertainties for the SC and OC albedo, respectively, and ${\rm cov}(\hat\sigma_{\rm sc},\hat\sigma_{\rm oc})$ is the covariance between the SC and OC albedo. 
In this calculation, we used only the statistical uncertainties for the SC and OC albedo, rather than the total uncertainties, because systematic effects common to both polarizations appear in both the numerator and denominator and largely cancel out.
The covariance term in Equation \ref{eq:error_propagation} is calculated as the covariance of two sample means, since each albedo recorded in Table \ref{tab:Goldstone_results} and \ref{tab:gbt_results} is essentially the mean of the albedos measured from the 10 sub-spectra of that epoch, as explained in the previous paragraph. Thus, we first calculated the covariance matrix of the SC and OC albedos for the 10 sub-spectra, then we divided the off-diagonal term by the number of sub-spectra (10) to obtain the covariance between the sample means.

\begin{figure}[ht!]
\plotone{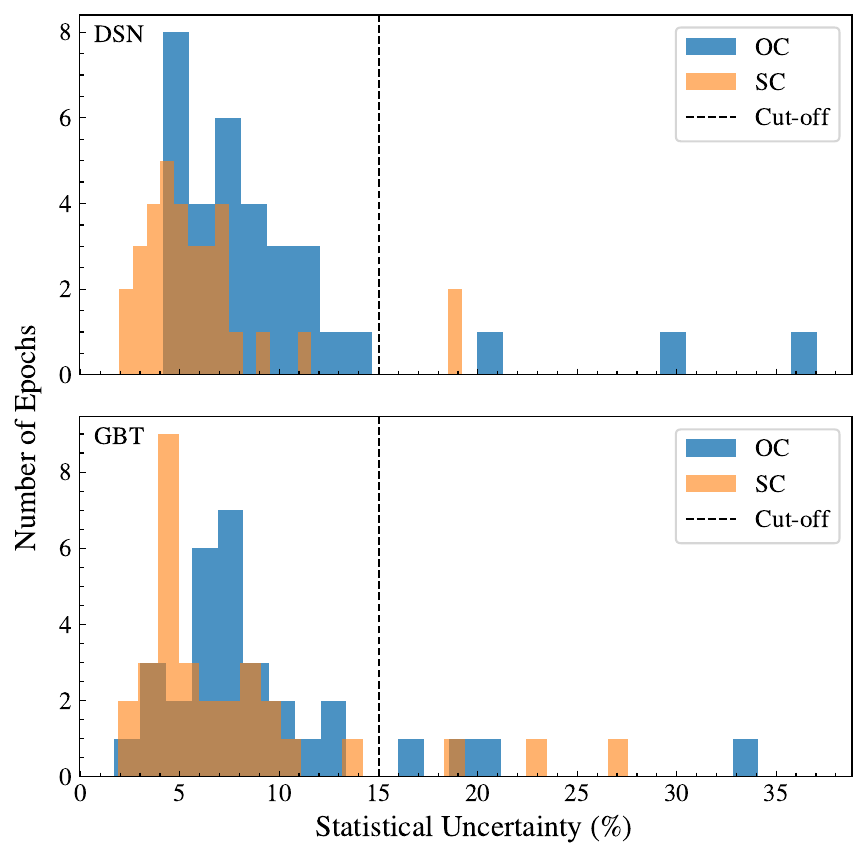}
\caption{Histogram of statistical uncertainties for Goldstone data (aka DSN, top) and GBT data (bottom). The histogram is not normalized so the y-axis shows the number of epochs.
  The $15\%$
  cut-off
  effectively separates the main distribution from the outliers.  
\label{fig:rel_unc}}
\end{figure}

For a few epochs, the statistical uncertainties are unusually large due to factors such as antenna tracking errors, transmitter performance issues, and weather fluctuations, all of which can degrade the data quality. As a result, these epochs do not provide reliable measurements of the radar scattering properties. We therefore identified epochs for which the statistical uncertainty of either the OC or SC albedo exceeded $15\%$, flagged them in Tables \ref{tab:Goldstone_results} and \ref{tab:gbt_results}, and excluded them from subsequent analyses. The $15\%$ cut-off was chosen because histograms of the statistical uncertainties for all epochs (Figure \ref{fig:rel_unc}) show that this value effectively separates the main distribution from the outliers. 

In addition, two Goldstone observing epochs (2023 OCT 18 and 2024 DEC 23) were excluded from subsequent analyses because a comparison of the circular polarization ratios measured from the Goldstone and GBT datasets reveals that the Goldstone polarization ratios for these two epochs are clear outliers relative to the rest of the epochs (Appendix \ref{app:pol_ratio_distribution}). We therefore suspect that these observations do not provide reliable measurements of Europa's radar scattering properties.

Finally, the backscattering behavior of planetary surfaces can be expressed by a global scattering law. For icy satellites, an appropriate scattering law takes the form \(d\sigma/dA \sim \cos^m(\theta)\), where $d\sigma$ is an infinitesimal increment of radar cross section, 
$dA$ represents an infinitesimal surface area element on the satellite, and $\theta$ is the angle of incidence between the incoming radio wave and the outward surface normal of $dA$ \citep{Ostro1992}. Integration of this law over all possible incidence angle values at each Doppler frequency yields an expression of the form
\begin{equation}
  S(f) \sim \left(1-\frac{f^2}{(B_{\rm ll}/2)^2}\right)^\frac{m}{2},
\label{eq:spectral_shape}
\end{equation}
where $f$ is the Doppler frequency, $B_{\rm ll}$ is the limb-to-limb bandwidth, and $m$ is the exponent that provides insight into the disk-integrated scattering properties of the target \citep{simp73,Ostro1992}. We estimated $m$ by fitting $S(f)$ to the normalized total (OC + SC) echo power spectra using
non-linear least-squares optimization (Figure \ref{fig:example_result}).
The values for $m$ and their root reduced chi-square (RRC) are listed in Tables \ref{tab:Goldstone_results} and \ref{tab:gbt_results}. 

\setlength{\tabcolsep}{3pt}
\begin{deluxetable*}{cccccccccccc}[!t]
\tabletypesize{\footnotesize}
\tablewidth{0pt}
\tablecaption{Goldstone Results for Europa \label{tab:Goldstone_results}}
\tablehead{
  \colhead{Date} & \colhead{$\hat\sigma_{\rm OC}$}  & \colhead{${\rm SNR}_{\rm OC}$} & \colhead{$\Delta\hat\sigma_{\rm OC,stat}$} & \colhead{$\hat\sigma_{\rm SC}$} &  \colhead{${\rm SNR}_{\rm SC}$} & \colhead{$\Delta\hat\sigma_{\rm SC,stat}$} & \colhead{$\hat\sigma_{\rm tot}$}  & \colhead{${\rm SNR}_{\rm tot}$} & \colhead{$\mu_c$} & \colhead{$m$} & \colhead{RRC}
}
\startdata
2011 SEP 24  &  0.700  &  8.49  &  9.97  &  1.17  &  14.3  &  4.87  &  1.87  &  22.8  &  1.67 $\pm$ 0.14  &  2.10 $\pm$ 0.04  &  1.89 \\
2011 SEP 27  &  1.071  &  12.7  &  7.82  &  1.46  &  17.4  &  3.49  &  2.53  &  30.1  &  1.37 $\pm$ 0.13  &  1.56 $\pm$ 0.03  &  1.48 \\
2011 SEP 29  &  0.816  &  12.5  &  5.58  &  1.25  &  19.0  &  4.42  &  2.06  &  31.4  &  1.53 $\pm$ 0.12  &  1.76 $\pm$ 0.03  &  1.86 \\
2011 OCT 01  &  0.994  &  13.5  &  5.08  &  1.46  &  19.8  &  1.95  &  2.45  &  33.3  &  1.47 $\pm$ 0.08  &  1.73 $\pm$ 0.03  &  2.30 \\
2011 OCT 03  &  0.924  &  12.1  &  5.75  &  1.33  &  17.3  &  3.50  &  2.26  &  29.3  &  1.44 $\pm$ 0.09  &  2.03 $\pm$ 0.04  &  2.20 \\
2011 OCT 06  &  0.811  &  8.91  &  7.90  &  1.13  &  12.3  &  7.18  &  1.94  &  21.2  &  1.39 $\pm$ 0.16  &  1.88 $\pm$ 0.04  &  1.22 \\
2011 OCT 10  &  0.970  &  10.7  &  7.06  &  1.50  &  16.6  &  3.26  &  2.47  &  27.2  &  1.54 $\pm$ 0.11  &  2.25 $\pm$ 0.06  &  2.76 \\
2011 OCT 13  &  1.006  &  10.8  &  4.76  &  1.59  &  16.9  &  5.27  &  2.59  &  27.7  &  1.58 $\pm$ 0.07  &  2.02 $\pm$ 0.04  &  1.47 \\
2011 OCT 15\textsuperscript{\textdagger}  &  0.668  &  7.98  &  21.0  &  0.92  &  11.1  &  19.0  &  1.59  &  19.1  &  1.38 $\pm$ 0.09  &  2.47 $\pm$ 0.07  &  1.94 \\
2011 OCT 17  &  0.870  &  10.5  &  5.35  &  1.27  &  15.6  &  5.76  &  2.14  &  26.1  &  1.47 $\pm$ 0.14  &  1.99 $\pm$ 0.04  &  1.68 \\
2011 OCT 19  &  0.908  &  10.3  &  10.3  &  1.33  &  15.1  &  5.17  &  2.24  &  25.5  &  1.46 $\pm$ 0.15  &  2.09 $\pm$ 0.06  &  2.39 \\
2011 OCT 20  &  0.842  &  10.5  &  6.55  &  1.29  &  16.4  &  4.65  &  2.13  &  26.9  &  1.53 $\pm$ 0.12  &  2.56 $\pm$ 0.05  &  2.17 \\
2011 OCT 24  &  0.812  &  10.2  &  8.71  &  1.24  &  15.9  &  5.78  &  2.05  &  26.1  &  1.53 $\pm$ 0.16  &  2.19 $\pm$ 0.05  &  1.91 \\
2011 OCT 26  &  0.966  &  11.5  &  7.16  &  1.35  &  16.4  &  3.16  &  2.31  &  27.8  &  1.40 $\pm$ 0.09  &  1.82 $\pm$ 0.03  &  1.46 \\
2011 OCT 29\textsuperscript{\textdagger}  &  0.308  &  2.52  &  30.1  &  0.61  &  4.99  &  11.3  &  0.92  &  7.51  &  1.99 $\pm$ 0.58  &  4.71 $\pm$ 0.21  &  1.20 \\
2011 OCT 31  &  0.964  &  11.8  &  4.13  &  1.36  &  17.0  &  2.62  &  2.32  &  28.8  &  1.41 $\pm$ 0.07  &  1.85 $\pm$ 0.02  &  1.39 \\
2011 NOV 04  &  0.770  &  8.98  &  7.84  &  1.15  &  14.0  &  4.76  &  1.92  &  23.0  &  1.50 $\pm$ 0.18  &  1.98 $\pm$ 0.05  &  2.03 \\
2011 NOV 05  &  0.896  &  10.6  &  8.19  &  1.47  &  17.3  &  3.89  &  2.36  &  27.8  &  1.64 $\pm$ 0.14  &  2.34 $\pm$ 0.05  &  2.56 \\
2011 NOV 07  &  0.809  &  6.57  &  12.2  &  1.21  &  9.68  &  9.51  &  2.02  &  16.2  &  1.50 $\pm$ 0.23  &  1.98 $\pm$ 0.06  &  1.70 \\
2011 NOV 11  &  0.920  &  8.64  &  11.4  &  1.54  &  14.5  &  4.03  &  2.46  &  23.1  &  1.68 $\pm$ 0.17  &  2.29 $\pm$ 0.05  &  1.81 \\
2011 NOV 16  &  1.014  &  11.9  &  4.91  &  1.41  &  16.6  &  3.58  &  2.42  &  28.4  &  1.39 $\pm$ 0.09  &  1.92 $\pm$ 0.04  &  1.61 \\
2011 NOV 18  &  0.703  &  8.43  &  9.85  &  1.18  &  14.3  &  6.19  &  1.88  &  22.7  &  1.68 $\pm$ 0.23  &  2.80 $\pm$ 0.09  &  2.95 \\
2011 NOV 23  &  1.097  &  8.51  &  8.18  &  1.52  &  12.0  &  7.16  &  2.62  &  20.5  &  1.39 $\pm$ 0.17  &  1.83 $\pm$ 0.04  &  1.47 \\
2023 OCT 18\textsuperscript{\textdagger}  &  0.661  &  4.56  &  14.1  &  1.39  &  9.29  &  6.63  &  2.05  &  13.8  &  2.11 $\pm$ 0.34  &  2.73 $\pm$ 0.08  &  1.33 \\
2023 OCT 25\textsuperscript{\textdagger}  &  0.191  &  1.25  &  37.1  &  0.59  &  3.87  &  19.2  &  0.78  &  5.12  &  3.10 $\pm$ 1.37  &  9.41 $\pm$ 0.42  &  1.72 \\
2023 OCT 30  &  0.881  &  5.55  &  11.6  &  1.46  &  9.18  &  7.91  &  2.34  &  14.7  &  1.66 $\pm$ 0.19  &  2.62 $\pm$ 0.10  &  2.10 \\
2023 NOV 10  &  0.915  &  6.60  &  11.2  &  1.41  &  10.1  &  5.75  &  2.33  &  16.7  &  1.54 $\pm$ 0.21  &  2.10 $\pm$ 0.06  &  1.38 \\
2023 NOV 12  &  1.097  &  7.95  &  8.87  &  1.47  &  10.5  &  7.18  &  2.56  &  18.4  &  1.34 $\pm$ 0.13  &  2.17 $\pm$ 0.06  &  2.20 \\
2024 DEC 03  &  1.091  &  12.3  &  5.18  &  1.40  &  15.9  &  4.27  &  2.50  &  28.2  &  1.29 $\pm$ 0.08  &  1.97 $\pm$ 0.05  &  1.79 \\
2024 DEC 07  &  1.000  &  12.4  &  4.74  &  1.37  &  17.3  &  4.37  &  2.37  &  29.8  &  1.37 $\pm$ 0.10  &  1.94 $\pm$ 0.05  &  1.95 \\
2024 DEC 12  &  0.931  &  11.6  &  6.32  &  1.32  &  16.6  &  3.21  &  2.26  &  28.2  &  1.42 $\pm$ 0.10  &  1.65 $\pm$ 0.03  &  1.67 \\
2024 DEC 21  &  0.997  &  11.8  &  4.43  &  1.21  &  15.0  &  6.59  &  2.21  &  26.7  &  1.22 $\pm$ 0.09  &  2.08 $\pm$ 0.06  &  2.74 \\
2024 DEC 23\textsuperscript{\textdagger}  &  0.973  &  10.6  &  7.11  &  0.99  &  10.9  &  7.47  &  1.96  &  21.6  &  1.02 $\pm$ 0.08  &  2.30 $\pm$ 0.09  &  2.77 \\
\hline
\multirow{2}{*}{\parbox[c]{2.3cm}{\centering Mean\\($G_0$=74.23 dBi)}}
& \multirow{2}{*}{\parbox[c]{1cm}{\centering 0.92\\$\pm$ 0.11}}
& \multirow{2}{*}{}
& \multirow{2}{*}{}
& \multirow{2}{*}{\parbox[c]{1cm}{\centering 1.35\\$\pm$ 0.13}}
& \multirow{2}{*}{}
& \multirow{2}{*}{}
& \multirow{2}{*}{\parbox[c]{1cm}{\centering 2.27\\$\pm$ 0.22}}
& \multirow{2}{*}{}
& \multirow{2}{*}{1.44 $\pm$ 0.12}
& \multirow{2}{*}{1.91 $\pm$ 0.31}
& \multirow{2}{*}{} \\
&&&&&&&&&&&\\
\enddata
\tablecomments{Radar scattering properties of Europa from Goldstone observations. Dates are given in UTC. $\hat\sigma_{\rm OC}$ and ${\rm SNR}_{\rm OC}$ are the radar albedo and signal-to-noise ratio in OC, $\hat\sigma_{SC}$ and ${\rm SNR_{SC}}$ are the radar albedo and signal-to-noise ratio in SC, and $\hat\sigma_{\rm tot}$ and ${\rm SNR}_{\rm tot}$ are the total radar albedo and total signal-to-noise ratio. 
$\Delta\hat\sigma_{\rm OC,stat}$ and $\Delta\hat\sigma_{\rm SC,stat}$ are the statistical uncertainties of the OC and SC albedos, respectively, expressed as percentages of the corresponding albedo values.
$\mu_c$ is the circular polarization ratio given by $\hat\sigma_{sc}/\hat\sigma_{oc}$. The uncertainties listed for $\mu_c$ are calculated using Equation \ref{eq:error_propagation}. $m$ is the exponent of the global scattering law that gives rise to Equation \ref{eq:spectral_shape}, and RRC is the root reduced chi-square of the fit of Equation \ref{eq:spectral_shape} to the normalized total echo power spectra. The uncertainty listed in $m$ is the formal uncertainty of the fit.
  The epochs with a dagger superscript either have an albedo with statistical uncertainty $> 15\%$ or have
    anomalous polarization ratios and are excluded from subsequent analyses, leaving 28 epochs in the Goldstone dataset.
The last row lists the means and root-mean-square (rms) dispersions of the radar albedos, $\mu_c$, and $m$. The albedo means are unweighted, while the means of $\mu_c$ and $m$ are weighted by their uncertainties listed in the table.}
\end{deluxetable*}

\setlength{\tabcolsep}{3pt}
\begin{deluxetable*}{cccccccccccc}[!t]
\tabletypesize{\footnotesize}
\tablewidth{0pt}
\tablecaption{GBT Results for Europa \label{tab:gbt_results}}
\tablehead{
\colhead{Date} & \colhead{$\hat\sigma_{\rm OC}$}  & \colhead{${\rm SNR}_{\rm OC}$} & \colhead{$\Delta\hat\sigma_{\rm OC,stat}$} & \colhead{$\hat\sigma_{\rm SC}$} &  \colhead{${\rm SNR}_{\rm SC}$} & \colhead{$\Delta\hat\sigma_{\rm SC,stat}$} & \colhead{$\hat\sigma_{\rm tot}$}  & \colhead{${\rm SNR}_{\rm tot}$} & \colhead{$\mu_c$} & \colhead{$m$} & \colhead{RRC}
}
\startdata
2011 SEP 24\textsuperscript{\textdagger}  &  0.799  &  3.24  &  34.1  &  1.13  &  4.63  &  13.6  &  1.92  &  7.87  &  1.41 $\pm$ 0.51  &  2.02 $\pm$ 0.09  &  1.20 \\
2011 SEP 27  &  0.821  &  15.6  &  8.12  &  1.21  &  23.4  &  3.55  &  2.03  &  38.9  &  1.48 $\pm$ 0.14  &  1.85 $\pm$ 0.04  &  2.83 \\
2011 SEP 29  &  0.607  &  11.2  &  9.04  &  1.01  &  18.9  &  6.34  &  1.61  &  30.0  &  1.66 $\pm$ 0.12  &  1.66 $\pm$ 0.02  &  1.53 \\
2011 OCT 01  &  0.739  &  12.6  &  4.57  &  1.16  &  20.2  &  4.02  &  1.90  &  32.8  &  1.57 $\pm$ 0.09  &  2.23 $\pm$ 0.05  &  2.39 \\
2011 OCT 03  &  0.829  &  12.8  &  3.64  &  1.14  &  18.0  &  5.79  &  1.97  &  30.9  &  1.38 $\pm$ 0.07  &  2.28 $\pm$ 0.04  &  2.29 \\
2011 OCT 06  &  0.623  &  9.00  &  7.78  &  1.01  &  14.8  &  9.37  &  1.63  &  23.8  &  1.62 $\pm$ 0.11  &  2.07 $\pm$ 0.05  &  1.91 \\
2011 OCT 10  &  0.806  &  10.5  &  6.64  &  1.11  &  14.7  &  4.90  &  1.92  &  25.2  &  1.38 $\pm$ 0.10  &  2.04 $\pm$ 0.04  &  1.39 \\
2011 OCT 13  &  0.984  &  9.43  &  8.00  &  1.56  &  14.7  &  4.48  &  2.54  &  24.2  &  1.58 $\pm$ 0.11  &  1.96 $\pm$ 0.05  &  1.99 \\
2011 OCT 15\textsuperscript{\textdagger}  &  0.578  &  8.44  &  20.3  &  0.85  &  12.7  &  18.9  &  1.43  &  21.2  &  1.48 $\pm$ 0.11  &  1.98 $\pm$ 0.04  &  1.66 \\
2011 OCT 17  &  0.797  &  11.3  &  3.52  &  1.13  &  16.2  &  3.55  &  1.93  &  27.5  &  1.42 $\pm$ 0.07  &  2.06 $\pm$ 0.05  &  2.31 \\
2011 OCT 19  &  0.811  &  11.4  &  7.15  &  1.32  &  18.8  &  1.87  &  2.13  &  30.2  &  1.63 $\pm$ 0.12  &  2.05 $\pm$ 0.04  &  2.13 \\
2011 OCT 20  &  0.855  &  11.8  &  4.61  &  1.35  &  19.1  &  2.76  &  2.21  &  30.9  &  1.58 $\pm$ 0.08  &  1.95 $\pm$ 0.05  &  3.00 \\
2011 OCT 24  &  0.695  &  9.50  &  5.72  &  1.21  &  16.7  &  3.96  &  1.90  &  26.2  &  1.74 $\pm$ 0.12  &  1.82 $\pm$ 0.04  &  2.12 \\
2011 OCT 26\textsuperscript{\textdagger}  &  0.396  &  5.51  &  19.4  &  0.61  &  8.56  &  23.3  &  1.00  &  14.1  &  1.53 $\pm$ 0.25  &  2.72 $\pm$ 0.11  &  2.17 \\
2011 OCT 29\textsuperscript{\textdagger}  &  0.381  &  2.80  &  12.4  &  0.41  &  3.07  &  27.6  &  0.80  &  5.88  &  1.09 $\pm$ 0.36  &  6.58 $\pm$ 0.43  &  1.91 \\
2011 OCT 31  &  0.830  &  12.3  &  5.60  &  1.31  &  19.7  &  3.06  &  2.14  &  32.0  &  1.57 $\pm$ 0.09  &  1.59 $\pm$ 0.03  &  1.71 \\
2011 NOV 04  &  0.774  &  11.1  &  4.27  &  1.16  &  16.9  &  4.75  &  1.94  &  28.0  &  1.50 $\pm$ 0.09  &  2.01 $\pm$ 0.04  &  1.49 \\
2011 NOV 05  &  0.860  &  12.2  &  1.71  &  1.30  &  18.7  &  4.47  &  2.16  &  30.8  &  1.52 $\pm$ 0.08  &  1.94 $\pm$ 0.04  &  2.31 \\
2011 NOV 07  &  N/A  &  N/A  &  N/A  &  N/A  &  N/A  &  N/A  &  N/A  &  N/A  &  N/A  &  2.04 $\pm$ 0.06  &  2.09 \\
2011 NOV 11  &  0.742  &  8.27  &  7.24  &  1.07  &  11.7  &  9.28  &  1.81  &  20.0  &  1.44 $\pm$ 0.16  &  2.11 $\pm$ 0.05  &  1.82 \\
2011 NOV 16  &  0.952  &  8.75  &  5.90  &  1.37  &  12.4  &  4.03  &  2.32  &  21.1  &  1.44 $\pm$ 0.10  &  2.18 $\pm$ 0.05  &  1.46 \\
2011 NOV 18  &  0.667  &  9.11  &  7.70  &  1.05  &  14.6  &  5.20  &  1.72  &  23.7  &  1.57 $\pm$ 0.13  &  3.01 $\pm$ 0.09  &  2.98 \\
2011 NOV 23  &  0.891  &  7.20  &  10.7  &  1.06  &  8.62  &  11.0  &  1.95  &  15.8  &  1.19 $\pm$ 0.11  &  2.05 $\pm$ 0.06  &  1.48 \\
2023 OCT 18  &  0.927  &  7.55  &  5.75  &  1.17  &  8.32  &  7.64  &  2.10  &  15.9  &  1.27 $\pm$ 0.08  &  2.35 $\pm$ 0.07  &  1.77 \\
2023 OCT 25  &  0.835  &  7.33  &  10.6  &  1.14  &  8.64  &  8.29  &  1.97  &  16.0  &  1.37 $\pm$ 0.19  &  1.98 $\pm$ 0.05  &  1.01 \\
2023 OCT 30  &  N/A  &  N/A  &  N/A  &  N/A  &  N/A  &  N/A  &  N/A  &  N/A  &  N/A  &  2.22 $\pm$ 0.10  &  2.84 \\
2023 NOV 10  &  1.212  &  7.07  &  7.93  &  1.56  &  8.26  &  6.17  &  2.77  &  15.3  &  1.28 $\pm$ 0.15  &  2.22 $\pm$ 0.07  &  1.40 \\
2023 NOV 12  &  0.876  &  8.32  &  9.01  &  1.20  &  9.93  &  7.26  &  2.08  &  18.3  &  1.37 $\pm$ 0.14  &  1.81 $\pm$ 0.04  &  1.11 \\
2024 DEC 03  &  1.166  &  8.35  &  5.91  &  1.46  &  11.7  &  4.50  &  2.63  &  20.0  &  1.25 $\pm$ 0.08  &  1.89 $\pm$ 0.04  &  1.54 \\
2024 DEC 07  &  1.030  &  7.54  &  11.3  &  1.31  &  10.8  &  8.11  &  2.34  &  18.3  &  1.27 $\pm$ 0.17  &  2.51 $\pm$ 0.08  &  2.79 \\
2024 DEC 12  &  0.844  &  7.72  &  8.86  &  1.23  &  12.5  &  5.58  &  2.07  &  20.2  &  1.46 $\pm$ 0.11  &  2.19 $\pm$ 0.06  &  1.88 \\
2024 DEC 21\textsuperscript{\textdagger}  &  1.012  &  5.54  &  16.4  &  1.30  &  7.99  &  8.24  &  2.32  &  13.5  &  1.29 $\pm$ 0.17  &  1.96 $\pm$ 0.06  &  1.22 \\
2024 DEC 23  &  0.692  &  5.86  &  12.7  &  1.12  &  10.7  &  4.23  &  1.81  &  16.5  &  1.62 $\pm$ 0.24  &  2.63 $\pm$ 0.08  &  0.87 \\
\hline
\multirow{2}{*}{\parbox[c]{2.3cm}{\centering Mean\\($G_0$=74.23 dBi)}}
& \multirow{2}{*}{\parbox[c]{1cm}{\centering 0.84\\$\pm$ 0.14}}
& \multirow{2}{*}{}
& \multirow{2}{*}{}
& \multirow{2}{*}{\parbox[c]{1cm}{\centering 1.22\\$\pm$ 0.15}}
& \multirow{2}{*}{}
& \multirow{2}{*}{}
& \multirow{2}{*}{\parbox[c]{1cm}{\centering 2.06\\$\pm$ 0.28}}
& \multirow{2}{*}{}
& \multirow{2}{*}{1.46 $\pm$ 0.14}
& \multirow{2}{*}{1.95 $\pm$ 0.32}
& \multirow{2}{*}{} \\
&&&&&&&&&&&\\
\enddata
\tablecomments{Radar scattering properties of Europa from the GBT observations. Dates are given in UTC. $\hat\sigma_{\rm OC}$ and ${\rm SNR}_{\rm OC}$ are the radar albedo and signal-to-noise ratio in OC, $\hat\sigma_{\rm SC}$ and ${\rm SNR}_{\rm SC}$ are the radar albedo and signal-to-noise ratio in SC, and $\hat\sigma_{\rm tot}$ and ${\rm SNR}_{\rm tot}$ are the total radar albedo and total signal-to-noise ratio. 
$\Delta\hat\sigma_{\rm OC,stat}$ and $\Delta\hat\sigma_{\rm SC,stat}$ are the statistical uncertainties of the OC and SC albedos, respectively, expressed as percentages of the corresponding albedo values.
$\mu_c$ is the circular polarization ratio given by $\hat\sigma_{sc}/\hat\sigma_{oc}$. The uncertainties listed for $\mu_c$ are calculated using Equation \ref{eq:error_propagation}. $m$ is the exponent of the global scattering law that gives rise to Equation \ref{eq:spectral_shape}, and RRC is the root reduced chi-square of the fit of Equation \ref{eq:spectral_shape} to the normalized total echo power spectra. The uncertainty listed in $m$ is the formal uncertainty of the fit. 
The epochs with a dagger superscript have an albedo with statistical uncertainty $> 15\%$ and are therefore excluded from subsequent analyses. There are also two epochs that do not have an estimate of system temperature, so their radar albedos are listed as ``N/A", and they are excluded from future analyses as well. In the end, 26 epochs remain in the GBT dataset.
The last row lists the means and root-mean-square (rms) dispersions of the radar albedos, $\mu_c$, and $m$. The albedo means are unweighted, while the means of $\mu_c$ and $m$ are weighted by their uncertainties listed in the table.}
\end{deluxetable*}

\begin{figure*}[ht!]
\centering
\includegraphics[width=\textwidth,height=0.35\textheight,keepaspectratio]{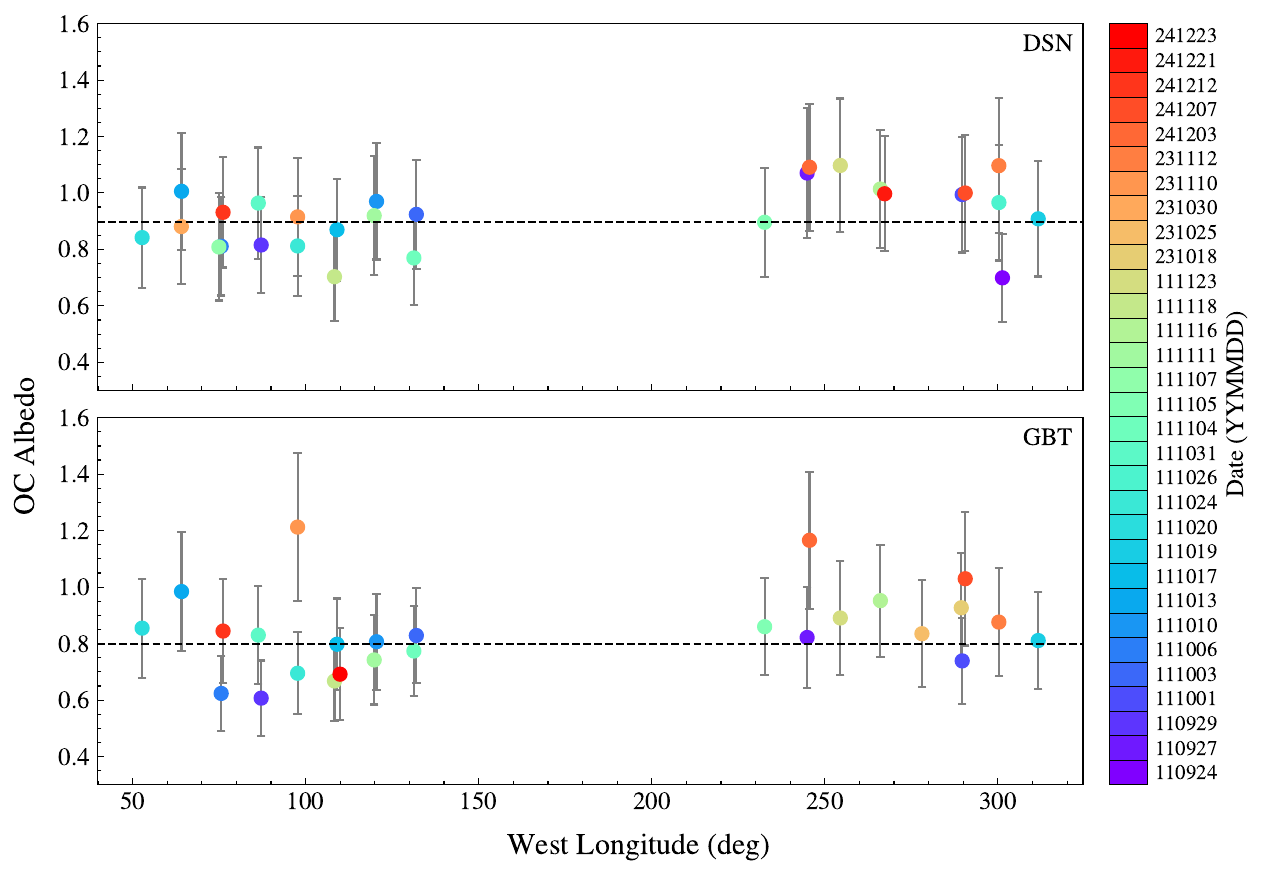}
\caption{(Top) Measured OC albedo values from Goldstone (from Table \ref{tab:Goldstone_results}) and (bottom) measured OC albedo values from the GBT (from Table \ref{tab:gbt_results}). The horizontal dashed lines indicate the weighted means of the albedos, which is 0.90 for Goldstone and 0.80 for GBT. The error bars on our albedo values are taken to be the total uncertainty calculated in Section \ref{sec:results}. The color bar on the right shows the dates of our observations in YYMMDD. The x-axis is the subradar point west longitude.
\label{fig:oc_results_long}}
\end{figure*}

\begin{figure*}[ht!]
\centering
\includegraphics[width=\textwidth,height=0.35\textheight,keepaspectratio]{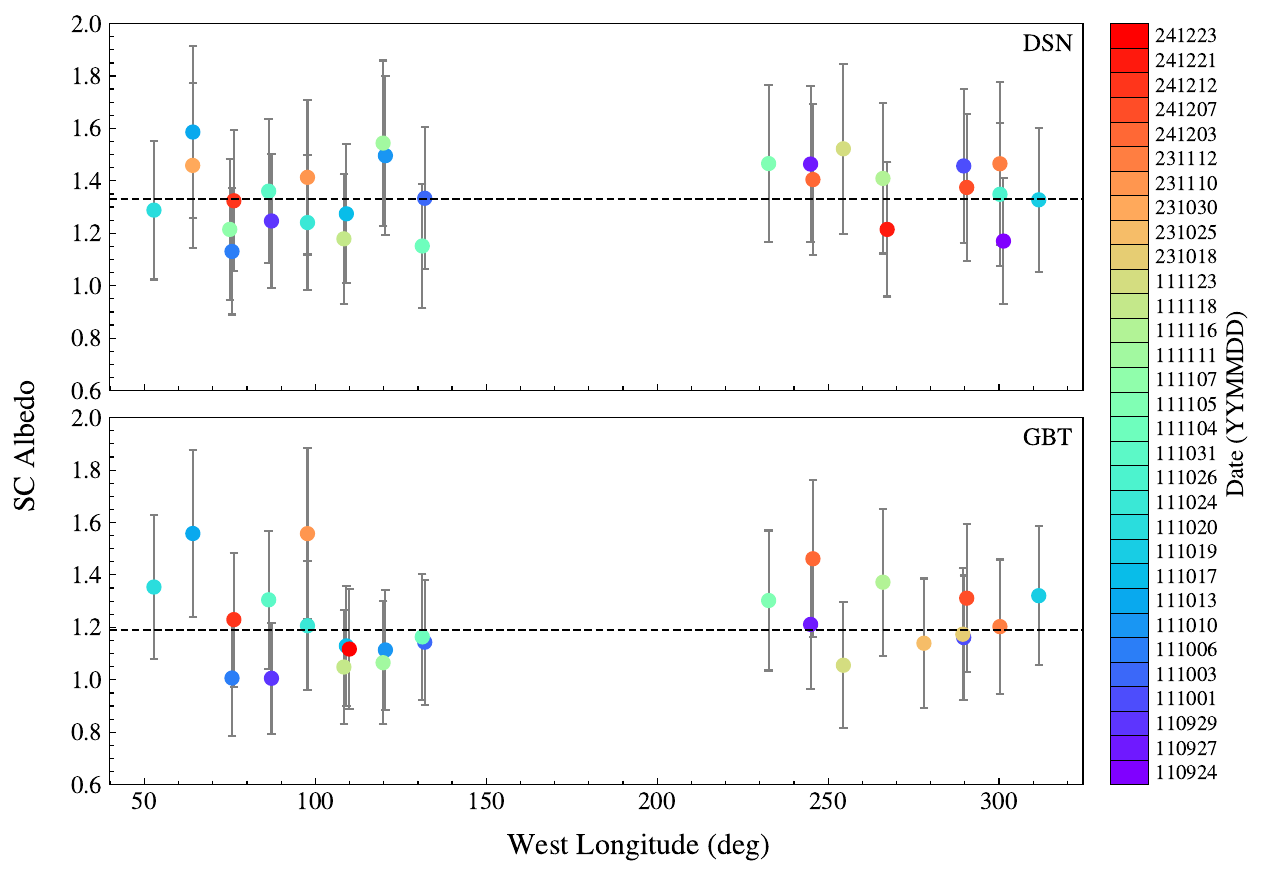}
\caption{(Top) Measured SC albedo values from Goldstone (from Table \ref{tab:Goldstone_results}) and (bottom) measured SC albedo values from the GBT (from Table \ref{tab:gbt_results}). The horizontal dashed lines indicate the weighted means of the albedos, which is 1.33 for Goldstone and 1.19 for GBT. The error bars on our albedo values are taken to be the total uncertainty calculated in Section \ref{sec:results}. The color bar on the right shows the dates of our observations in YYMMDD. The x-axis is the subradar point west longitude.
  \label{fig:sc_results_long}}
\end{figure*}

\begin{figure*}[ht!]
\centering
\includegraphics[width=\textwidth,height=0.35\textheight,keepaspectratio]{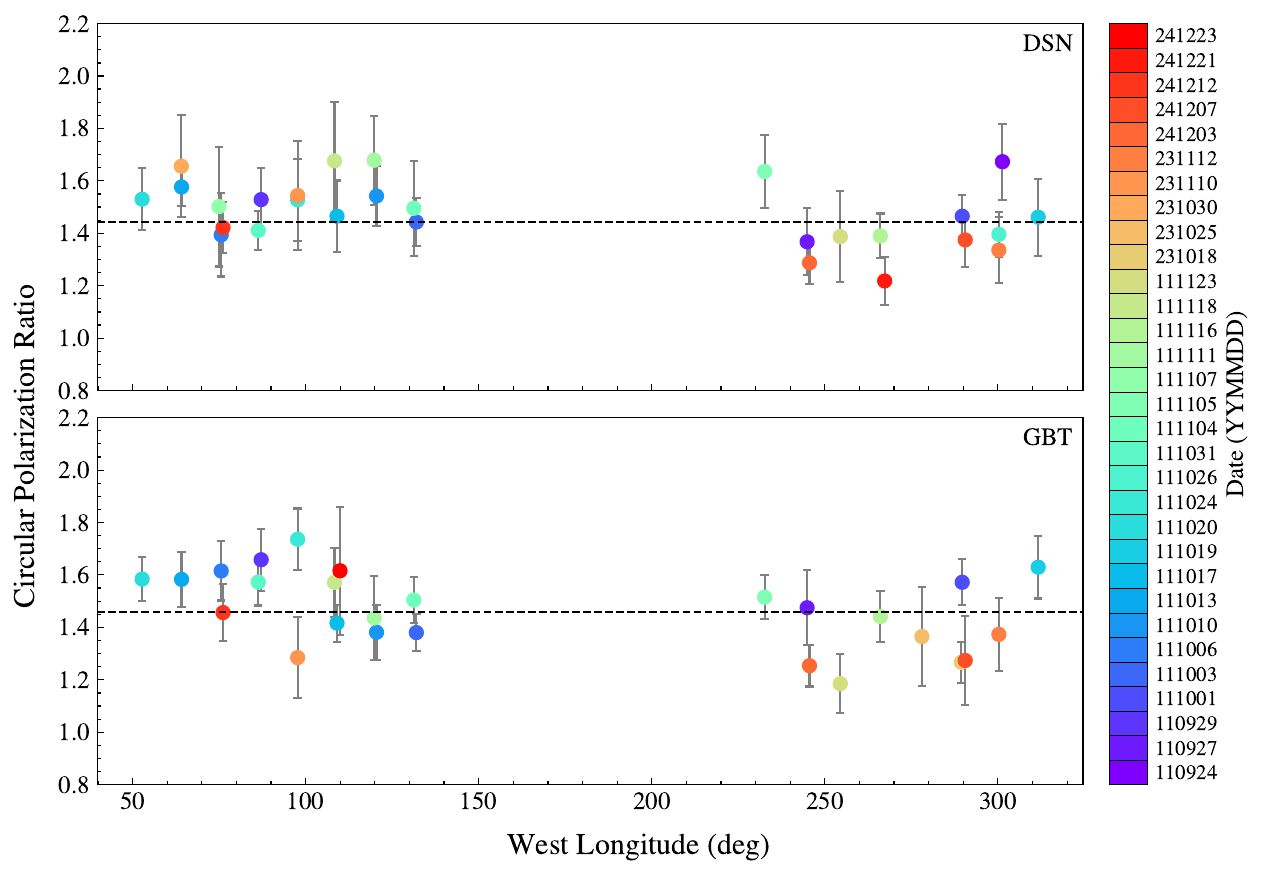}
\caption{(Top) Measured circular polarization ratio from Goldstone (from Table \ref{tab:Goldstone_results}) and (bottom) measured circular polarization ratio from the GBT (from Table \ref{tab:gbt_results}). The horizontal dashed lines indicate the weighted means of the ratios, which is 1.44 for Goldstone and 1.46 for GBT. The error bars on our polarization ratios are taken to be the uncertainty calculated in Equation \ref{eq:error_propagation}. The color bar on the right shows the dates of our observations in YYMMDD. The x-axis is the subradar point west longitude.
  \label{fig:pol_ratio_long}}
\end{figure*}

The OC and SC radar albedo values and the circular polarization ratios are shown as a function of west longitudes on Figures \ref{fig:oc_results_long}, \ref{fig:sc_results_long}, and \ref{fig:pol_ratio_long}.
For Goldstone, the unweighted means and root-mean-square (rms) dispersions of our radar albedos are $\hat\sigma_{\rm OC}$ = 0.92 $\pm$ 0.11 and $\hat\sigma_{\rm SC}$ = 1.35 $\pm$ 0.13 for OC and SC, respectively. The weighted mean and rms dispersion of our circular polarization ratio measurements is $\mu_c$ = 1.44 $\pm$ 0.12.  For GBT, the unweighted means and rms dispersions of our radar albedos are $\hat\sigma_{\rm OC}$ = 0.84 $\pm$ 0.14 and $\hat\sigma_{\rm SC}$ = 1.22 $\pm$ 0.15 for OC and SC, respectively. The weighted mean and rms dispersion of our circular polarization ratio measurements is $\mu_c$ = 1.46 $\pm$ 0.14. 
Our results confirm the unusually high ($>1$) circular polarization ratios of icy satellites, which led to the early recognition that their icy crusts enable an unusual scattering mechanism~\citep[e.g.,][]{camp78,ostr80}.  
The weighted mean and rms dispersion of our estimates of the exponent in the radar scattering law is $m$ = 1.91 $\pm$ 0.31 for Goldstone and 1.95 $\pm$ 0.32 for GBT.  Our result confirms the highly diffuse scattering behavior of icy satellites, unlike the more specular scattering observed on terrestrial planets \citep{RevModPhys.65.1235}.

\section{Discussion} \label{sec:discussion}

\subsection{Disk-integrated properties}
\begin{figure*}[ht!]
\centering
\includegraphics[width=\textwidth,height=0.35\textheight,keepaspectratio]{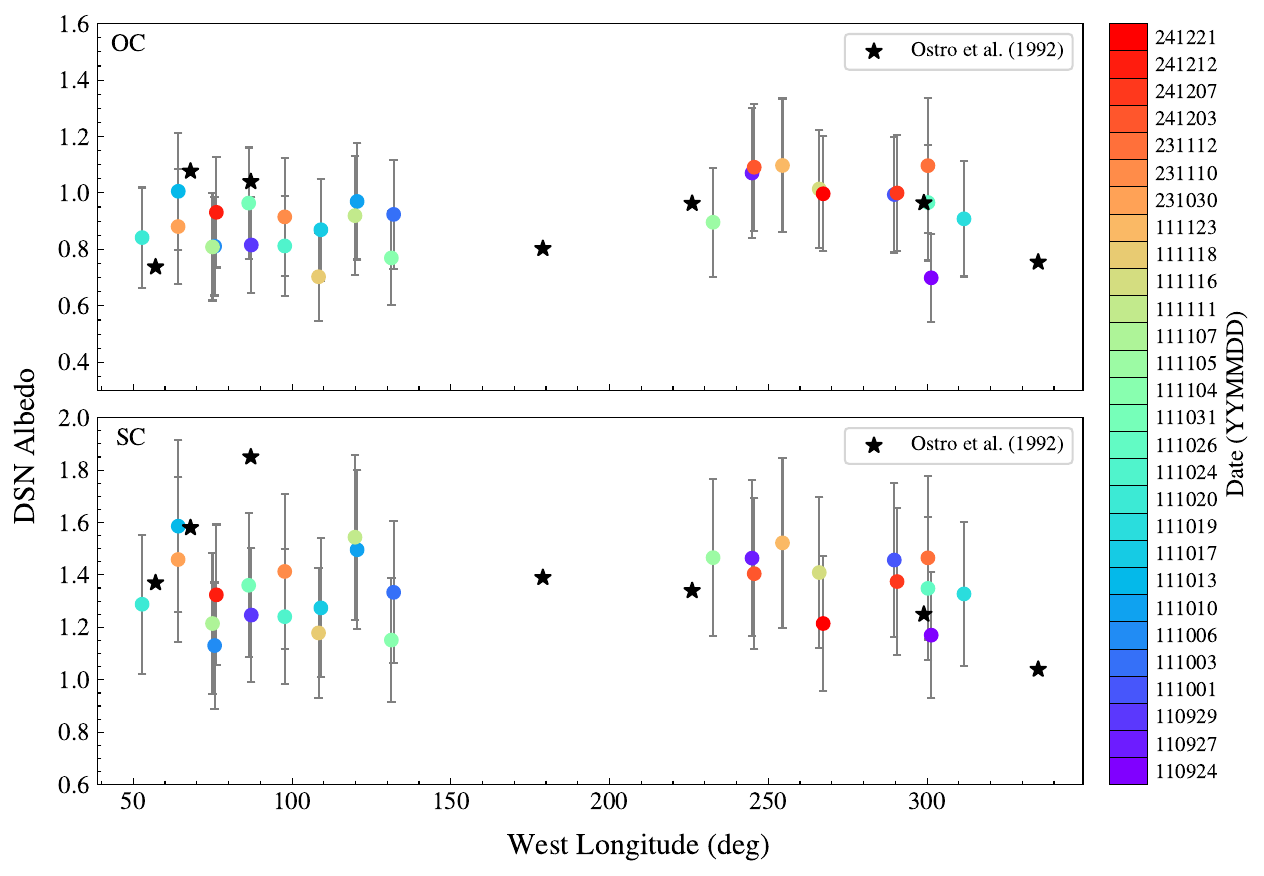}
\caption{ Measured OC (top) and SC (bottom) albedo values from Goldstone (from Table \ref{tab:Goldstone_results}) as well as \citet{Ostro1992}'s OC and SC albedo values (black stars). At comparable longitudes, our values agree well with \citet{Ostro1992}'s, except for epochs at west longitude $\sim$87 degrees, where \citet{Ostro1992}'s SC albedo appears larger than ours, but they still agree within uncertainty, given that the uncertainty of \citet{Ostro1992}'s values are around 35\%, the average of their published range of uncertainties (20\%--50\%). The error bars on our values are taken to be the total uncertainty calculated in Section \ref{sec:results}. The color bar on the right indicates the dates of our observations in YYMMDD. The x-axis is the subradar point west longitude.
\label{fig:dsn_results_ostro}}
\end{figure*}

We compare our results to the most recent 3.5-cm radar results of Europa published in 1992 \citep{Ostro1992}. In that study, the 3.5-cm Goldstone OC radar albedo for Europa has an unweighted mean and root-mean-square (rms) dispersion of 0.91 $\pm$ 0.13, and the corresponding SC radar albedo has an unweighted mean and rms dispersion of 1.40 $\pm$ 0.23. These values are consistent with our results of 0.92 $\pm$ 0.11 for OC and 1.35 $\pm$ 0.13 for SC.
The single-epoch values of \citet{Ostro1992} are also consistent with ours at comparable longitudes
(Figure \ref{fig:dsn_results_ostro}). 
Moreover, the weighted mean and rms dispersions of our measurements of the circular polarization ratio $\mu_c$ = 1.44 $\pm$ 0.12 is in excellent agreement with the 1.43 $\pm$ 0.24 value obtained by \citet{Ostro1992}. 
This agreement likely arises because most miscalibrations that may affect radar albedo estimates cancel out when taking the ratio of SC and OC albedos. Therefore, both their and our $\mu_c$ values are expected to be more robust than the radar albedo values.
There is also reasonable agreement between our estimate of the scattering law exponent, $m$ = 1.91 $\pm$ 0.31, and the value of 1.7 $\pm$ 0.4 published by \cite{Ostro1992}.
This comparison is restricted to the Goldstone data because the 1992 study did not include observations from the GBT. 

The close agreement between our results and those of \citet{Ostro1992}
  suggests that Europa has not experienced global, extensive resurfacing processes that would produce detectable changes in its radar properties
between 1987--1991 and 2011--2024.

\begin{figure}[ht!]
\plotone{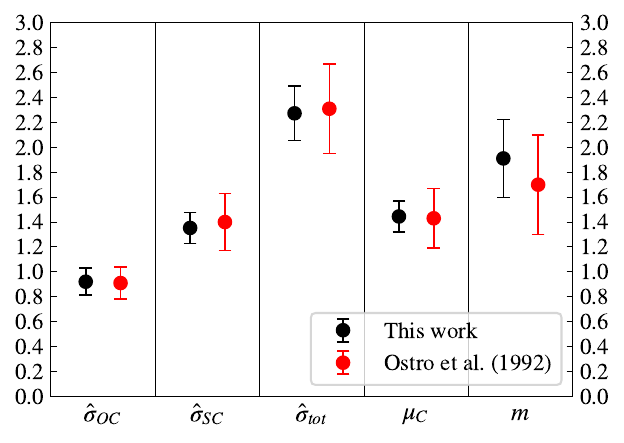}
\caption{Estimates of $\hat\sigma_{\rm OC}$, $\hat\sigma_{\rm SC}$, $\hat\sigma_{\rm tot}$, $\mu_c$, and $m$ from Goldstone 3.5 cm observations in this study and \citet{Ostro1992}'s study. For $\hat\sigma_{\rm OC}$, $\hat\sigma_{\rm SC}$, and $\hat\sigma_{\rm tot}$, we used unweighted means and rms dispersions, whereas for $\mu_c$ and $m$, we used weighted means and rms dispersions. Our values are in black and \citet{Ostro1992}'s values are in red.
\label{fig:comparison_to_1992}}
\end{figure}

In Table~\ref{tab:1992_comparison} and Figure~\ref{fig:comparison_to_1992}, we list and visualize the means and rms dispersions of our values with \citet{Ostro1992}'s values for $\hat\sigma_{OC}$, $\hat\sigma_{SC}$, $\hat\sigma_{tot}$, $\mu_c$, and $m$. It is immediately apparent that our values are very consistent with \citet{Ostro1992}'s.

\begin{deluxetable}{ccc}[t!]
\tabletypesize{\scriptsize}
\tablewidth{0pt}
\label{tab:1992_comparison}
\tablecaption{Comparison of disk-integrated radar properties}
\tablehead {
  \colhead{Dataset} & \colhead{This work} & \colhead{\citet{Ostro1992}}} %
\startdata
$\hat\sigma_{\rm OC}$ &  0.92 $\pm$ 0.11 & 0.91 $\pm$ 0.13   \\
$\hat\sigma_{\rm SC}$ &  1.35 $\pm$ 0.13 & 1.40 $\pm$ 0.23  \\
$\hat\sigma_{\rm tot}$ &  2.27 $\pm$ 0.22  & 2.31 $\pm$ 0.36 \\
$\mu_c$ & 1.44 $\pm$ 0.12  & 1.43 $\pm$ 0.24 \\
$m$ & 1.91 $\pm$ 0.31  & 1.7 $\pm $0.4 \\
$N_{\rm obs}$ & 28 & 7 \\
\enddata
\tablecomments{Estimates of $\hat\sigma_{\rm OC}$, $\hat\sigma_{\rm SC}$, $\hat\sigma_{\rm tot}$, $\mu_c$, and $m$ from Goldstone 3.5 cm observations in this study and \citet{Ostro1992}'s study.  For $\hat\sigma_{\rm OC}$, $\hat\sigma_{\rm SC}$, and $\hat\sigma_{\rm tot}$, we used unweighted means and rms dispersions, whereas for $\mu_c$ and $m$, we used weighted means and rms dispersions. $N_{\rm obs}$ lists the total number of epochs that went into calculating the means and rms dispersions.}
\end{deluxetable}

\subsection{Disk-resolved properties}
Because our dataset includes the most longitudinally comprehensive set of radar measurements of Europa to date, we are able to investigate how radar albedo and circular polarization ratio vary as a function of longitude (Figures~\ref{fig:oc_results_long}, \ref{fig:sc_results_long}, and \ref{fig:pol_ratio_long}) or geological unit. If a prominent radar feature were present on Europa, we would expect to observe a substantial difference in albedo or polarization ratio at the corresponding longitude.
However, visual inspection of the figures suggests that the albedo and polarization ratio values vary little with longitude, which motivated us to test the hypothesis that 
the measured albedos
are consistent with a single constant value. For each polarization and each telescope, we therefore adopted the null hypothesis that the
data are consistent with their weighted mean.
To evaluate whether this hypothesis can be rejected, we performed a chi-squared ($\chi^2$) goodness-of-fit test
\begin{equation} \label{eq: chi-square}
  \chi^2 = \sum_i \frac{(x_i - \bar{x})^2}{\sigma_i^2},
\end{equation}
where we set $x_i$ to the radar albedo
at epoch $i$, $\bar{x}$ to the weighted mean of 
the radar albedos for that polarization and telescope, 
and $\sigma_i$ to the total uncertainty for that epoch.  
The number of degrees of freedom (DOF) is the number of available epochs minus one.
We performed a similar test for polarization ratios, where in this case $\sigma_i$ is the uncertainty calculated by Equation \ref{eq:error_propagation}.

The results (Table \ref{tab:hypothesis_testing}) show that the $\chi^2$ values for the albedos and the Goldstone polarization ratio are well below the critical $\chi^2$ values, indicating that the null hypotheses cannot be rejected. These albedo and polarization ratio measurements are therefore statistically consistent with their weighted means, in agreement with \citet{Ostro1992}, who also reported that disk-integrated properties have little dependence on rotation phase. However, the $\chi^2$ value for the GBT polarization ratio exceeds the critical value, suggesting that the GBT polarization ratios
  exhibit statistically significant intrinsic variability.  We note, however, that one false positive is expected to occur in $\sim$20 hypothesis tests conducted at the 95\% confidence level. We conducted 6 separate hypothesis tests. We discuss the possible sources of this variability in the next subsection.

\begin{deluxetable}{ccccc}[t!]
\tabletypesize{\scriptsize}
\tablewidth{0pt}
\tablecaption{Testing the consistency of radar albedos and circular polarization ratios \label{tab:hypothesis_testing}}
\tablehead {
  \colhead{Dataset} & \colhead{Weighted mean} & \colhead{$\chi^2$} & \colhead{DOF}  & \colhead{$\chi^2_{\rm crit}$}}
\startdata
Goldstone $\hat\sigma_{\rm OC}$ & 0.90 & 9.21 & 27 & 40.11  \\
Goldstone $\hat\sigma_{\rm SC}$ & 1.33 & 5.70 & 27 & 40.11  \\
Goldstone $\mu_c$ & 1.44 & 26.86 & 27 & 40.11 \\
GBT $\hat\sigma_{\rm OC}$ & 0.80 & 14.23 & 25 & 37.65  \\
GBT $\hat\sigma_{\rm SC}$ & 1.19 & 7.88 & 25 & 37.65 \\
GBT $\mu_c$ & 1.46 & 45.11 & 25 & 37.65 \\
\enddata
\tablecomments{Testing the null hypothesis that radar albedo and polarization ratio values are consistent with their weighted mean. The chi-squared ($\chi^2$) and the number of degrees of freedom (DOF) are listed for each fit of the weighted mean to the data. %
  Also listed is the critical value $\chi^2_{\rm crit}$ for a confidence level of 95\%.  To reject the null hypothesis, $\chi^2 > \chi^2_{\rm crit}$ is required.
With the exception of the GBT $\mu_c$ measurements, all $\chi^2$ values are well below $\chi^2_{\rm crit}$, indicating that the null hypotheses cannot be rejected. Only the GBT $\mu_c$ measurements are inconsistent with
  a constant value at the 95\% confidence level.}
\end{deluxetable}

\subsection{Hemispherical dichotomy}

Although \citet{Ostro1992} found little dependence of disk-integrated properties on rotation phase, they were able to point out some radar features on the icy satellites by constructing radar reflectivity maps from their more sensitive Arecibo 13-cm observations. For instance, they found out that the equatorial region on Europa's leading side (centered on 90 degrees west longitude) is more radar-dark than the equatorial region on Europa's trailing side (centered on 270 degrees west longitude).
A
potential explanation for this observation invokes the
possible existence of sublimation-sculpted, east-west aligned, bladed ice structures known on Earth as penitentes.
The existence of these structures has been proposed by \citet{Hobley2018Formation}, but there is some debate about their viability on Europa \citep{Hand2020-jy}. 
If such structures do indeed exist, they would develop primarily in the equatorial region and reach dimensions much larger than the wavelength \citep{Hobley2018Formation}. As such, they may be capable of efficiently absorbing incoming radar waves in the equatorial regions, with possibly a greater impact on the OC polarization because of large incidence angles with respect to the penitente blades.
\citet{Hobley2018Formation} further suggested that the leading hemisphere appears more radar-dark because the trailing hemisphere is more heavily contaminated by particulates transported through Jupiter’s magnetosphere, which may suppress penitente formation.

Visual inspection of Figures \ref{fig:oc_results_long} and \ref{fig:sc_results_long} suggests that the radar albedos of the trailing side may be systematically higher than the radar albedos of the leading side. To test this apparent hemispherical dichotomy, we performed two-sample t-tests on the radar albedo measurements. All values with subradar longitudes between 0 and 180 degrees were grouped into one sample, while values with subradar longitudes between 180 and 360 degrees were grouped into a second sample. The null hypothesis is that the two independent samples have identical mean values. Assuming a $95\%$ confidence level, a p-value $\leq$ 0.05 would allow us to reject the null hypothesis. 
For Goldstone data, the p-values for the OC and SC polarizations are 0.01 and 0.22, respectively. For GBT data, the p-values for the OC and SC polarizations are 0.06 and 0.43, respectively. Only the Goldstone OC albedo yields a statistically significant result,
  suggesting the presence of two distinct means in the albedo measurements. The remaining datasets do not reach statistical significance. Nevertheless, the OC albedo yields lower p-values than the SC albedo for both telescopes, suggesting that
  a dichotomy in scattering properties is more likely in the OC polarization, possibly related to the presence of penitente-like structures or other scattering mechanisms that affect OC more strongly than SC.
The ultimate cause of the apparent leading-vs-trailing side dichotomy may be related to surface deposition, sputtering, or impact gardening in Jupiter's magnetosphere.

We also performed a two-sample t-test on the circular polarization ratios. If the dichotomy is apparent in the OC echoes, it should also be reflected in the polarization ratios. The results yield p-values of 0.02 and 0.03 for the Goldstone and GBT polarization ratios, respectively. This result confirms that, at the 95\% confidence level, the polarization ratio measurements from both telescopes are consistent with a leading-vs-trailing side dichotomy in radar scattering properties.

\subsection{New bounds on the coherent backscatter effect}

All of our results show that the radar albedo values for Europa (both SC and OC) are much greater than the
$\sim$0.1 values recorded for terrestrial planets
and asteroids, and the circular polarization ratios $\mu_c$ are also all greater than 1, which is anomalous \citep{RevModPhys.65.1235}. These scattering properties are consistent with the coherent backscatter opposition effect (CBOE), as described in Section \ref{sec:introduction}.
Because the CBOE is predicted to have a strong dependence on the transmitter-target-receiver ($\beta$) angle, bistatic radar observations at a variety $\beta$ angles could provide more robust evidence that the CBOE is indeed responsible for the radar scattering properties of the icy satellites.  With bistatic radar observations at our disposal, we have an opportunity to observe the behavior of the CBOE peak at a variety of $\beta$ angles.

\begin{figure}[ht!]
\plotone{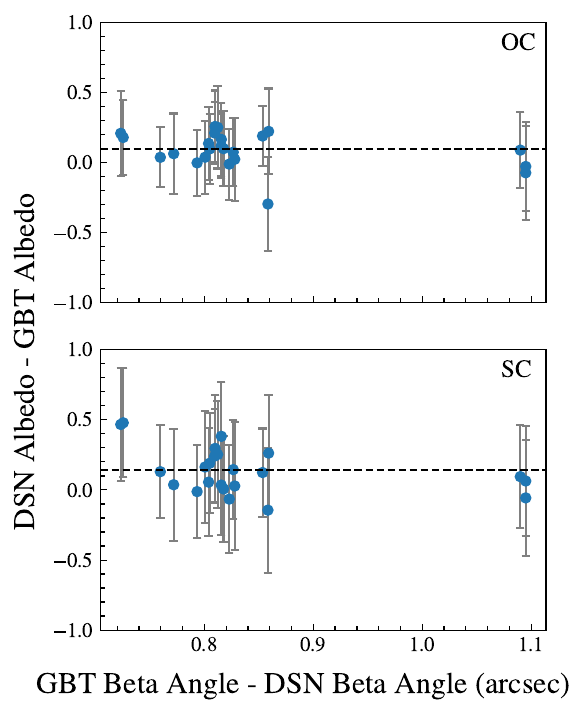}
\caption{Difference between the Goldstone and GBT radar albedos as a function of the difference in their $\beta$ angles, showing values for OC (top) and SC (bottom).  The error bars were calculated using the standard error propagation formula and the total uncertainties of the OC and SC albedo.
The covariance between the Goldstone and GBT albedos are calculated in the same way as described in Section \ref{sec:results}.
The dashed horizontal lines are at 0.10 (OC) and 0.14 (SC), the weighted means of the differences.
\label{fig:cboe_diff}}
\end{figure}

\begin{figure*}[t!]
\centering
\includegraphics[width=\textwidth,height=0.35\textheight,keepaspectratio]{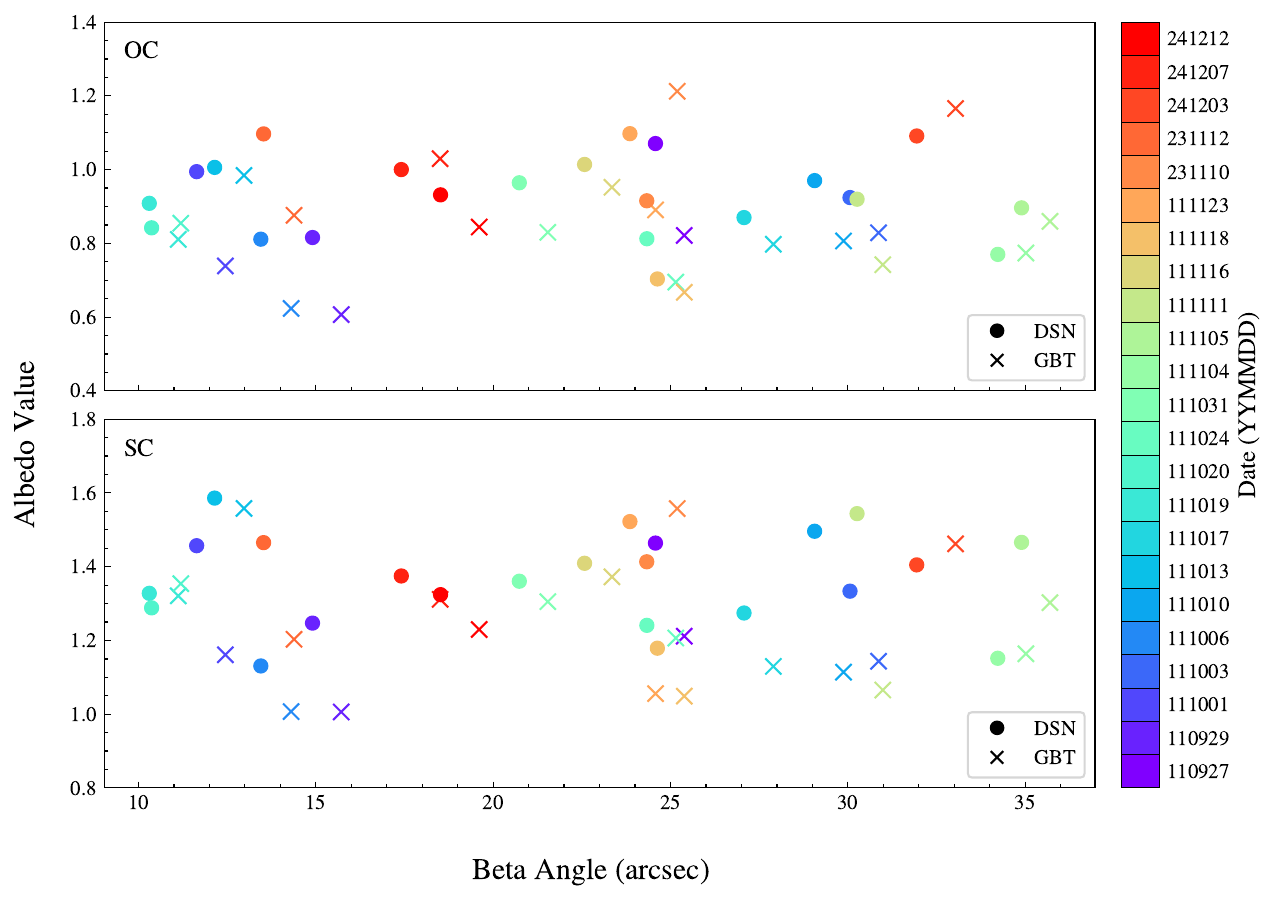}
\caption{ Measured OC (top panel) and SC (bottom panel) albedo values from Goldstone (dots) and GBT (crosses) plotted as a function of $\beta$ angle. Each epoch is marked by a distinct color. For both telescopes and both polarizations, the albedo values remain approximately constant across the range of $\beta$ values. However, within individual epochs, the GBT albedo values appear systematically lower than the corresponding Goldstone values, likely reflecting systematic biases in the telescope sensitivity estimates discussed earlier in this section. 
\label{fig:cboe_full}}
\end{figure*}

CBOE theory predicts that the intensity of the received light in the SC polarization falls off like a bell curve as a function of $\beta$.
The CBOE can increase the SC intensity by up to a factor of 2 when $\beta$ equals zero. At greater $\beta$ angles, the increase due to the CBOE is smaller \citep{hapk90}.
For our observations, because GBT is located to the East of Goldstone, the Goldstone to GBT light rays always trace out a larger $\beta$ angle than the Goldstone to Goldstone light rays. Therefore, for each epoch, the Goldstone observations should give us a higher received intensity than the GBT observations. To verify whether the GBT observations exhibit a drop in intensity due to CBOE, we plotted the difference between Goldstone and GBT albedos with respect to the {\em difference} in $\beta$ angles between the two telescopes (Figure \ref{fig:cboe_diff}).
If our observations sampled the descending part of the CBOE gain curve, we would expect the SC albedo difference to increase as the {\em difference} in $\beta$ angles between the two telescopes increases, because the SC albedo measured at GBT should exhibit a larger decrease. We would also expect the OC albedo differences to remain relatively constant because the CBOE is expected to affect the OC intensity to a much lesser extent than the SC intensity. According to Figure \ref{fig:cboe_diff}, however, the differences for our SC values do not
appear to increase as a function of the difference in $\beta$ angle, and our OC differences seem to remain relatively constant. 
The weighted means and rms dispersions of our OC and SC differences are 0.10 $\pm$ 0.12 and 0.14 $\pm$ 0.16 respectively, indicating that there is a slight loss of intensity at GBT of about $10\%$, but this loss is observed in both polarizations and
does not appear to vary as a function of $\beta$ angle difference.
To further test the hypothesis of constancy of albedo difference as a function of $\beta$ angle difference, we performed a chi-squared ($\chi^2$) goodness-of-fit test on the
albedo differences. For each polarization, we adopted the null hypothesis that the
albedo differences are consistent with their weighted mean. To evaluate whether this hypothesis can be rejected, we calculated the $\chi^2$ using Equation \ref{eq: chi-square}, where we set $x_i$ to the radar albedo difference at epoch $i$, $\bar{x}$ to the weighted mean, and $\sigma_i$ to the difference's propagated uncertainty. The number of degrees of freedom is 22, because we are using
only
the epochs with reliable %
data at both Goldstone and GBT.

We found $\chi^2$ values of 3.99 for OC and 3.83 for SC, respectively. Assuming a 95\% confidence level, a $\chi^2$ value $>$ 33.92 is required to reject the null hypothesis, and our chi-square values are well below the threshold, indicating that our
albedo differences are consistent with their weighted means. 
The fact that we do not observe a change in SC albedo 
difference 
across $\beta$ angle difference 
and that the albedo differences 
of both polarizations are consistent with $\approx0.1$ suggests that the loss of intensity at GBT is likely due to systematic biases in estimates of telescope sensitivity
at either DSN or GBT, and not due to the CBOE.

We also plotted the radar albedo values for all epochs as a function of $\beta$ angle (Figure \ref{fig:cboe_full}).
If our observations sampled the descending part of the CBOE gain curve, we would expect the albedo values
to decrease with increasing $\beta$ angle in a bell-shaped profile, but instead we
observe that the albedo values remain approximately constant across the range of $\beta$ angles. This result further supports the conclusion that we are not observing any measurable loss in intensity due to the CBOE effect and suggests that even the largest $\beta$ angle, 36 arcsec, lies well within the main lobe of the CBOE peak. Consequently, the width of the CBOE peak is likely much larger than 36 arcsec.

The width of the CBOE peak at half maximum is predicted to be $\sim\lambda/2\pi L$, where $\lambda$ is the wavelength and $L$ is the mean depth the photons can diffuse inside the icy shell before being absorbed \citep{mackintosh1988}. Because we have established a lower bound (36 arcsec) for the width of the CBOE peak, we can deduce an upper bound for $L$.  For $\lambda = 0.035$~m, we find $L < 32$ m,
which corresponds to $\sim$1000 wavelengths. This constraint provides insights into the potential penetration depth of 3.5 cm wavelength radio waves within the icy shells of Europa and may help interpret the radar measurements from NASA’s Europa Clipper and ESA's JUICE mission, which carry ice-penetrating radar instruments, albeit at much longer wavelengths of $\sim$33 m.

\subsection{Alternative Goldstone antenna gain values}
As mentioned in the previous
section, we suspect that the apparent
discrepancy between DSN and GBT cross-section measurements is due to systematic biases in telescope sensitivity estimates, either at DSN or GBT.
One plausible source for this bias is a miscalibration of the DSS-14 antenna gain appropriate for X-band radar observations.
Throughout this paper,
we have used Goldstone antenna gain values calculated with
$G_0=74.23$~dBi,
as described in Section \ref{sec:observations}.  This $G_0$ value yields excellent agreement between our cross-section results and \citet{Ostro1992}'s results
because we adjusted the $G_0$ to match the 74 dBi gain used in \citet{Ostro1992}'s study. 
However, it is possible that the 74 dBi gain used in \citet{Ostro1992}'s study is in error.  We found that another choice of $G_0$ yields excellent agreement between the Goldstone and GBT cross-section measurements, at the expense of a reduced agreement between our results and \citet{Ostro1992}'s results.

The DSN handbook \citep{jpl_dsn_handbook} lists $G_0=73.17$ dBi for transmission at 7145 MHz and $G_0=74.55$ dBi for reception at 8420 MHz for the telemetry cone of the DSS-14 tri-cone configuration.  Although radar observations use a different cone and may require different gain values, we adjusted these $G_0$ values to match our observing frequency (8560 MHz) with the standard gain equation $G=4\pi A_{\rm eff}/\lambda^2$, assuming that $A_{\rm eff}$ -- the effective area -- remains constant at X band. This adjustment yields $G_0=74.74$ dBi for transmit and $G_0=74.69$ dBi for receive. Using these revised $G_0$ values, we recalculated the radar albedo estimates for both Goldstone and GBT observations. With these alternate $G_0$ values, the differences between the GBT and Goldstone albedos have weighted mean differences of 0.0 for both the OC and SC polarizations.
For the 28 valid Goldstone measurements, we obtained an unweighted mean and root-mean-square (rms) dispersion of 0.74 $\pm$ 0.09  for the Goldstone OC albedo, and 1.08 $\pm$ 0.10 for the corresponding SC albedo.  For the 26 valid GBT measurements, we obtained 0.75 $\pm$ 0.13 for the GBT OC albedo, and 1.08 $\pm$ 0.13 for the corresponding SC albedo. The weighted mean and rms dispersion for the polarization ratios remains 1.47 $\pm$ 0.12 for Goldstone and 1.45 $\pm$ 0.14 for GBT, similar to the results from Section \ref{sec:results}.  The fact that the polarization ratios are unchanged is expected because systematic biases in telescope sensitivity estimates cancel out when computing the ratio of SC over OC cross-sections.  

In summary, we can adopt a DSS-14 gain value that reproduces the average gain used by \citet{Ostro1992}, in which case we obtain excellent agreement with their results but puzzling differences between Goldstone and GBT cross-section values.  Alternatively, we can adopt a gain value that applies to the DSS-14 X-band telemetry cone, in which case we obtain excellent agreement between Goldstone and GBT measurements but radar albedo values that are approximately 20\% lower than those reported by \citet{Ostro1992}.

\subsection{Future Work}
Futher analysis of these observations will be used to measure Europa's spin axis orientation and to place bounds on the amplitude of its longitude librations. These quantities can provide important information about the moment of inertia, rheology, and thickness of the icy shell, while also facilitating Europa Clipper and JUICE mission operations.

Additional observations scheduled between January 5--23, 2026 were cancelled following the September 16, 2025 DSS-14 azimuth over-rotation mishap that caused extensive damage to 11 hoses and 85 cables in the cable wrap structure~\citep{mishap}.

After GSSR returns to operations, observations with a Goldstone-Madrid baseline could help refine bounds on the width of the CBOE peak.
In addition, bistatic experiments that sample a wide range of $\beta$ angles may be possible with the X-band receivers onboard the Europa Clipper and JUICE spacecraft.  Although they cannot receive Goldstone's high-power, 8560 MHz transmissions, they can receive signals from the DSN's lower-power telemetry transmitters near 7.2 GHz~\citep{jpl_dsn_handbook}. 

\section{Conclusions} \label{sec:conclusion}

In this work, we described the most longitudinally comprehensive dataset of Europa's radar measurements to date, covering a substantially wider range of rotational phases than previous studies.  We measured disk-integrated radar properties, such as radar albedo, circular polarization ratio, and exponent of a diffuse scattering law. During the observations, the radar signal was transmitted from the Deep Space Network at Goldstone, and the echoes were received simultaneously at both Goldstone and the Green Bank Telescope. The 33 observations span 2011 to 2024.
After eliminating points with large statistical uncertainties and two additional outliers, 28 epochs remain in the Goldstone dataset and 26 epochs remain in the GBT dataset.  Disk-integrated properties are listed in Tables \ref{tab:Goldstone_results} and \ref{tab:gbt_results}, and reduced spectra for all epochs are shown in Appendices \ref{app:dsn_spectra} and \ref{app:gbt_spectra} as well as online (Appendix \ref{app:data_availability}).  They are in excellent agreement with \citet{Ostro1992}'s work, provided that we use a DSS-14 antenna gain value adjusted to match the 74 dBi average gain used by these authors.

We also performed a statistical analysis of radar albedos and circular polarization ratios as a function of subradar point longitude (Table~\ref{tab:hypothesis_testing}). The results indicate that the disk-integrated radar albedo of Europa is statistically consistent with a constant value, supporting the conclusion that disk-integrated radar albedo does not vary much with subradar longitude. 
However, the disk-integrated polarization ratio measured using GBT has statistically significant variability.

Radar reflectivity maps from \citet{Ostro1992} and our albedo values (Figure \ref{fig:oc_results_long} and \ref{fig:sc_results_long}) suggest that there may be a hemispherical dichotomy in radar scattering properties on Europa. Results from a two-sample t-test performed on our dataset indicate that the hemispherical dichotomy is more apparent in our OC polarization data than in our SC polarization data. 
More specifically, the Goldstone OC albedos and the circular polarization ratios measured at both telescopes reach the level of statistical significance required to
  reject the hypothesis of identical means for the leading and trailing sides of Europa at the $95\%$ confidence level.

Finally, our results are consistent with the coherent backscatter opposition effect (CBOE), which is currently the leading physical explanation for the unusual radar reflectivities of the icy satellites. Our bistatic observations allow us to place a lower bound of 36 arcsec on the width of the CBOE peak and an upper bound of 32~m on the mean photon penetrating depth of X-band radar waves within the icy shell.

\begin{acknowledgments}
TX was supported in part by NASA FINESST grant 80NSSC26K0201 and NSF Astronomy and Astrophysics Research Grant 2408493.  JLM was supported in part by NASA grants NNX12AG34G, 80NSSC19K0870, and NSF Astronomy and Astrophysics Research Grant 2408493.

We thank reviewers for judicious comments that improved the manuscript.  We thank 
Seth Truitt, Emmanuel Acuna, Jose Hernandez, Larry Snedeker, Lane Kast, Travis Tennyson, Dan Kelley, Marc Silva, Randy Pritchett, Jeff LaGrange, Randy Heuser, 
David Rose, Rob Taggart, Zack Graham, Aaron Lovato, Brandon Moore, Donna Stricklin, Greg Monk, Catherine Tounzen, Barry Sharp, David Curry, Ron Maddalena, Kevin Gum, and Anish Roshi
for assistance with the observations.

This work was supported in part by the Goldstone Solar System Radar (GSSR), which is operated by the Jet Propulsion Laboratory, California Institute of Technology, under a contract with the National Aeronautics and Space Administration. We gratefully acknowledge the support of the NASA Deep Space Network (DSN) staff in facilitating these observations.

The National Radio Astronomy Observatory and Green Bank Observatory are facilities of the U.S. National Science Foundation operated under cooperative agreement by Associated Universities, Inc.

\end{acknowledgments}

\clearpage

\appendix
\onecolumngrid
\section{Goldstone System Temperature Estimation} \label{app:sysT_estimation}

\begin{figure}[ht!]
\centering
\includegraphics[width=\textwidth,height=0.4\textheight,keepaspectratio]{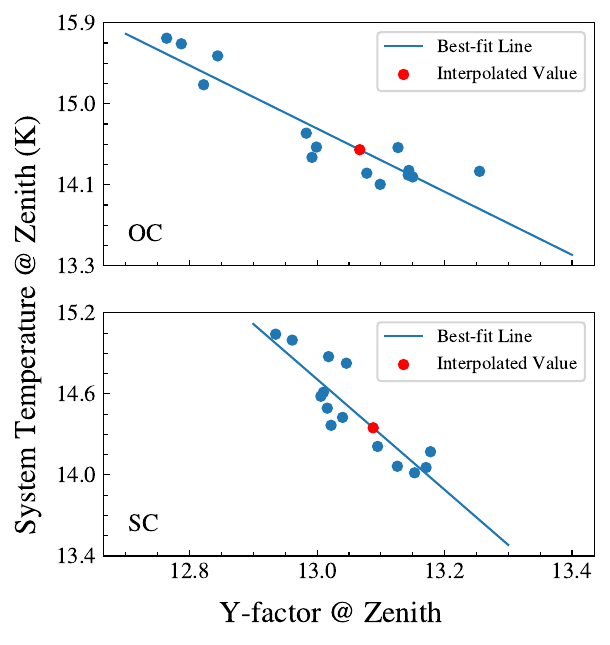}

\caption{Scatter plot of Goldstone zenith system temperatures in
  kelvins vs.\ Goldstone zenith y-factor. The top panel is for channel 1, or the OC signal, and the bottom panel is for channel 2, or the SC signal. The blue dots represent the epochs
  where we have both the zenith system temperature and the zenith y-factor measurements. The blue line is the least squares fit to the blue dots. The red dots represent the epoch (2011 NOV 23)
  for which we only have the zenith y-factor measurement.
  We interpolated its zenith system temperatures using the fitted linear relations.
\label{fig:sysT_calibration}}
\end{figure}

To estimate the Goldstone on-source system temperatures for certain epochs, we implemented the following procedure. First, for epochs that we have both the zenith system temperature and the zenith y-factor measurements, we plotted the zenith system temperature as a function of the zenith y-factor. The results showed a clear linear relationship, with a Pearson correlation coefficient equal to -0.92 for channel 1 and -0.88 for channel 2 (Figure \ref{fig:sysT_calibration}). The corresponding p-values for these coefficients are $2.84\times10^{-6}$ and $2.68\times10^{-5}$ respectively, indicating that such observed correlations are highly unlikely to occur by chance. Therefore, for the epoch (2011 NOV 23) with only the zenith y-factor measurement, we interpolated its corresponding zenith system temperatures using these linear relationships. The zenith system temperatures were then converted to on-source system temperatures by adding the noise contributions from Jupiter and from the additional Earth atmosphere traversed by the radar signal. The equations used to calculate these contributions were taken from the Deep Space Network (DSN) Telecommunications Link Design Handbook \citep{jpl_dsn_handbook}, and the results of the calculations are listed in Table \ref{tab:Goldstone_sysT_estimation}. The final deduced Goldstone on-source system temperatures are included in Table \ref{tab:Goldstone_observation}, along with the other measured on-source system temperatures. 

\begin{deluxetable*}{ccccccc}[t!]
\tabletypesize{\scriptsize}
\tablewidth{0pt}
\tablecaption{Estimates for the Missing Goldstone On-Source System Temperatures \label{tab:Goldstone_sysT_estimation}}
\tablehead {
\colhead{Date} & \colhead{Zenith OC $T_{\rm sys}$} & \colhead{Zenith SC $T_{\rm sys}$} & \colhead{$T_{\rm Jup}$} & \colhead{$T_{\rm atm}$} & \colhead{On-Source OC $T_{\rm sys}$} & \colhead{On-Source SC $T_{\rm sys}$} \\ [-7pt] \colhead{(UTC)} & \colhead{(K)} & \colhead{(K)} & \colhead{(K)} & \colhead{(K)} & \colhead{(K)} & \colhead{(K)} \\ [-7pt]
\colhead{(1)} & \colhead{(2)} & \colhead{(3)} & \colhead{(4)} & \colhead{(5)} & \colhead{(6)} & \colhead{(7)}
}
\startdata
2011 NOV 16  &  14.26  &  14.18  &  0.000309 & 0.613 & 14.87 & 14.79 \\
2011 NOV 23  &  14.51  &  14.32  &  0.000889  &  0.637 & 15.15 & 14.96 \\
\enddata
\tablecomments{Calculations of on-source system temperatures on the basis of zenith y factors or zenith system temperatures.  Columns (2) and (3) are the zenith system temperatures. The zenith $T_{\rm sys}$ for 2011 NOV 16 were measured, and the zenith $T_{\rm sys}$ for 2011 NOV 23 were interpolated (Figure \ref{fig:sysT_calibration}).
Column (4) is the estimation of Jupiter's contribution to $T_{\rm sys}$ per \citet{jpl_dsn_handbook}. Column (5) is the estimation of the additional Earth atmosphere's contribution to $T_{\rm sys}$. Columns (6) and (7) are the final estimated on-source system temperatures.}
\end{deluxetable*}

\onecolumngrid
\section{Baseline Fitting} \label{app:baseline_fitting}
\begin{figure*}[h!]
\plotone{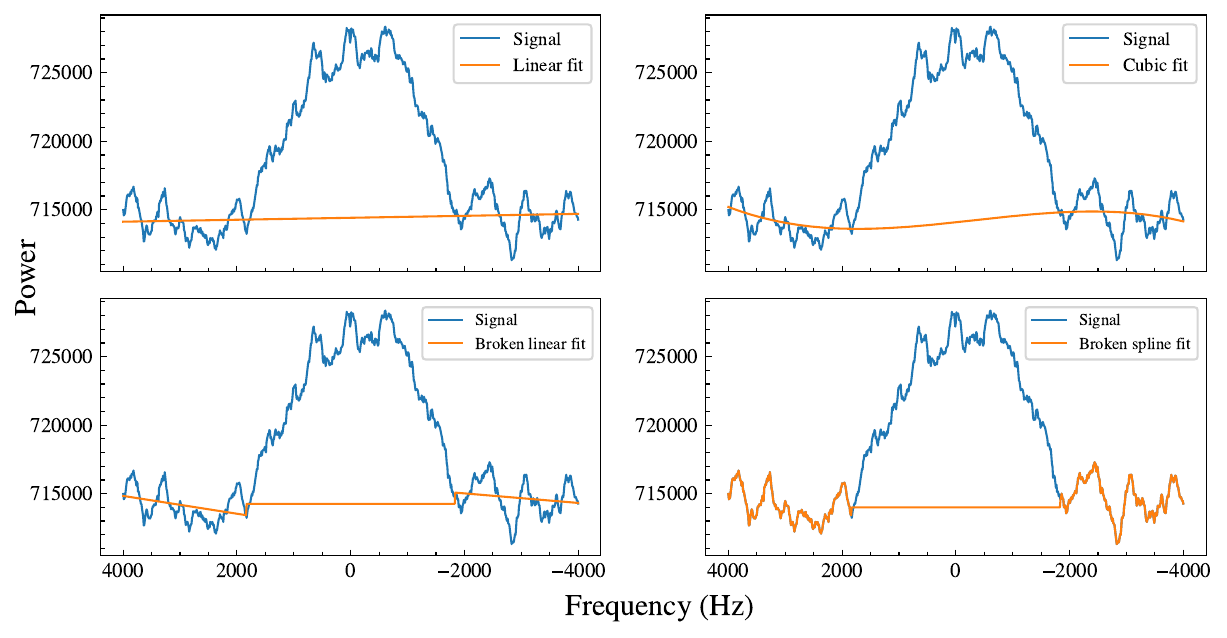}
\caption{Illustration of four different fits of the noise baseline for the Goldstone OC data at epoch 2011 OCT 03. The upper left panel shows the linear fit, the upper right panel shows the cubic fit, the lower left panel shows the broken linear fit, and the lower right panel shows the broken spline fit. For each panel, the blue line represents the data (after smoothing by the Savitszky-Golay filter) and the orange line represents the fitted noise baseline. For the bottom two panels, there are abrupt changes in the fitted baselines at the edges of the echo because we took the average of the endpoint of the left fit and the start point of the right fit for the baseline right beneath the echo, as described in the text.  
\label{fig:fitted_baselines}}
\end{figure*}

\begin{figure*}[t!]
\centering
\includegraphics[width=\textwidth,height=0.4\textheight,keepaspectratio]{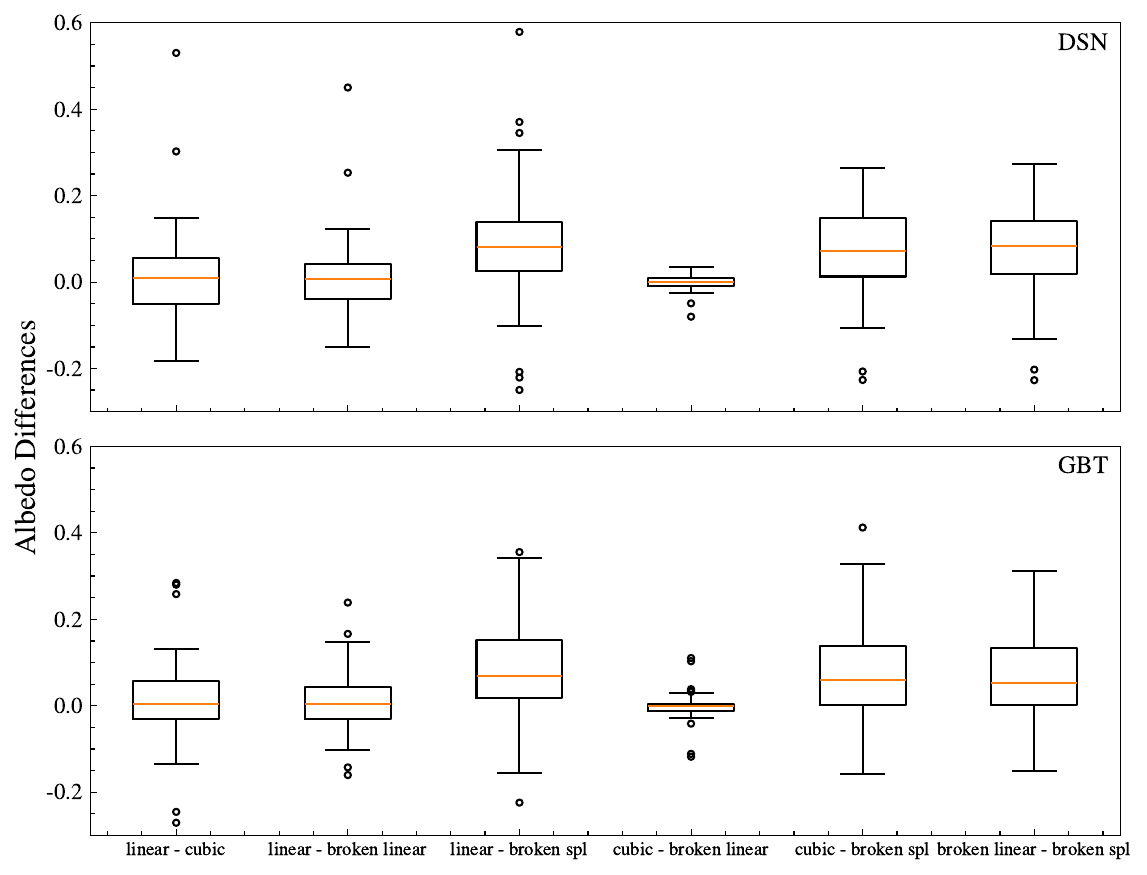}
\caption{Box-and-whisker plot showing pairwise albedo differences obtained with different baseline removal techniques. On the x-axis, ``linear'' represents the linear fit, ``cubic'' represents the cubic polynomial fit, ``broken linear'' represents the broken linear fit, and ``broken spl'' represents the broken spline fit. Thus, ``linear - cubic'' represents the difference between the albedos calculated with the linear fit and those calculated with the cubic polynomial fit.  Other x-axis labels follow the same pattern. The top panel is for the DSN antenna at Goldstone, and the bottom panel is for the GBT.  For both telescopes, the difference between the cubic fit and the broken linear fit is the smallest, with a median value of -0.0008. The orange line indicates the median, the box indicates the inter-quartile range (IQR), and the whiskers indicate the minimum and maximum (defined as the first quartile minus 1.5$\times$IQR and the third quartile plus 1.5$\times$IQR, respectively). The dots outside the box and whisker are the outliers. 
\label{fig:baseline_comparison}}
\end{figure*}

We experimented with several different methods for modeling the receiver noise baseline before ultimately adopting the broken linear fit. The methods we have tested include a linear fit, a cubic polynomial fit, a broken linear fit, and a broken spline fit. In the linear or cubic polynomial fit, a polynomial of the corresponding degree was fitted directly to the baseline using least squares. In the broken linear fit, linear polynomials were fitted separately to the baseline at the left and right side of the echo using least squares, and the portion of the baseline beneath the echo was set to a constant value equal to the average of the endpoint of the left fit and the start point of the right fit. The broken spline fit followed a similar procedure, but it used cubic splines instead of linear polynomials. Figure \ref{fig:fitted_baselines} shows the four different methods that we implemented.  The calculated albedo values resulting from the cubic polynomial and the broken linear fits are much more consistent with each other than those obtained using other methods (Figure \ref{fig:baseline_comparison}), suggesting that they are the two most reliable methods.
However, because the choice of a third-degree polynomial is somewhat arbitrary, we adopted the broken linear fit as our baseline model. 

\section{Circular Polarization Ratio Comparison} \label{app:pol_ratio_distribution}
\begin{figure*}[h!]
\centering
\includegraphics[width=\textwidth,height=0.4\textheight,keepaspectratio]{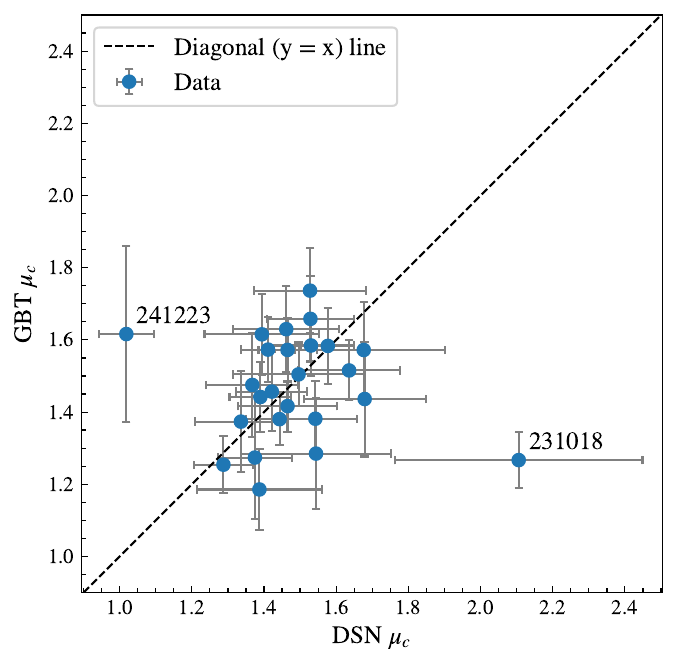}
\caption{Comparison of Goldstone and GBT circular polarization ratios ($\mu_c$) for 25 epochs where both quantities are available. The error bars represent the uncertainties in the polarization ratios listed in Tables \ref{tab:Goldstone_results} and \ref{tab:gbt_results}. The dashed diagonal (y = x) line denotes the expected one-to-one relation if the polarization ratios measured at the two telescopes were identical. The two outliers are labeled by their observing epoch (YYMMDD).  Possible explanations for the outliers include unusually large errors in receiver gain or system temperature estimates, undetected internal or external radio frequency interference, unidentified data-taking anomalies, imperfect baseline subtraction, or a combination of factors.
\label{fig:pol_ratio_distribution}}
\end{figure*}

Because systematic biases that may affect the albedo calculations largely cancel out when computing the ratio of SC and OC albedos, the circular polarization ratio ($\mu_c$) provides a more robust metric for comparing radar scattering properties. Figure \ref{fig:pol_ratio_distribution} compares the $\mu_c$ values derived from the Goldstone and GBT datasets.
  Because the $\mu_c$ values measured at Goldstone and GBT are expected to be similar, we expect the data points to lie close to the dashed one-to-one line.
  This behavior is indeed observed for the majority of the observations, but two epochs (2023 OCT 18 and 2024 DEC 23) stand out as clear outliers, exhibiting anomalous Goldstone $\mu_c$ values.

\clearpage

\section{Goldstone Spectra} \label{app:dsn_spectra}

\begin{figure*}[b!]  
    \centering
    \includegraphics[width=\textwidth,height=0.96\textheight,keepaspectratio]{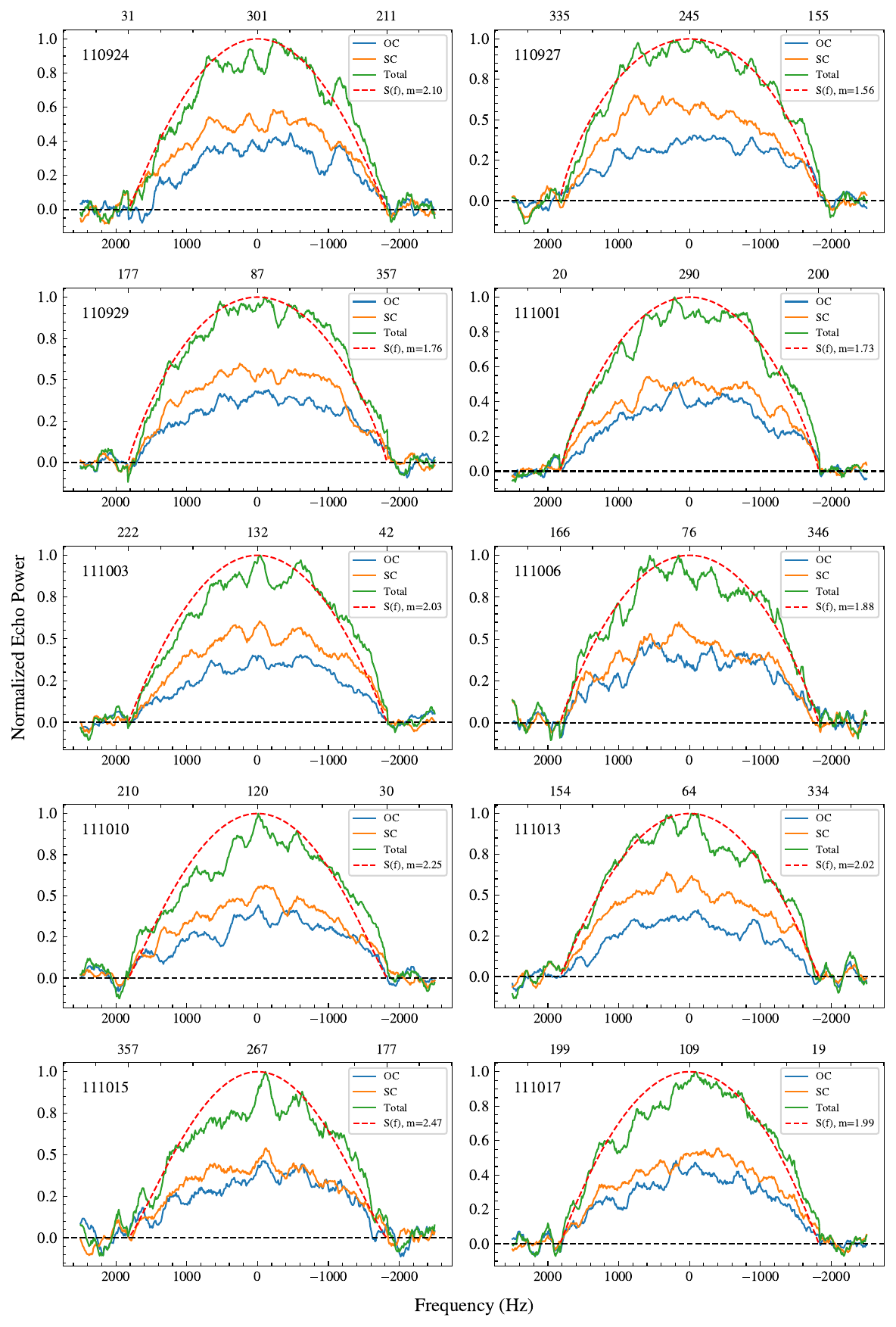}
    \label{fig:dsn_spectra_1}
\end{figure*}

\begin{figure*}[t!]  
    \centering
    \includegraphics[width=\textwidth,height=\textheight,keepaspectratio]{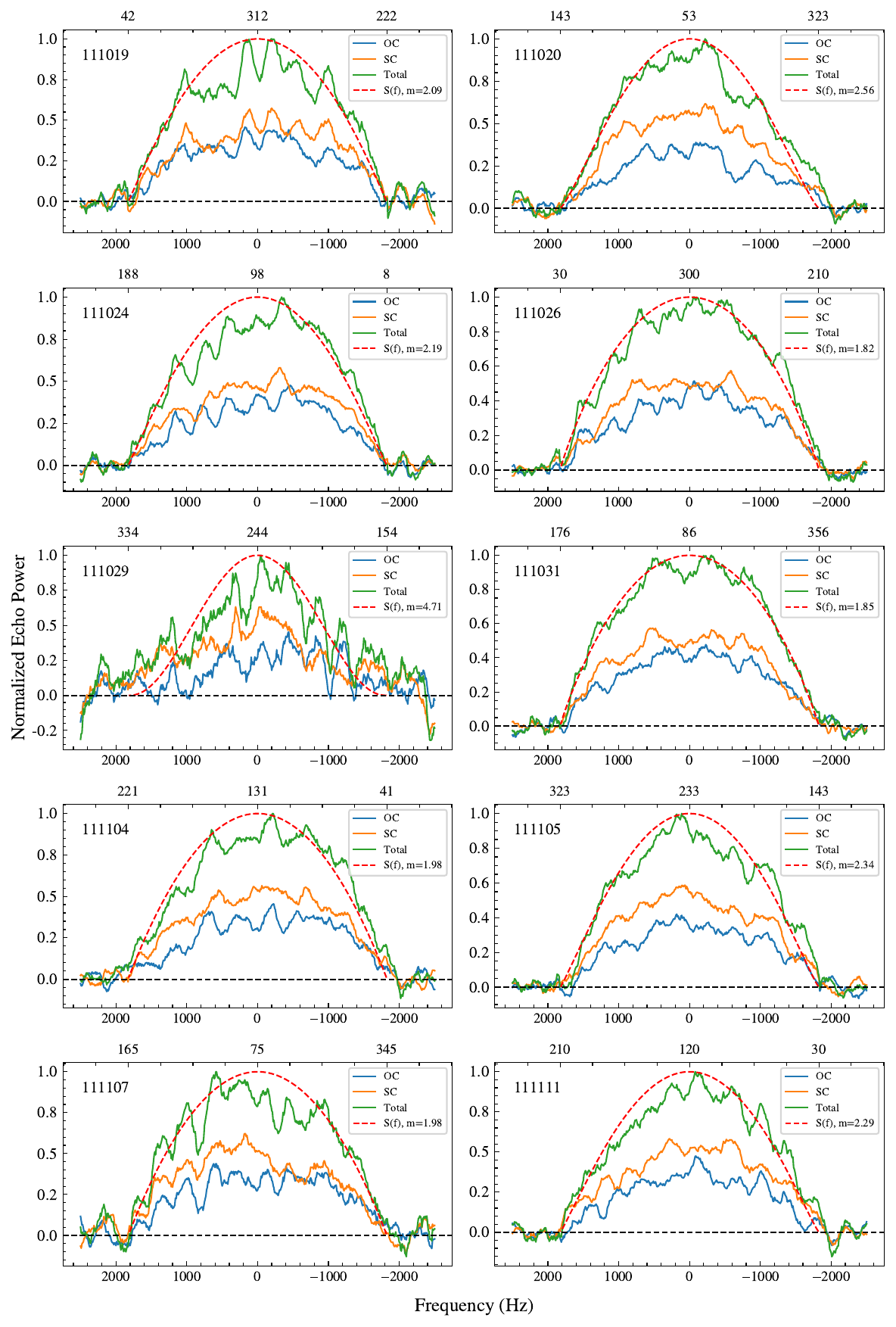}
    \label{fig:dsn_spectra_2}
\end{figure*}
\clearpage

\begin{figure*}[t!]  
    \centering
    \includegraphics[width=\textwidth,height=\textheight,keepaspectratio]{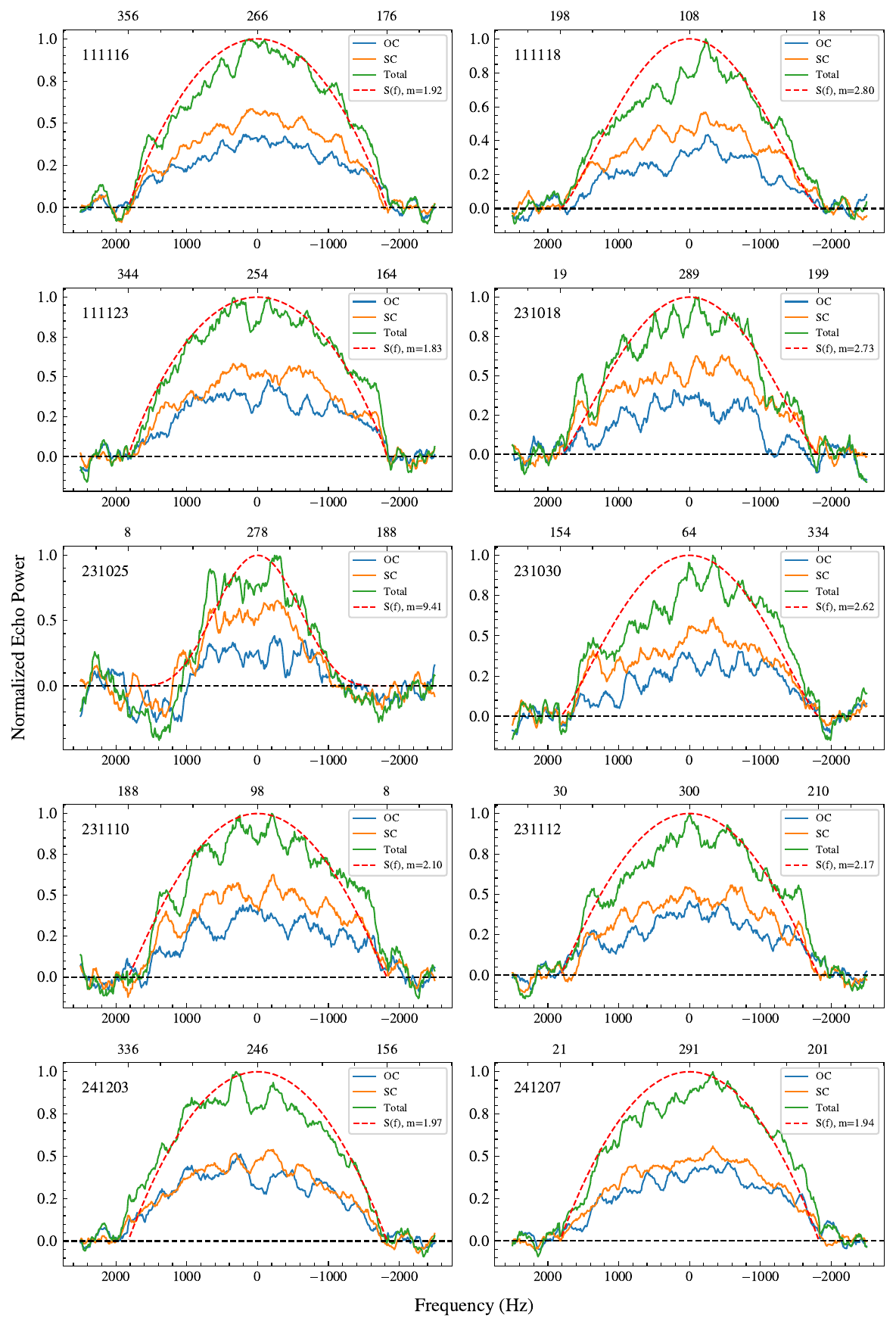}
    \label{fig:dsn_spectra_3}
\end{figure*}
\clearpage

\begin{figure*}[t!]  
    \centering
    \includegraphics[width=\textwidth,height=0.4\textheight,keepaspectratio]{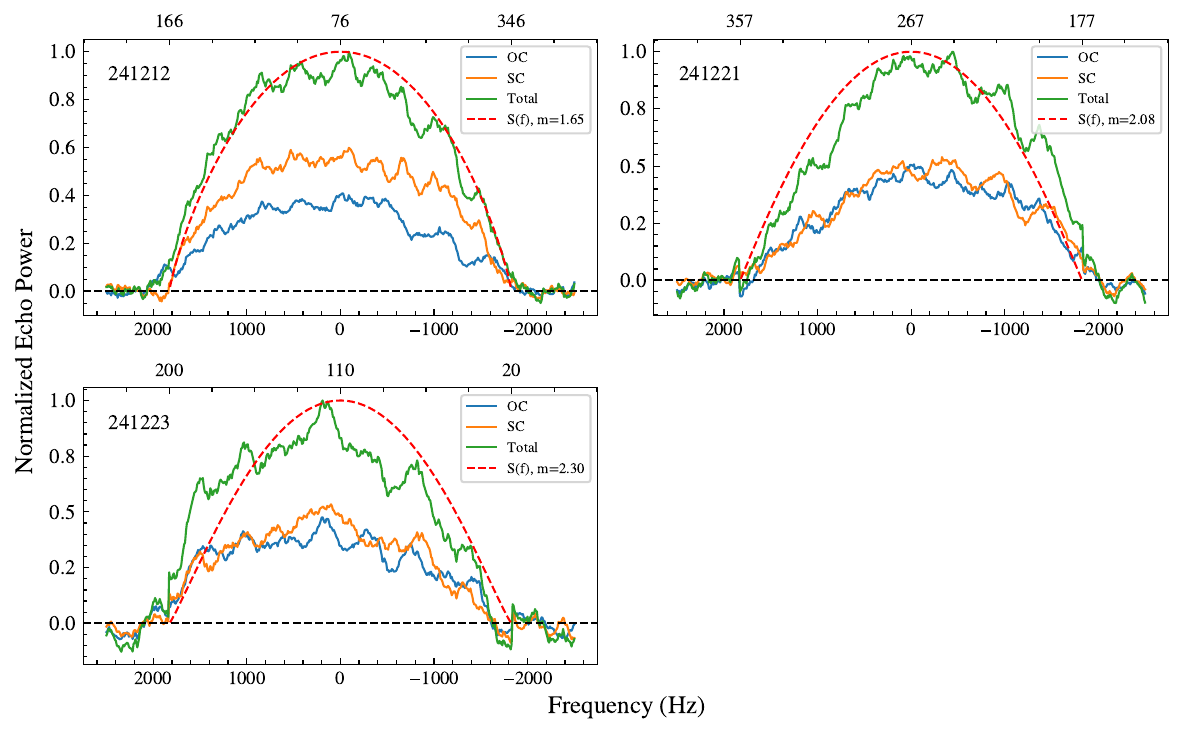}
    \caption{Reduced echo power spectra for Goldstone observations, with the epoch shown on each panel as YYMMDD. 
      The blue, orange, and green lines represent the OC, SC, and total (OC+SC) echo power, respectively. The power is normalized to the total echo power's maximum.   The red dashed line represents the scattering law $S(f)$ with the fitted exponent $m$.  The number at the top center of the panel is the subradar west longitude in degrees, whereas the numbers on the left and right are the west longitudes of the target's approaching (left) and receding (right) limbs, respectively. Note that frequency on the x-axis increases from right to left.
    \label{fig:dsn_spectra_4}}
\end{figure*}
\clearpage

\section{GBT Spectra} \label{app:gbt_spectra}

\begin{figure*}[b!]  
    \centering
    \includegraphics[width=\textwidth,height=0.95\textheight,keepaspectratio]{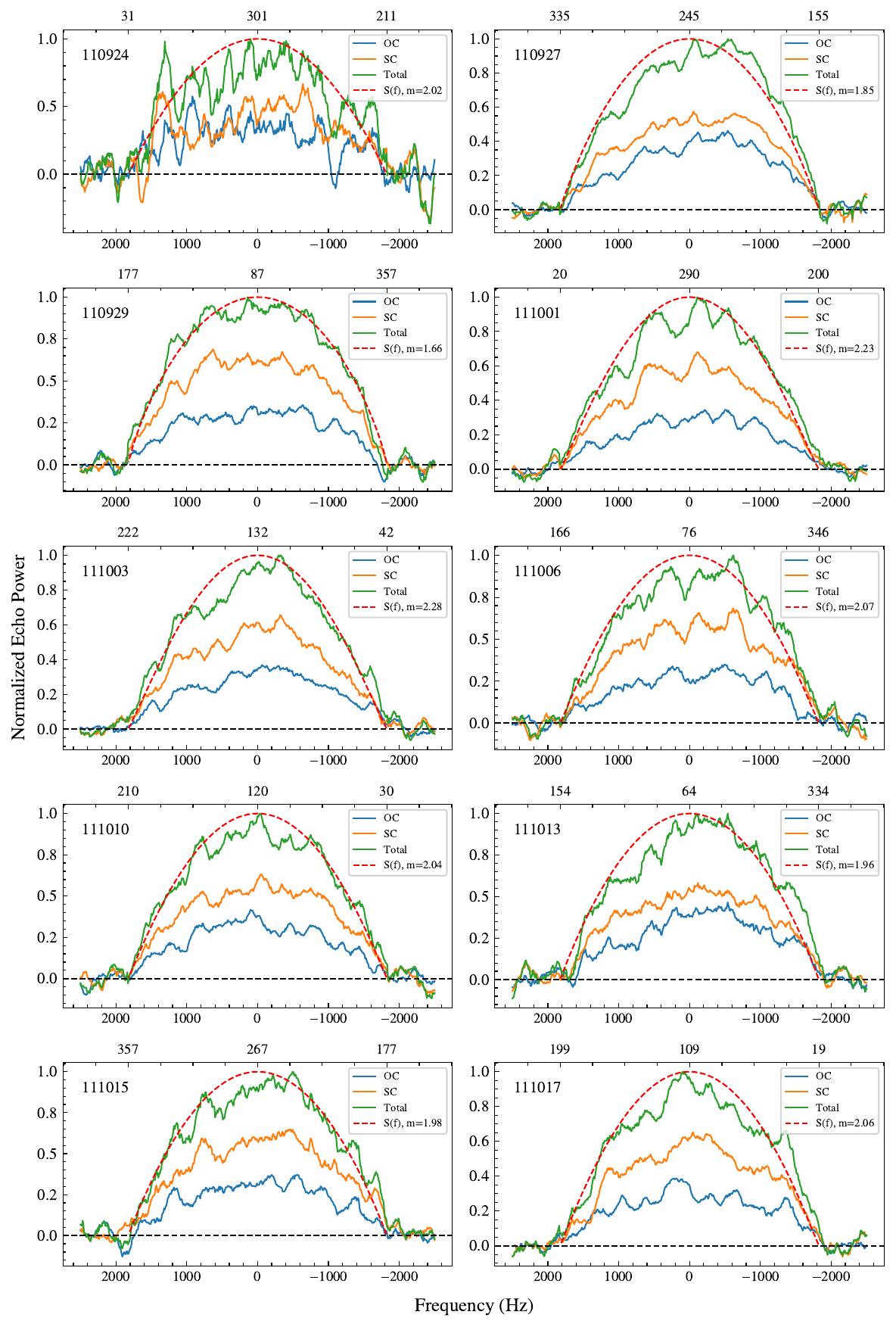}
    \label{fig:gbt_spectra_1}
\end{figure*}

\begin{figure*}[t!]  
    \centering
    \includegraphics[width=\textwidth,height=\textheight,keepaspectratio]{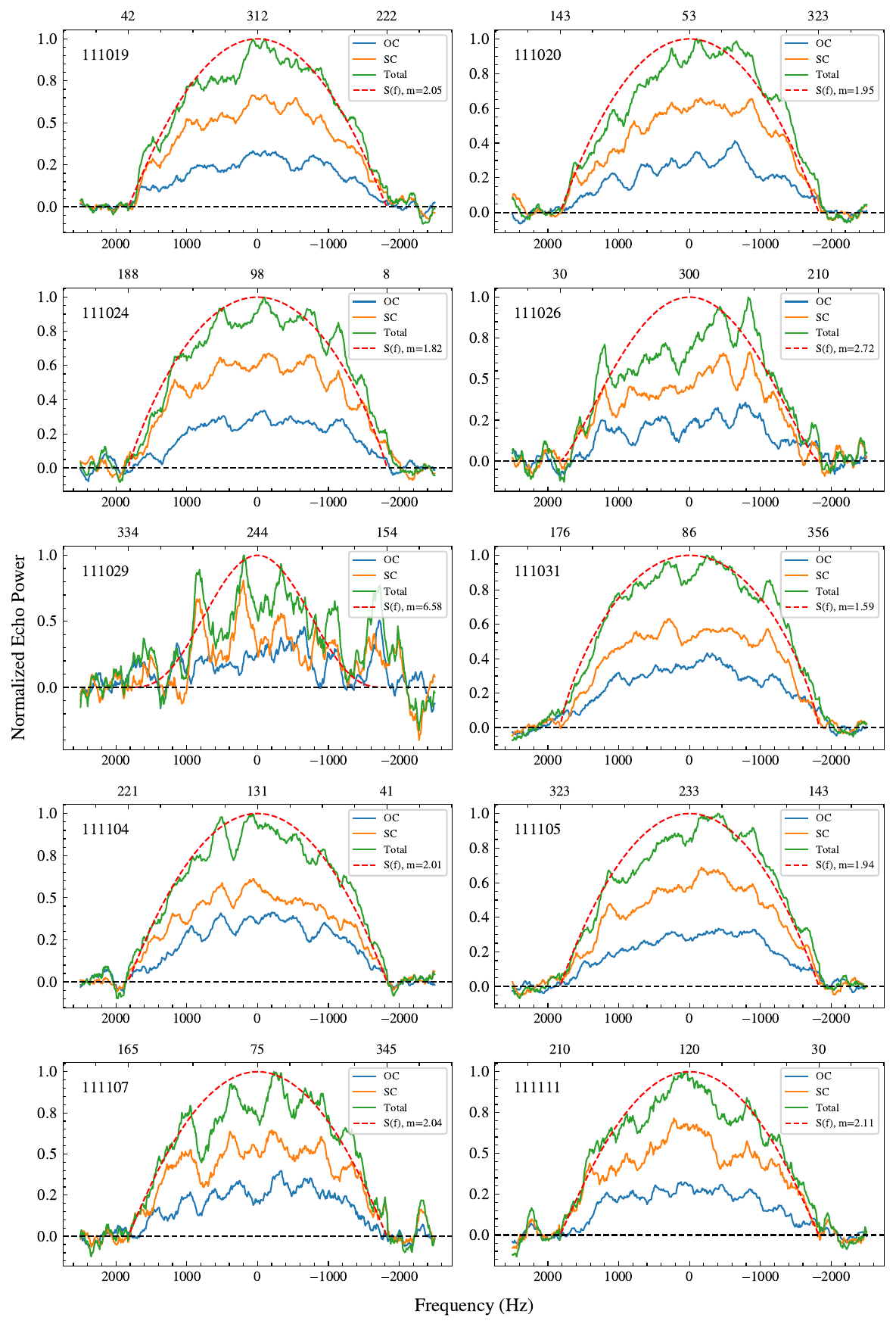}
    \label{fig:gbt_spectra_2}
\end{figure*}
\clearpage

\begin{figure*}[t!]  
    \centering
    \includegraphics[width=\textwidth,height=\textheight,keepaspectratio]{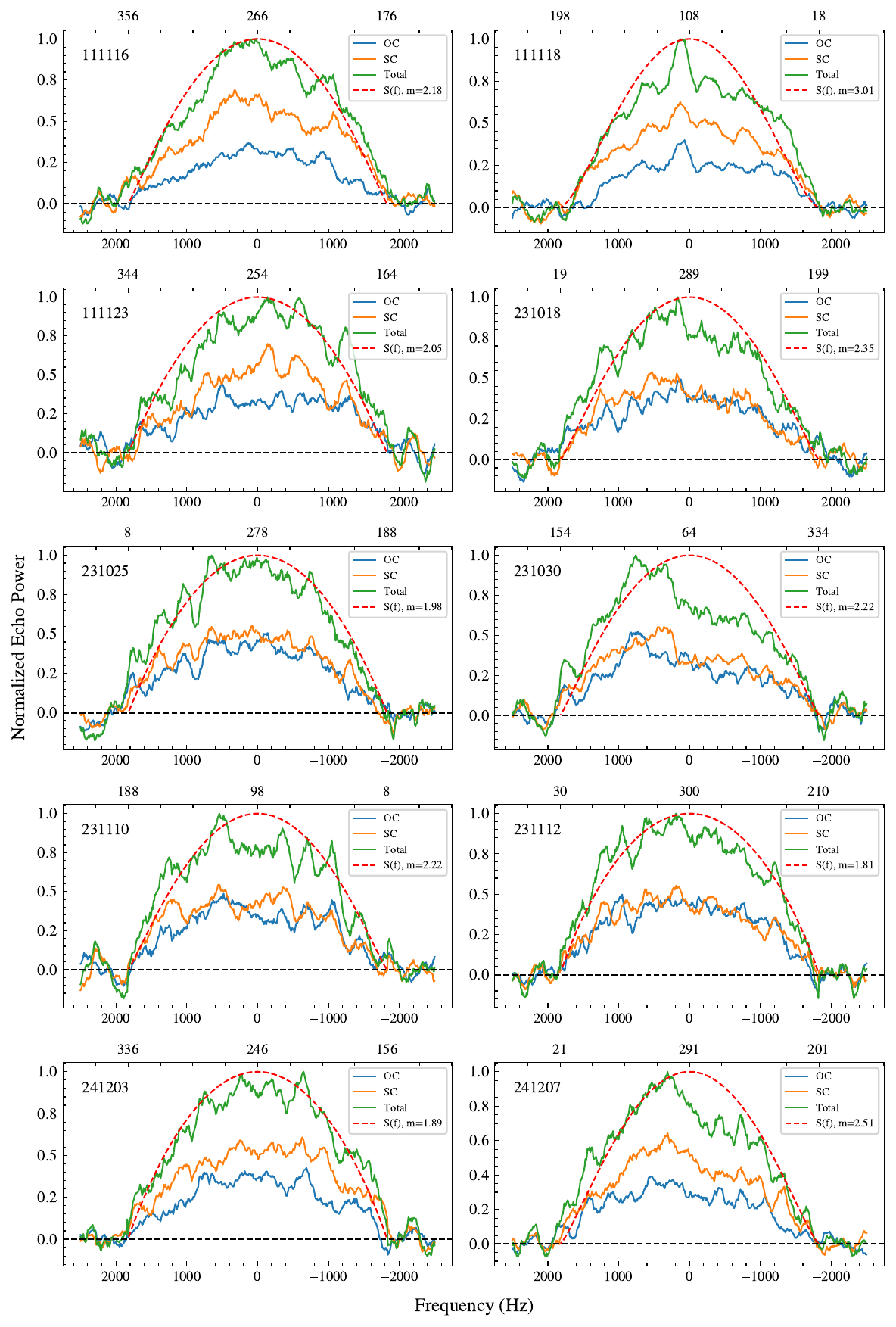}
    \label{fig:gbt_spectra_3}
\end{figure*}
\clearpage

\begin{figure*}[t!]  
    \centering
    \includegraphics[width=\textwidth,height=0.4\textheight,keepaspectratio]{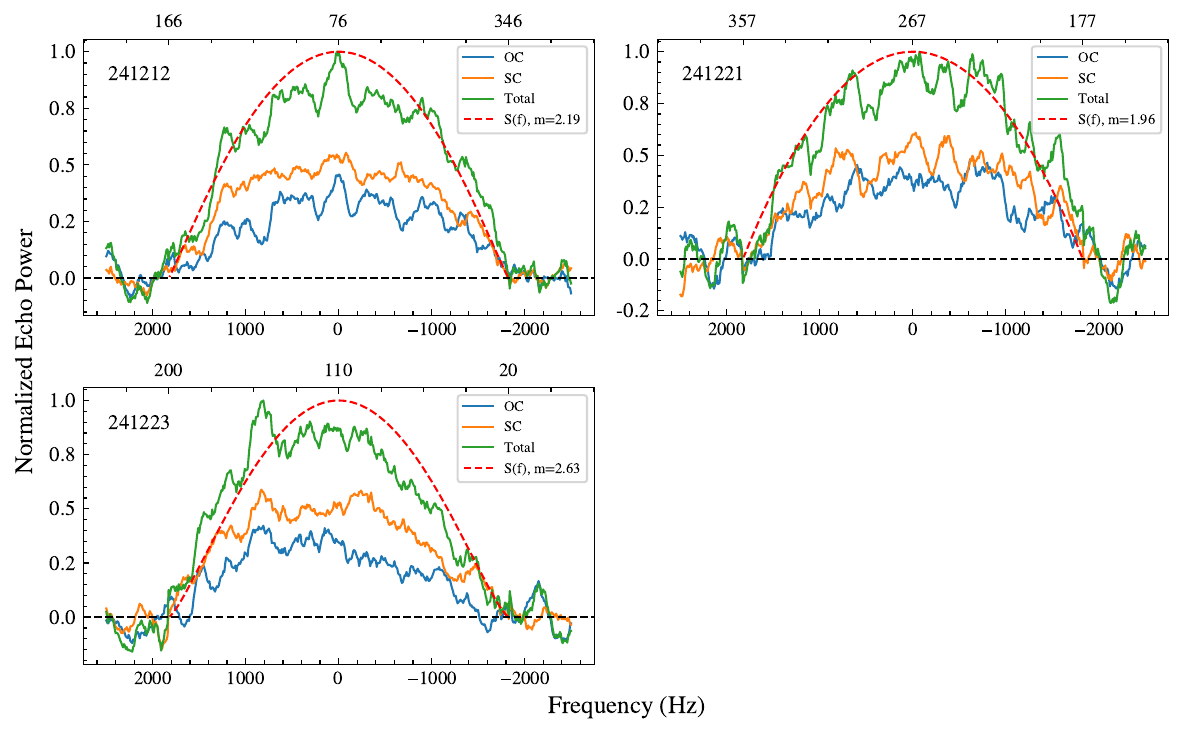}
    \caption{Reduced echo power spectra for the GBT observations, with the epoch shown on each panel as YYMMDD. 
      The blue, orange, and green lines represent the OC, SC, and total (OC+SC) echo power, respectively. The power is normalized to the total echo power's maximum.   The red dashed line represents the scattering law $S(f)$ with the fitted exponent $m$.  The number at the top center of the panel is the subradar west longitude in degrees, whereas the numbers on the left and right are the west longitudes of the target's approaching (left) and receding (right) limbs, respectively. Note that frequency on the x-axis increases from right to left.
      \label{fig:gbt_spectra_4}}
\end{figure*}

\section{Data Availability} \label{app:data_availability}
Goldstone and GBT power spectra are available in comma-separated-values (CSV) format as online supplementary material.  
Each file has 100 columns and 502 rows. The first column contains the frequencies, and the remaining 99 columns contain the total, OC, and SC normalized power for each one of the 33 epochs. The labels of the 99 columns include the epoch date in YYMMDD and the polarization (either total, OC, or SC).

Because the power spectra were normalized to the total echo power’s maximum, we also provide the normalization factor for each epoch in a separate CSV file, which also contains a scale factor to convert the power from arbitrary units to physical units.  This CSV file includes 34 rows and 7 columns. The first column lists the epoch date in YYMMDD, and the remaining columns list the normalization factors and the scale factors for OC and SC.  For two epochs, the GBT scale factors are listed as 0 because we do not have a record of system temperatures for these epochs, as explained in Section \ref{sec:observations}.

For each polarization, multiplication of the values of the spectra by the normalization factor yields the received raw power in arbitrary units per frequency bin. Then, multiplication of the received raw power by the scale factor yields the received power in units of watts per frequency bin. Summation of the powers in the frequency bins that fall within the limb-to-limb bandwidth yields the total echo received power in units of watts.

\clearpage
\bibliographystyle{aasjournalv7}
\bibliography{references}{}

\end{document}